# The Two-Dimensional Infinite Heisenberg Classical Square Lattice: Exact Theory and Experimental Results

Jacques Curély

*Laboratoire Ondes et Matière d'Aquitaine (LOMA), UMR 5798, Université de Bordeaux*
*351 Cours de la Libération, 33405 Talence Cedex, France*

We rigorously examine $2d$-infinite square lattices composed of classical spins isotropically coupled between first-nearest neighbors. Each local exchange Hamiltonian associated with each site $(i,j)$ is expanded on the basis of its eigenfunctions played by spherical harmonics $Y_{l_{i,j}, m_{i,j}}$. The corresponding eigenvalues are modified Bessel functions of the first kind. In the thermodynamic limit a numerical study allows one to select the higher-degree term of the characteristic polynomial associated with the zero-field partition function $Z_N(0)$. It is characterized by $m_{i,j} = 0$ and $l_{i,j} = l$ for the whole lattice so that a very simple exact closed-form expression can be then derived, thus permitting to express the free energy $F$ and the specific heat $C_V$, for any temperature. Then we report a thermal study of the basic term appearing in the higher-degree term of $Z_N(0)$. We show that it appears crossovers between two consecutive terms. Coming from high temperatures where the term characterized by $l = 0$ is dominant, near absolute zero, eigenvalues showing increasing $l$-values are more and more selected when the temperature is cooling down. When $T$ reaches zero, all the successive dominant eigenvalues become equivalent so that the critical temperature is $T_c = 0$ K. Owing to a similar method we derive an exact expression *valid for any temperature* for the spin-spin correlations, the correlation length $\xi$ and the susceptibility $\chi$ which show the same thermal crossovers as $Z_N(0)$. Near $T_c = 0$ K we obtain a diagram of magnetic phases which is similar to the one derived through a renormalization approach. The low-temperature magnetic properties are described in terms of universal parameters $k_B T/\rho_s$ and $k_B T/\Delta$ where $\rho_s$ and $\Delta$ are the spin stiffness and the $T=0$-energy gap between the ground state and the first exited one, respectively. We obtain the same expressions for $\xi$ and the surface density of $F$, $C_V$ and $\chi$ (for a compensated antiferromagnet) as the corresponding ones derived through a renormalization process, for each zone of the magnetic phase diagram. We show that, near $T_c = 0$ K, the lattice is composed of quasi rigid quasi independent Kadanoff blocks of length $\xi$ and magnetic moment $M(T)$, the unit cell moment, so that $\chi k_B T \sim \xi^2 M(T)^2$. Finally we compare experimental susceptibilities to the theoretical expression of $\chi$ derived in this article, for two types of $2d$-compounds (showing or not organic ligands inside and between sheets of $Mn^{2+}$ ions). We obtain a remarkable good agreement between the $J$-values of the exchange energy derived from the fits and the corresponding ones previously measured as well as a value of the Landé factor close to the theoretical one.

## I. INTRODUCTION

Since the middle of the eighties with the discovery of high-temperature superconductors [1], the nonlinear σ-model in 2+1 dimensions has known a new interest for it allows to describe the properties of two-dimensional ($2d$) quantum antiferromagnets such as $La_2CuO_4$[2-10]. Indeed these antiferromagnets, when properly doped, become superconductors up to a critical temperature $T_c$ notably high compared to other types of superconducting materials. In this respect, the nonlinear σ-model has been conjectured to be equivalent at low temperatures to the $2d$-Heisenberg model [11,12], which in turn can be derived from the Hubbard model in the large $U$-limit [13]. In a seminal paper, Chakravarty *et al.* [6] have studied this model using the method of one-loop renormalization group (RG) improved perturbation theory initially developed by Nelson and Pelkovits [14]. These authors have related the σ-model to the spin-1/2 Heisenberg model by simply considering: (i) A nearest-neighbor $s=1/2$ antiferromagnetic Heisenberg Hamiltonian on a square lattice characterized by a large exchange energy; (ii) very small interplanar couplings and spin anisotropies. In addition, they have pointed out that the long-wavelength, low-energy properties are well described by a mapping to a *2d-classical Heisenberg* magnet because all the effects of quantum fluctuations can be resorbed by means of adapted renormalizations of the coupling constants. Thus, the particular and important conclusion of Chakravarty's work motivated us to focus again on the $2d$-classical O(3) model developed on a square lattice [15-18].

In this article we reconsider this study but, now, we give the *complete* treatment for a $2d$ square lattice. $d$ represents the crystallographic dimension and $D = d + 1$ is the space-time one. The lattice is composed of classical spins isotropically coupled between first-nearest neighbors. If $H^{ex}_{i,j}$ is the corresponding local exchange Hamiltonian associated with each lattice site $(i,j)$ the evaluation of the zero-field partition function $Z_N(0)$ necessitates to expand each local operator $\exp(-\beta H^{ex}_{i,j})$ on the infinite basis of spherical harmonics.

Each harmonics is characterized by a couple of integers $(l,m)$, with $l \geq 0$ and $m \in [-l, +l]$. The corresponding eigenvalue of each operator $\exp(-\beta H^{ex}_{i,j})$ is nothing but the modi-fied Bessel function of the first kind $(\pi/2\beta J)^{1/2} \times I_{l+1/2}(-\beta J)$ where $\beta = 1/k_B T$ is the Boltzmann factor and $J$ the exchange energy between consecutive spin neighbors whereas each spherical harmonics is the eigenfunction. A polynomial expansion *valid for any temperature* can be obtained for the zero-field partition function $Z_N(0)$. Each basic term of the expansion appears as a product of two parts: i) A radial part involving itself the product of all the $l_{i,j}$-eigenvalues characterizing all the operators $\exp(-\beta H_{i,j}^{ex})$ for the whole lattice (with one eigenvalue per bond); ii) an angular part composed of a product of all the integrals $F_{i,j}$ (with one integral per lattice site $(i,j)$). Each of these integrals $F_{i,j}$ is composed of a product of spherical harmonics with one spherical harmonics per bond connecting the site $(i,j)$ to its first nearest neighbors. As a result the polynomial expansion describing the zero-field partition function $Z_N(0)$ is nothing but a characteristic polynomial.

The non-vanishing condition of each integral $F_{i,j}$ leads to a first set of selection rules with a first subset for the $l$'s and a second one for the $m$'s involved. For the whole lattice and for each type of integer $l$ or $m$, we obtain a system of imbricate equations (with one equation per lattice site) i.e., equations involving $l$'s on the one hand and $m$'s on the other one due to the fact that each classical spin interacts with its first-nearest neighbors $(i+1,j)$, $(i,j-1)$, $(i-1,j)$ and $(i,j+1)$. At first sight, when the lattice is finite, it seems that the solution of this system is not unique. In other words no unique closed-form expression exists for the characteristic polynomial associated with the zero-field partition function $Z_N(0)$ so that the statistical problem seems unsolvable from a mathematical point of view [19]. The examination of the case of a finite lattice is out of the scope of the present article.

In the thermodynamic limit (i.e., in the physical case of an infinite lattice), we numerically show that the value $m = 0$ is selected, all the $l$'s remaining equal to the common value $l_0$. We have previously seen that this is the highest-degree term of the $l$-polynomial expansion of the characteristic polynomial associated with the zero-field partition function $Z_N(0)$ [19]. In other words we surprisingly obtain a very simple closed-form expression for $Z_N(0)$, valid for any temperature.

In addition, in this important physical case for which $m = 0$, we report a further thermal numerical study of the $l$-highest-degree term. For sake of simplicity we restrict to the case of similar exchange energies $J = J_1 = J_2$ but the result can be applied to the general case $J_1 \neq J_2$. As a result, if $J_1 = J_2$, the basic highest $l$-term reduces to $F_{i,j} [(\pi/2\beta J)^{1/2} I_{l+1/2}(-\beta J)]^2$ where the integral $F_{i,j}$ is composed of four spherical harmonics $Y_{l,0}$. Thus we show that it appears *crossovers* between two consecutive terms inside this highest-degree term. It means that the characteristic polynomial expansion can be consequently reduced to this single term within a given temperature range: This is due to the presence of integrals $F_{i,j} \neq 1$ (whereas for spin chains we always have $F_{i,j} = 1$).

Coming from high temperatures, the value $l = 0$ characterizes the dominant term for reduced temperatures such as $k_B T/|J| \geq 0.255$. For $0.255 \geq k_B T/|J| \geq 0.043$ we have $l = 1$ and so on. As a result, $l$-eigenvalues, with increasing $l > 0$, are successively dominant when temperature is decreasing. In the vicinity of absolute zero the dominant term is characterized by $l \to +\infty$ whereas, for consecutive terms characterized by integers associated with the ranks $l$ and $l + 1$, the ratio of integrals $F_{i,j}$ tends towards unity. As all the modified Bessel functions of the first kind $(\pi/2\beta J)^{1/2} I_{l+1/2}(-\beta J)$ have a close behavior when $\beta \to +\infty$ (or $T \to 0$) we derive that, near absolute zero, all the $l$-eigenvalues become equivalent, confirming the fact that the critical temperature is $T_c = 0$ K, in agreement with Mermin-Wagner's theorem [20].

If considering the thermodynamic functions of interest they are all obtained by deriving $Z_N(0)$ with respect to temperature (specific heat) or to the magnitude of the applied external induction $B$ in the vanishing $B$-limit (spin-spin correlations, correlation length and susceptibility). As a result all these functions *valid for any temperature* show the same polynomial structure as $Z_N(0)$. It means that they are composed of a common factor (the same highest $l$-term appearing in $Z_N(0)$ whose basic term is $F_{i,j}[(\pi/2\beta J)^{1/2} I_{l+1/2}(-\beta J)]^2$) multiplied by another factor coming from the adequate derivation of $Z_N(0)$. As a result each of these thermodynamic function is characterized by a dominant term (i.e., a new $l$-eigenvalue) showing the same $l$-value in the same given temperature range appearing in the numerical study of $Z_N(0)$. In other words the polynomial associated with this thermodynamic function can be truncated in the same temperature range.

Owing to a work similar to the one leading to write a closed-form expression for $Z_N(0)$ in the thermodynamic limit we show that the spin correlation $|<S_{k,k'}>|$ vanishes for any lattice site $(k,k')$ so that the correlation function $|\Gamma_{k,k'}|$ reduces to the spin-spin correlation. A similar work allows one to derive a closed-form expression for the spin-spin correlation $|<S_{0,0}\cdot S_{k,k'}>|$ valid for any temperature. Meanwhile we have shown that all the correlation paths between the correlated sites $(0,0)$ and $(k,k')$ are confined within a correlation rectangle whose vertices involve sites $(0,0)$, $(0,k')$, $(k,k')$ and $(k,0)$ (*theorem* 1): This is the *correlation domain*. In addition we show that, for an infinite square lattice and for any couple of correlated sites $(0,0)$ and $(k,k')$, all the correlation paths have the same weight exclusively depending on $k$ and $k'$ i.e., the same length characterized by $k$ vertical bonds and $k'$ horizontal ones. This length corresponds to the shortest possible one between any couple of sites $(0,0)$ and $(k,k')$. This illustrates Maupertuis' least action principle (*theorem* 2). As a result an exact expression can be derived for the correlation length and the susceptibility.

In the vicinity of $T_c$ we define scaling parameters $x_1$ and $x_2$ previously introduced by several authors [6,10] through the *spin stiffness* $\rho_s$ for $x_1$ i.e., $x_1 = k_B T/2\pi\rho_s$ and $\Delta$ the *T=0-energy gap between the ground state and the first exited state* (very small due to the fact that the spectrum is nearly continuous) i.e., $x_2 = k_B T/\Delta$. In the low-temperature study of $Z_N(0)$ the argument $\beta|J|$ of each $l$-eigenvalue (where $J$ is the exchange energy between first-nearest neighbor sites) must be replaced by a scaling parameter $\zeta l$ characterized by a closed-form expression expressed vs $x_1$ or $x_2$ which gives the analytic expression of branches separating the various magnetic phases. The corresponding diagram exhibits three magnetic phases: It is exactly similar to the one derived from the renormalization group technique [6,10].

If considering the low-temperature limit of the spin-spin correlation concerning correlated sites located at a quasi-infinite distance, we show that the Ornstein-Zernike law is fulfilled [22]. We verify that equal-time correlations decay with the anomalous power law $\exp(-r/\xi)/r^{-(D-2+\eta)}$ [10,21]. When $D = 3$ we find $\eta = -1$ whereas $\eta = 0$ through a renormalization technique [10]. We explain this discrepancy between these values. In addition we derive four low-temperature behaviors for the correlation length i.e., one per magnetic phase of the diagram, in perfect agreement with previous results obtained from a renormalization technique [10]. These behaviors can be also directly obtained from the definition of the correlation length. At $T_c = 0$ K the critical exponent is $\nu = 1$, as previously shown [6,10]. As a result we identify three regimes: The Renormalized Classical Regime (RCR, $x_1 \ll 1$), the Quantum Disordered Regime (QDR, $x_2 \ll 1$) and the Quantum Critical Regime (QCR, $x_1 \gg 1$ and $x_2 \gg 1$).

In the vicinity of $T_c$ we show that the square root of the convergence ratio of the characteristic polynomial associated with $Z_N(0)$ is equal to the spin-spin correlation $|\langle \mathbf{S}_{0,0} \cdot \mathbf{S}_{0,1}\rangle|$ (or equivalently $|\langle \mathbf{S}_{0,0} \cdot \mathbf{S}_{1,0}\rangle|$ if $J_1 = J_2$) between first-nearest neighbors belonging to the same lattice line (or row). This result is only due to the fact that $|\langle \mathbf{S}_{i,j}\rangle| = 0$ because $T_c = 0$.

Finally, from the exact closed-form expression of $Z_N(0)$, we express the surface density of free energy $\mathcal{F}$. Its expression obeys the *hyperscaling law* introduced by de Gennes and Fisher [31]. Near $T_c = 0$ K and for each magnetic phase, we respectively obtain the same expression as the corresponding one derived from the renormalization group approach [10]. As a result we derive the surface density of specific heat and we show that it can be written as the product of a universal function $\Psi_i(x_i)$ ($i = 1, 2$) and the specific heat of a 2$d$-gapless Bose gas.

We also focus on the surface density susceptibility. We show that, near the critical point, the lattical is composed of quasi rigid quasi independent rectangular ($J_1 \neq J_2$) or square ($J_1 = J_2 = J$) Kadanoff blocks whose side lengths are the correlation lengths $\xi_1$ and $\xi_2$ ($J_1 \neq J_2$) or $\xi$ ($J_1 = J_2 = J$). If $M(T)$ is the magnetic moment per unit cell the product $\chi k_B T$ can be written as $\xi_1 \xi_2 M(T)^2$ (if $J_1 \neq J_2$) or $\xi^2 M(T)^2$ (if $J_1 = J_2 = J$) so that, in the critical domain i.e., if $T > 0$ K, the low-temperature behavior of $\chi k_B T$ is ruled by the competition between the divergence of $\xi_1 \xi_2$ (or $\xi^2$) and the behavior of $M(T)$.

For a ferromagnet or a 2$d$-noncompensated antiferromagnet, near $T_c = 0$ K, the dominant $l$-eigenvalue is that for which $l \to +\infty$. As a result, it is directly possible to obtain the exact low-temperature behavior from the closed-form polynomial expression of the susceptibility $\chi$: $M(T) \neq 0$ and $\chi k_B T$ diverges as $\xi_1 \xi_2$ (or $\xi^2$). Under these conditions we derive the critical exponent $\gamma = 3$. As the law $\gamma = \nu(2 - \eta)$ is universal this result allows one to validate the value $\eta = -1$ because $\nu = 1$.

For a 2$d$-compensated antiferromagnet $M(T) = 0$ as $T \to T_c$. The situation becomes more complicated because $\chi k_B T$ vanishes so that we must take into account all the contributions brought by each of the $l$-eigenvalues due to fact that they have a close behavior as $l \to +\infty$. As a result a specific treatment must be set on. We obtain a similar result as the one derived by Chubukov *et al*. [10]: $\chi$ can be written as a universal function $\Omega_i(x_i)$ ($i = 1, 2$) and the corresponding $T$-vanishing law characterizes the Renormalized Classical Regime (RCR, $x_1 \ll 1$), the Quantum Disordered Regime (QDR, $x_2 \ll 1$) or the Quantum Critical Regime (QCR, $x_1 \gg 1$ and $x_2 \gg 1$). Finally Chubukov *et al*. [10] have found that, if using a Monte Carlo method, near $T_c = 0$ K for which $x_1$ and $x_2$ diverge, $\Omega_i(\infty) = 0.25 \pm 0.04$ whereas $\Omega_i(x_i) > \Omega_i(\infty)$. As $\Omega_i(x_i) - \Omega_i(\infty) > 0.04$ we rigorously show that the value $\Omega_i(\infty) = 0.25$ is the lower bound of $\Omega_i(\infty)$ because it corresponds to the value obtained when the $l$-polynomial expansion of the susceptibility is restricted to the first term $l = 0$.

The low-temperature expressions of the respective surface densities of specific heat and susceptibility allows one to write the Wilson ration $R_W$ for each zone of the magnetic diagram. This ratio compares the magnetic fluctuations described by the low-temperature behavior of $\chi k_B T$ and the thermal ones described by the low-temperature behavior of the specific heat. Near $T_c = 0$ K, for a compensated antiferromagnet, we find that $R_W$ diverges for the Renormalized Classical and Quantum Disordered Regimes: The magnetic fluctuations strongly dominate the thermal ones, as expected. Surprisingly, in the Quantum Critical Regime $R_W < 1$: The thermal fluctuations slightly dominate the magnetic ones.

We conclude the present article with a comparison between theoretical and experimental susceptibilities characterizing 2$d$-compensated antiferromagnets. We have considered two classes of compounds. In class I we deal with inorganic compounds characterized by sheets of magnetic ions separated by nonmagnetic ions or organic ligands. These compounds are characterized by a relatively high 3$d$-ordering temperature $T_{3d}$ with respect to the Néel temperature $T_N$ so that the 2$d$-magnetic behavior is valid in a restricted temperature range [$T_{3d}$, $T_N$]. In class II compounds are characterized by sheets of similar magnetic ions separated by inorganic ligands of adjustable length. But, now, there are extra organic ligands between in-plane magnetic ions. $T_{3d}$ is pushed down to 2 K so that the range [$T_{3d}$, $T_N$] is plainly broadened, as expected. In the case of compounds characterized by a single exchange energy $J_1 = J_2 = J$ we derive a numerical relation at the maximum of susceptibility so that the experimental value of $J$ can be directly derived if knowing the experimental temperature value of this maximum. The reasoning can be extended to the general case $J_1 \neq J_2$.

For both classes I and II the fits of experimental susceptibilities have respectively given values of $J$ closed to the values measured by magnetochemists. Similarly for $G$, the Landé factor, we have found a value closed to the theoretical one 2 $\mu_B/\hbar$. For all the considered compounds of classes I and II we have shown that, in the critical domain, they are characterized by a Quantum Critical Regime (QCR).

## II. THEORETICAL BACKGROUND

### A. Definitions

The general Hamiltonian describing a lattice characterized by a square unit cell composed of $(2N+1)^2$ sites, each one being the carrier of a classical spin $\mathbf{S}_{i,j}$, is given by:

$$H = \sum_{i=-N}^{N} \sum_{j=-N}^{N} (H_{i,j}^{\text{ex}} + H_{i,j}^{\text{mag}}), \qquad (2.1)$$

with:

$$H_{i,j}^{\text{ex}} = (J_1 \mathbf{S}_{i,j+1} + J_2 \mathbf{S}_{i+1,j}) \cdot \mathbf{S}_{i,j}, \qquad (2.2)$$

$$H_{i,j}^{\text{mag}} = -G_{i,j} S_{i,j}^z B, \qquad (2.3)$$

where:

$G_{i,j} = G$ if $i+j$ is even or null,   $G_{i,j} = G'$ if $i+j$ is odd.  (2.4)

In Eq. (2.2) $J_1$ and $J_2$ refer to the exchange interaction between first-nearest neighbors belonging to the horizontal lines and vertical rows of the lattice, respectively. $J_i > 0$ (respectively, $J_i < 0$, with $i = 1, 2$) denotes an antiferromagnetic (respectively, ferromagnetic) coupling. $G_{i,j}$ is the Landé factor characterizing each spin $\mathbf{S}_{i,j}$ and expressed in $\mu_B/\hbar$ unit. In the case of a lattice showing edges, when dealing with outside bonds, the exchange Hamiltonian $H_{i,j}^{\text{ex}}$ must be corrected by assuming vanishing exchange energies. Finally we consider that the classical spins $\mathbf{S}_{i,j}$ are unit vectors so that the exchange energy $JS(S+1) \sim JS^2$ is written $J$. It means that we do not take into account the number of spin components in the normalization of $\mathbf{S}_{i,j}$'s so that $\mathbf{S}^2 = 1$.

The field-dependent partition function $Z_N(B)$ is defined as:

$$Z_N(B) = \prod_{i=-N}^{N} \prod_{j=-N}^{N} \int d\mathbf{S}_{i,j} \exp\left(-\beta \sum_{i=-N}^{N} \sum_{j=-N}^{N} \left(H_{i,j}^{\text{ex}} + H_{i,j}^{\text{mag}}\right)\right), \qquad (2.5)$$

where $\beta = 1/k_B T$ is the Boltzmann factor (which must not be confused with the critical exponent $\beta_c$). At this step it must be noticed that the calculation of the field-dependent partition function $Z_N(B)$ is plainly more complicated because of the presence of the further term $H_{i,j}^{\text{mag}}$ in the exponential argument, for each site $(i,j)$. This aspect is not examined in the present article. Under these conditions, the zero-field partition $Z_N(0)$ is simply obtained by integrating the operator $\exp(-\beta H^{\text{ex}})$ over all the angular variables characterizing the states of all the classical spins belonging to the lattice.

We finally recall the thermodynamic functions of main interest i.e., the correlation length and the total susceptibility per lattice site which can be defined as:

$$\xi = \left(\frac{\sum_k \sum_{k'} (k^2 + k'^2) |\Gamma_{k,k'}|}{\sum_k \sum_{k'} |\Gamma_{k,k'}|}\right)^{1/2}. \qquad (2.6)$$

$$\chi_{i,j} = \beta \sum_k \sum_{k'} G_{i,j} G_{i+k,j+k'} \Gamma_{k,k'} \qquad (2.7)$$

where $\Gamma_{k,k'}$ is the correlation function:

$$\Gamma_{k,k'} = <\mathbf{S}_{i,j} \cdot \mathbf{S}_{i+k,j+k'}> - <\mathbf{S}_{i,j}><\mathbf{S}_{i+k,j+k'}> \qquad (2.8)$$

if $(i,j)$ is the site of reference. In this article we choose $(0,0)$. In the previous equation, the bracket notation $<\cdots>$ means that we deal with a thermodynamic average. In other words, if we consider a lattice showing edges or wrapped on a torus, characterized by a square unit cell and composed of $(2N+1)^2$ sites, each one being the carrier of a classical spin $\mathbf{S}_{i,j}$, we can define the spin-spin correlation between any couple of spins in the vanishing-field limit:

$$<\mathbf{S}_{i,j} \cdot \mathbf{S}_{i+k,j+k'}> = \frac{1}{Z_N(0)} \int d\mathbf{S}_{-N,-N} \cdots \int d\mathbf{S}_{i,j} \mathbf{S}_{i,j} \cdots \int d\mathbf{S}_{i+k,j+k'} \mathbf{S}_{i+k,j+k'} \cdots \int d\mathbf{S}_{N,N} \exp\left(-\beta \sum_{i=-N}^{N} \sum_{j=-N}^{N} H_{i,j}^{\text{ex}}\right). \qquad (2.9)$$

The zero-field spin correlation $<\mathbf{S}_u>$, with $u = (i,j)$ (respectively, $(i+k,j+k')$) can be derived from Eq. (2.9) by replacing $\mathbf{S}_{i+k,j+k'}$ (respectively, $\mathbf{S}_{i,j}$) by unity. As we deal with isotropic (Heisenberg) couplings, we have the following properties:

$$<S_{i,j}^v \cdot S_{i+k,j+k'}^v> = \frac{1}{3}<\mathbf{S}_{i,j} \cdot \mathbf{S}_{i+k,j+k'}>, \quad v = x, y \text{ or } z,$$

$$<S_u^v> = \frac{1}{\sqrt{3}}<\mathbf{S}_u>, \quad u = (i,j) \text{ or } (i+k,j+k'), \qquad (2.10)$$

from which we immediately derive for the correlation function:

$$\Gamma_{k,k'}^v = \frac{1}{3}\Gamma_{k,k'}, \quad v = x, y \text{ or } z. \qquad (2.11)$$

Finally we can define the *self spin-spin correlation* $<(S_u^v)^2>$, with $v = x, y$ or $z$. We have $<\mathbf{S}^2> = 1$ due to the fact that the classical spin is considered as a unit vector. Consequently, as we deal with isotropic spin-spin couplings, we can write:

$$<(S_u^x)^2> = <(S_u^y)^2> = <(S_u^z)^2> = \frac{1}{3},$$

$$\xi = \xi^\nu, \quad \chi_{i,j}^\nu = \frac{\chi_{i,j}}{3}, \quad \chi = \chi_{i,j}^\nu, \quad \nu = x, y \text{ or } z. \quad (2.12)$$

Thus $\chi$ is the susceptibility per lattice site measured along one of the three equivalent axes $x$, $y$ or $z$.

## B. Calculation of the zero-field partition function of an infinite lattice

### B.1. Generalities

In this subsection we briefly recall the calculation of the zero-field partition function detailed in a previous paper [19] because it will plainly facilitate the calculation of the spin and spin-spin correlations. Due to the presence of classical spin moments, all the operators $H_{i,j}^{\text{ex}}$ commute and the exponential factor appearing in the integrand of Eq. (2.5) considered in the zero-field limit can be written:

$$\exp\left(-\beta \sum_{i=-N}^{N} \sum_{j=-N}^{N} H_{i,j}^{\text{ex}}\right) = \prod_{i=-N}^{N} \prod_{j=-N}^{N} \exp\left(-\beta H_{i,j}^{\text{ex}}\right). \quad (2.13)$$

As a result, the particular nature of $H_{i,j}^{\text{ex}}$ given by Eq. (2.2) allows one to separate the contributions corresponding to the exchange coupling involving classical spins belonging to the same horizontal line $i$ of the layer (i.e., $S_{i,j-1}$, $S_{i,j+1}$ and $S_{i,j}$) or to the same vertical row $j$ (i.e., $S_{i-1,j}$, $S_{i+1,j}$ and $S_{i,j}$). In fact, for each of the four contributions (one per bond connected to the site $(i,j)$ carrying the spin $S_{i,j}$), we have to expand a term such as $\exp(-A S_1.S_2)$ where $A$ is $\beta J_1$ or $\beta J_2$ (the classical spins $S_1$ and $S_2$ being considered as unit vectors). If we call $\Theta_{1,2}$ the angle between vectors $S_1$ and $S_2$, characterized by the couples of angular variables $(\theta_1, \varphi_1)$ and $(\theta_2, \varphi_2)$, it is possible to expand the operator $\exp(-A\cos\Theta_{1,2})$ on the infinite basis of spherical harmonics which are eigenfunctions of the angular part of the Laplacian operator on the sphere of unit radius $S^2$:

$$\exp(-A\cos\Theta_{1,2}) = 4\pi \sum_{l=0}^{+\infty} \left(\frac{\pi}{2A}\right)^{1/2} I_{l+1/2}(-A) \times$$

$$\times \sum_{m=-l}^{+l} Y_{l,m}^*(S_1) Y_{l,m}(S_2). \quad (2.14)$$

In the previous equation the $(\pi/2A)^{1/2} I_{l+1/2}(-A)$'s are modified Bessel functions of the first kind and $S_1$ and $S_2$ symbolically represent the couples $(\theta_1, \varphi_1)$ and $(\theta_2, \varphi_2)$. If we set:

$$\lambda_l(-\beta j) = \left(\frac{\pi}{2\beta j}\right)^{1/2} I_{l+1/2}(-\beta j), \quad j = J_1 \text{ or } J_2, \quad (2.15)$$

each operator $\exp(-\beta H_{i,j}^{\text{ex}})$ is finally expanded on the infinite basis of eigenfunctions (the spherical harmonics), whereas the $\lambda_l$'s are nothing but the associated eigenvalues. Under these conditions, the zero-field partition function $Z_N(0)$ directly appears as a characteristic polynomial and can be written for a lattice wrapped on a torus showing $(2N)^2$ sites and $2(2N)^2$ bonds:

$$Z_N(0) = (4\pi)^{8N^2} \sum_{l_{N,-N}=0}^{+\infty} \lambda_{l_{N,-N}}(-\beta J_1) \sum_{l'_{N,-N}=0}^{+\infty} \lambda_{l'_{N,-N}}(-\beta J_2) \times \ldots$$

$$\times \sum_{l_{-N,N-1}=0}^{+\infty} \lambda_{l_{-N,N-1}}(-\beta J_1) \sum_{m_{N,-N}=-l_{N,-N}}^{+l_{N,-N}} \sum_{m'_{N,-N}=-l'_{N,-N}}^{+l'_{N,-N}} \ldots \sum_{m_{-N,N-1}=-l_{-N,N-1}}^{+l_{-N,N-1}} \prod_{i=-N}^{N} \prod_{j=-N}^{N} F_{i,j}, \quad (2.16)$$

$$F_{i,j} = \int dS_{i,j} Y_{l'_{i+1,j}, m'_{i+1,j}}(S_{i,j}) Y_{l_{i,j-1}, m_{i,j-1}}(S_{i,j}) Y_{l_{i,j}, m_{i,j}}^*(S_{i,j}) Y_{l'_{i,j}, m'_{i,j}}^*(S_{i,j}) \quad (2.17)$$

where $F_{i,j}$ is the current integral per site (with one spherical harmonics per bond). Using the following decomposition of any pair of spherical harmonics appearing in the integrand of $F_{i,j}$ [23]

$$Y_{l_1, m_1}(S) Y_{l_2, m_2}(S) = \sum_{L=|l_1-l_2|}^{l_1+l_2} \sum_{M=-L}^{+L} \left[\frac{(2l_1+1)(2l_2+1)}{4\pi(2L+1)}\right]^{1/2} \times$$

$$\times C_{l_1 \ 0 \ l_2 \ 0}^{L \ 0} C_{l_1 \ m_1 \ l_2 \ m_2}^{L \ M} Y_{L,M}(S) \quad (2.18)$$

$F_{i,j}$ can be expressed as a Clebsch-Gordan (C.G.) series. If the lattice is wrapped on a torus $F_{i,j}$ is always given by Eq. (2.17) because we always have in-sites characterized by four spherical harmonics:

$$F_{i,j} = \frac{1}{4\pi} \left[(2l'_{i+1,j}+1)(2l_{i,j-1}+1)(2l_{i,j}+1)(2l'_{i,j}+1)\right]^{1/2} \times$$

$$\times \sum_{L_{i,j}=L_<}^{L_>} \frac{1}{2L_{i,j}+1} \sum_{M_{i,j}=-L_{i,j}}^{+L_{i,j}} C_{l'_{i+1,j} \ 0 \ l_{i,j-1} \ 0}^{L_{i,j} \ 0} \times$$

$$\times C_{l'_{i+1,j} \ m'_{i+1,j} \ l_{i,j-1} \ m_{i,j-1}}^{L_{i,j} \ M_{i,j}} C_{l_{i,j} \ 0 \ l'_{i,j} \ 0}^{L_{i,j} \ 0} C_{l_{i,j} \ m_{i,j} \ l'_{i,j} \ m'_{i,j}}^{L_{i,j} \ M_{i,j}}.$$

$$(2.19)$$

In the previous equations $C_{l_1 \, m_1 \, l_2 \, m_2}^{l_3 \, m_3}$ is a Clebsch-Gordan (C.G.) coefficient. The C.G. coefficients $C_{l_{i,j} \, m_{i,j} \, l'_{i,j} \, m'_{i,j}}^{L_{i,j} \, M_{i,j}}$ and $C_{l'_{i+1,j} \, m'_{i+1,j} \, l_{i,j-1} \, m_{i,j-1}}^{L_{i,j} \, M_{i,j}}$ (with $M_{i,j} \neq 0$ or $M_{i,j} = 0$) appearing in Eq. (2.19) do not vanish if the triangular inequalities $|l_{i,j} - l'_{i,j}| \leq L_{i,j} \leq l_{i,j} + l'_{i,j}$ and $|l'_{i+1,j} - l_{i,j-1}| \leq L_{i,j} \leq l'_{i+1,j} + l_{i,j-1}$ are fulfilled, respectively. As a result, we must have $L_< = \max(|l'_{i+1,j} - l_{i,j-1}|, |l_{i,j} - l'_{i,j}|)$ and $L_> = \min(l'_{i+1,j} + l_{i,j-1}, l_{i,j} + l'_{i,j})$.

### B.2. Principles of construction of the characteristic polynomial associated with the zero-field partition function

We have previously shown that the zero-field partition function can be written under the general form [19]

$$Z_N(0) = (4\pi)^{8N^2} \left[ \sum_{l=0}^{+\infty} \left( \lambda_l(-\beta J_1) \lambda_l(-\beta J_2) \right)^{4N^2} \sum_{\{m\}} \right.$$

$$\left. + \sum_{\substack{\{l_1,...,l_K\}, \\ \{l'_1,...,l'_K\}}} \left( \lambda_{l_1}(-\beta J_1) \right)^{a_{1,l_1}} ... \left( \lambda_{l'_K}(-\beta J_2) \right)^{a_{K,l'_K}} \sum_{\substack{\{m_1,...,m_K\}, \\ \{m'_1,...,m'_K\}}} \right] \times$$

$$\times \prod_{i=-N}^{N} \prod_{j=-N}^{N} F_{i,j} \,. \tag{2.20}$$

In the first term, the symbolical notation $\{m\}$ refers to the $2(2N)^2$ summations over the relative integers $m_{i,j}$. For the second term the notation $\{l_1,...,l_K\}, \{l'_1,...,l'_K\}$ (respectively, $\{m_1, ..., m_K\}, \{m'_1, ..., m'_K\}$) symbolically represents the set of authorized integers $l_{i,j}$ and $l'_{i,j}$ (respectively, $m_{i,j}$ and $m'_{i,j}$) on which the $l_1$-, ..., $l_K$-, $l'_1$-, ..., $l'_K$ (respectively, $m_1$-, ..., $m_K$-, $m'_1$-, ..., $m'_K$-) summations must be achieved.

The examination of Eq. (2.20) giving the polynomial expansion of the zero-field partition function $Z_N(0)$ allows one to say that its writing is nothing but that one derived from the formalism of the transfer-matrix technique. Each term of Eq. (2.20) appears as a product of two subterms: (i) A *temperature-dependent radial factor* containing a product of the various eigenvalues $\lambda_l(-\beta j)$, $j = J_1$ or $J_2$ (with one eigenvalue per bond); (ii) an *angular factor* containing a product of integrals $F_{i,j}$ describing all the spin states of all the lattice sites (with one integral per site). At first sight this factor seems to be temperature-independent. This vision is wrong because the magnetic ordering is described by the set of all the couples of angles $(\theta_{i,j}, \varphi_{i,j})$ characterizing all the spin states and is highly temperature-dependent (through the set of coefficients $l_{i,j}$ and $l'_{i,j}$, cf Eq. (2.14)).

As the square lattice contains $2(2N)^2$ bonds (when wrapped on a torus) i.e., $(2N)^2$ horizontal and vertical bonds, each eigenvalue $\lambda_l(-\beta j)$ appearing in each radial factor is characterized by a superscript giving the number of similar bonds showing the *same* positive (or null) value $l$ or $l'$. Thus, the sum of all these superscripts is equal to $(2N)^2$ for the total contribution of the horizontal (respectively, vertical) lattice lines (respectively, rows). When $J_1 = J_2$ this sum then becomes equal to $2(2N)^2$. Under these conditions the evaluation of the set of integers $(..., l_{i,j},..., l'_{i,j}, ...)$ for the whole lattice allows the complete determination of each radial factor of the characteristic polynomial.

The angular factor is determined by the simultaneous knowledge of the sets $(..., l_{i,j},..., l'_{i,j}, ...)$ and $(..., m_{i,j},..., m'_{i,j}, ...)$, with $m_{i,j} \in [-l_{i,j}, +l_{i,j}]$ and $m'_{i,j} \in [-l'_{i,j}, +l'_{i,j}]$, due to the fact that each current integral $F_{i,j}$ closely depends on these integers. The set of integers $(..., l_{i,j},..., l'_{i,j}, ...)$ is shared with the radial factor, thus ensuring the link with temperature for the local magnetic ordering characterized by the couple of angles $(\theta_{i,j}, \varphi_{i,j})$, as previously noted. In fact, $F_{i,j}$ is described by the datum of four couples of integers. Two couples $(l_{i,j}, m_{i,j})$ and $(l'_{i,j}, m'_{i,j})$ characterize the current site $(i,j)$ whereas the two other ones $(l'_{i+1,j}, m'_{i+1,j})$ and $(l_{i,j-1}, m_{i,j-1})$ refer to the first-nearest neighbor sites, thus conferring an imbricate character to the product of integrals $F_{i,j}$ constituting the final angular factor. In other words, it means that the determination of the set $(..., l_{i,j},..., l'_{i,j}, ...)$ must be achieved in a first step, for the whole lattice, even if this latter is infinite, and the set of relative integers $(..., m_{i,j},..., m'_{i,j}, ...)$ must be then obtained in a separate second step. Thus, the non-vanishing conditions of each current integral $F_{i,j}$ (one for integers $l_{i,j}$, $l'_{i,j}$, $l'_{i+1,j}$ and $l_{i,j-1}$ and one for relative integers $m_{i,j}$, $m'_{i,j}$, $m'_{i+1,j}$ and $m_{i,j-1}$) are going to permit the introduction of *selection rules* over the various sets of integers $(..., l_{i,j},..., l'_{i,j}, ...)$ and $(..., m_{i,j},..., m'_{i,j}, ...)$.

A numerical argument can be then used to classify the various terms of the characteristic polynomial in order to write them in the decreasing modulus order. Notably this is an important turning point in the infinite-lattice limit where the term of higher degree must be selected. Due to the lattice structure previously described the corresponding basic contribution to each current term of the characteristic polynomial can be written as

$$u_{l_{i,j}, l'_{i,j}} = F_{i,j} \lambda_{l_{i,j}}(-\beta J_1) \lambda_{l'_{i,j}}(-\beta J_2) \tag{2.21}$$

with $F_{i,j} \neq 1$ given by Eq. (2.17).

### B.3. General selection rules for the whole lattice; consequences

For any finite or infinite lattice size, the non-vanishing condition of each current integral $F_{i,j}$ is mainly due to that of C.G. coefficients. It allows one to derive two types of universal *selection rules* which are *temperature-independent*.

The *first selection rule* concerns the coefficients $m$ and $m'$ appearing in Eq. (2.19). We have:

$$m_{i,j-1} + m'_{i+1,j} - m_{i,j} - m'_{i,j} = 0. \quad (SRm) \tag{2.22}$$

At this step we must note that, if each spherical harmonics $Y_{l,m}(\mathbf{S}) = Y_{l,m}(\theta,\varphi)$ appearing in the integrand of $F_{i,j}$ is replaced by its own definition i.e., $C_l^m \exp(im\varphi) P_l^m(\cos\theta)$ where $C_l^m$ is a constant depending on coefficients $l$ and $m$ and $P_l^m(\cos\theta)$ is the associated Legendre polynomial, the non-vanishing condition of the $\varphi$-part directly leads to Eq. (2.22). Finally, we can make two remarks. *(i) The SRm relation is unique. (ii) Due to the fact that the $\varphi$-part of the $F_{i,j}$-integrand is null, $F_{i,j}$ is a pure real number*.

More generally, for finding the $m_{i,j}$'s and $l_{i,j}$'s characterizing a lattice showing edges or a lattice wrapped on a torus, we have to solve a linear system of $(2N)^2$ equations (2.22) (one per site) but with $2(2N)^2$ unknowns $m_{i,j}$ and $m'_{i,j}$ (respectively $l_{i,j}$ and $l'_{i,j}$). As it remains $2(2N)^2 - (2N)^2 = (2N)^2$ independent solutions over the set $\mathbb{Z}$ of relative integers $m_{i,j}$ (respectively, over the set $\mathbb{N}$ of integers for the $l_{i,j}$'s), *it means that there are $(2N)^2$ different expressions for each local angular factor (respectively, radial factor) appearing in each term of the characteristic polynomial giving $Z_N(0)$ so that the statistical problem remains unsolved*. Thus, at first sight, this result means that *there is no unique expression for $Z_N(0)$, except if $|m_{i,j}| = |m'_{i,j}| \neq 0$ or $m_{i,j} = m'_{i,j} = 0$. As a result it would mean that there is no analytical expression of the zero-field partition function for a finite lattice*. This problem has been examined in a previous paper and is out of the scope of the present article [19].

The *second selection rule* is derived from the fact that the various coefficients $l$ and $l'$ appearing in Eq. (2.19) obey triangular inequalities $|l_{i,j} - l'_{i,j}| \leq L_{i,j} \leq l_{i,j} + l'_{i,j}$ and $|l_{i,j-1} - l'_{i+1,j}| \leq L_{i,j} \leq l_{i,j-1} + l'_{i+1,j}$ respectively followed by the C.G. coefficients $C_{l_{i,j}\ m_{i,j}\ l'_{i,j}\ m'_{i,j}}^{L_{i,j}\ M_{i,j}}$ and $C_{l'_{i+1,j}\ m'_{i+1,j}\ l_{i,j-1}\ m_{i,j-1}}^{L_{i,j}\ M_{i,j}}$ (with $M_{i,j} \neq 0$ or $M_{i,j} = 0$) [19]. If $M_{i,j} = 0$ we have a more restrictive vanishing condition [23]:

$$C_{l_1\ 0\ l_2\ 0}^{l_3\ 0} = 0, \text{ if } l_1 + l_2 + l_3 = 2g + 1,$$

$$C_{l_1\ 0\ l_2\ 0}^{l_3\ 0} = (-1)^{g-l_3}\sqrt{2l_3+1}\ K, \text{ if } l_1 + l_2 + l_3 = 2g, \quad (2.23)$$

with:

$$K = \frac{g!}{(g-l_1)!(g-l_2)!(g-l_3)!} \times$$
$$\times \left[\frac{(2g-2l_1)!(2g-2l_2)!(2g-2l_3)!}{(2g+1)!}\right]^{1/2}. \quad (2.24)$$

We can note that $K$ is unchanged under the permutation of integers $l_1$, $l_2$ and $l_3$. $C_{l_{i,j}\ 0\ l'_{i,j}\ 0}^{L_{i,j}\ 0}$ does not vanish if $l_{i,j} + l'_{i,j} + L_{i,j} = 2A_{i,j} \geq 0$ whereas, for $C_{l'_{i+1,j}\ 0\ l_{i,j-1}\ 0}^{L_{i,j}\ 0}$, we must have $l_{i,j-1} + l'_{i+1,j} + L_{i,j} = 2A'_{i,j} \geq 0$. Thus, if summing or substracting the two previous equations over $l$ and $l'$, we have $(2N)^2$ equations (one per lattice site) such as:

$$l_{i,j-1} + l'_{i+1,j} + l_{i,j} + l'_{i,j} = 2g_{i,j}, \quad (SRl1)$$
$$l_{i,j-1} + l'_{i+1,j} - l_{i,j} - l'_{i,j} = 2g'_{i,j}, \quad (SRl2) \quad (2.25)$$

(or equivalently $l_{i,j} + l'_{i,j} - l_{i,j-1} - l'_{i+1,j} = 2g''_{i,j}$, with $g''_{i,j} = -g'_{i,j}$) where $g_{i,j}$ or $g'_{i,j}$ is a relative integer. We obtain two types of equations which are similar to Eq. (2.22) but now, in Eq. (2.25), instead of having a null right member like in Eq. (2.22), we can have a positive, null or negative but always even second member $2g_{i,j}$.

### B.4. Zero-field partition function in the thermodynamic limit

The case of thermodynamic limit is less restrictive than that of a finite lattice studied in a previous article [19] because we deal with a purely numerical problem. We have previously seen that $Z_N(0)$ is directly expressed as a characteristic polynomial whose structure has been previously examined (*cf* Eq. (2.20)). For simplifying the discussion we restrict the study to the case $J = J_1 = J_2$ without loss of generality.

This polynomial is composed of two parts. The first part contains the general term $[F_{i,j}\lambda_l(-\beta J)^2]^{(2N)^2}$ with one integral $F_{i,j}$ per site (giving a spatial average of the spin orientation) and one eigenvalue $\lambda_l(-\beta J)$ per bond (*cf* Eq. (2.15)) i.e., all the bonds are characterized by the same integer $l$. But, that supposes that the separate numerical study of integrals $F_{i,j}$ allows one to select a unique $m$-value so that $F_{i,j}$ is maximum. In that case, if setting $F_{i,j} = F(l_i, l_j, m)$, we have

$$F_{i,j} = F(l_i, l_j, m) = \frac{(2l_i+1)(2l_j+1)}{4\pi} \times$$
$$\times \sum_{L=|l_i-l_j|}^{l_i+l_j} \frac{1}{2L+1}\left[C_{l_i\ 0\ l_j\ 0}^{L\ 0} C_{l_i\ m\ l_j\ m}^{L\ 2m}\right]^2. \quad (2.26)$$

The second part of the polynomial is a product of second-rank terms such as $[F_{i,j}\lambda_l(-\beta J)^2]^n$ with $n < (2N)^2$ so that only $n$ bonds are characterized by the same integer $l$. For the other ones we have $n_1, n_2, \ldots n_k$ bonds characterized by different integers $l_1, l_2, \ldots l_k$, respectively, and we must have $n + n_1 + n_2 + \ldots + n_k = (2N)^2$. In addition some integrals $F_{i,j}$ can contain different $m_{i,j}$'s and $m'_{i,j}$'s so that the knowledge of the exact structure of these second-rank terms is complicated.

In the infinite-lattice limit the highest eigenvalue must naturally arise in the first-rank term $[F_{i,j}\lambda_l(-\beta J)^2]^{(2N)^2}$. If using Eq. (2.21), with $l_{i,j} = l'_{i,j} = l$, the corresponding contribution $u_{\max} = F_{i,j}\lambda_L(-\beta J)^2$ where $L = l_{\max}$, dominates all the other ones inside this term as well as all the terms composing the second-rank ones [19]. This can occur in the whole temperature range or in a smaller temperature range if there exist thermal crossover phenomena among the set of eigenvalues.

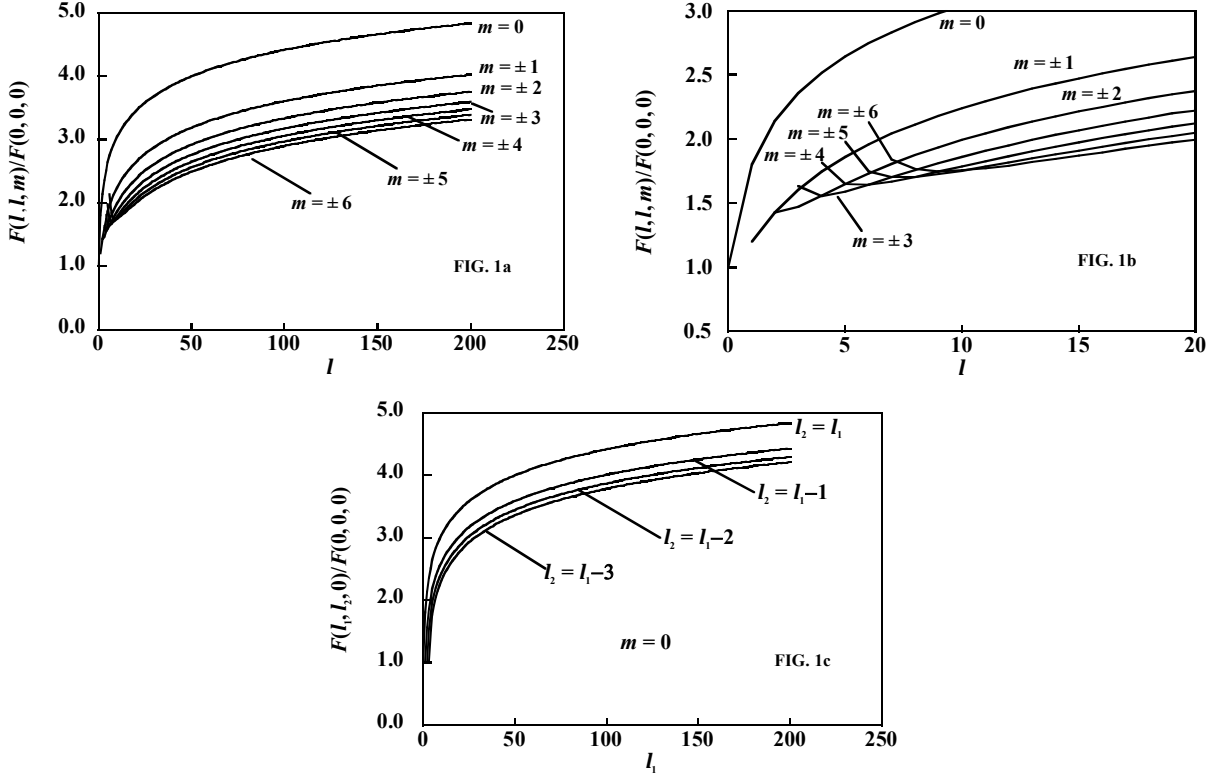

FIG. 1. a) Numerical study of the ratio $F(l, l, m)/F(0, 0, 0)$ vs $l$ for various values of $m$; ($F(l, l, m)$ is given by Eq. (2.26) and $F(0, 0, 0) = 1/4\pi$); b) zoom of the previous study; c) numerical study of the ratio $F(l_1, l_2, 0)/F(0, 0, 0)$ for various values of $l_2 \leq l_1$.

In this case the dominant eigenvalue $\lambda_l(-\beta J) = \lambda_l(-J/k_B T)$ over a temperature range becomes subdominant when the temperature $T$ is outside this range and a new eigenvalue previously subdominant becomes dominant and so on. In the case of 1$d$ spin chains, we always have the same highest eigenvalue (within the framework described previously).

But, we are going to see that, for 2$d$ isotropic couplings, there exist thermal crossover phenomena within the set of eigenvalues. This behavior is detailed and explained below.

A numerical study of the thermal behavior of the general term $F(l_{i,j}, l'_{i,j})\lambda_{l_{i,j}}(-\beta J)\lambda_{l'_{i,j}}(-\beta J)$ allowing to construct the characteristic polynomial giving $Z_N(0)$ is then necessary for determining the dominant term characterized by $l_{i,j} = l'_{i,j} = L$ over a given temperature range.

In this respect we have studied the integral $F_{i,j} = F(l, l, m)$ given by Eq. (2.26). In Fig. 1a we have reported the ratio $F(l, l, m)/F(0,0,0)$ vs $l$ for various $m$-values such as $|m| \leq l$ (with $F(0,0,0) = 1/4\pi$). We immediately observe that this ratio rapidly decreases for increasing $|m|$-values, for any $l$. However, we have zoomed the beginning of each curve corresponding to the case $|m| = l$. This trend is not followed but we always have $F(l, l, m) < F(l, l, 0)$ (see Fig. 1b). In addition, in Fig. 1c, for $m = 0$, we observe that $F(l_i, l_j, 0)$ decreases for $l_j < l_i$. As a result, when $N \to +\infty$, the integral $F(l, l, 0)$ appears as the dominant one i.e.,

$$[F(l,l,0)]^{(2N)^2} \gg [F(l,l,m)]^{(2N)^2}, \text{ as } N \to +\infty, (2.27)$$

so that the value $m = 0$ is selected. Under these conditions Eq. (2.20) can be rewritten in the thermodynamic limit if $J = J_1 = J_2$:

$$Z_N(0) = (4\pi)^{8N^2}\left[\sum_{l=0}^{+\infty}[F(l,l,0)\lambda_l(-\beta J)^2]^{4N^2}\right.$$
$$\left. + \prod_{i=-(N-1)}^{N}\prod_{j=-(N-1)}^{N}\sum_{l_i=0}^{+\infty}{}'\sum_{l_j=0}^{+\infty}{}'F(l_i,l_j,0)\lambda_{l_i}(-\beta J)\lambda_{l_j}(-\beta J)\right],$$

as $N \to +\infty$.(2.28)

The notation $\sum_{l_i=0}^{+\infty}{}'\sum_{l_j=0}^{+\infty}{}'$ means that $l_i$ and $l_j$ are chosen so that the corresponding current second-rank term cannot give back the first rank one in which $l_i = l_j = l$.

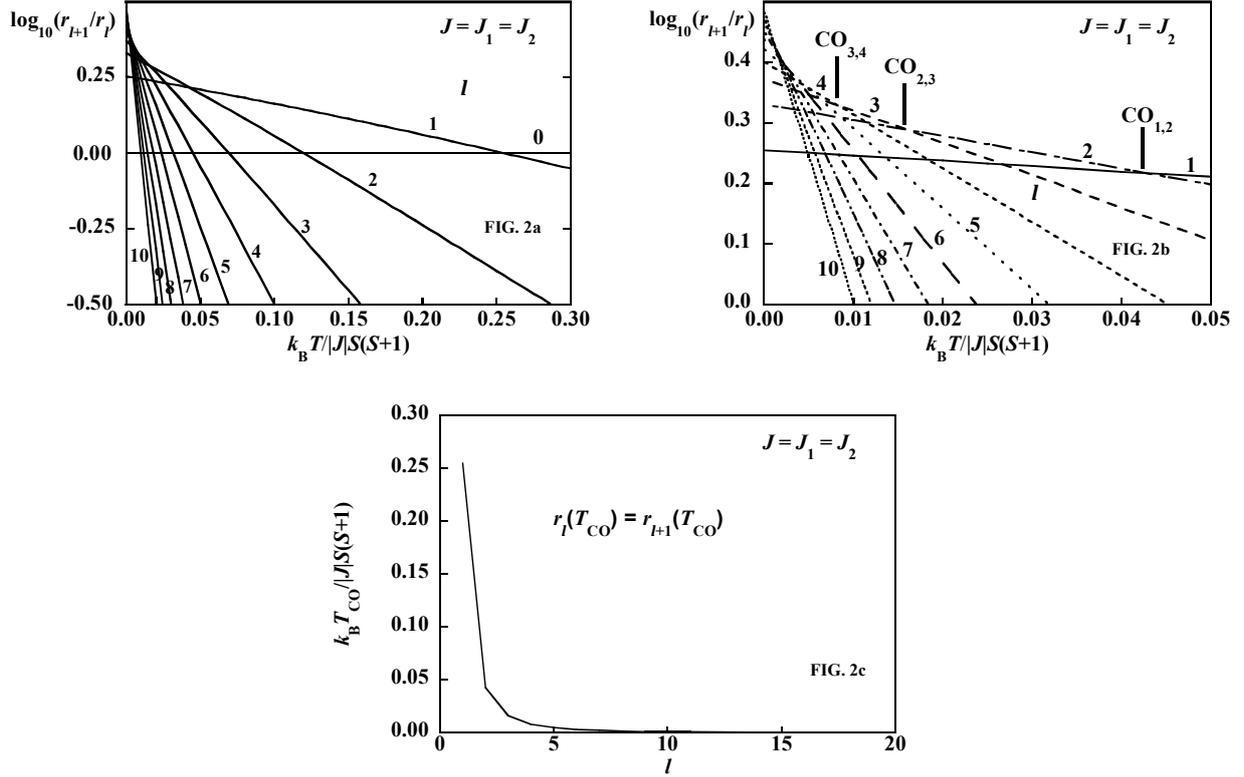

FIG. 2. a) Thermal variations of the ratio $\log_{10}(r_{l+1}/r_l)$ for various values of $l$ where $r_{l+1}/r_l$ is defined by Eq. (2.30); b) zoom of the plot allowing to have a better insight of the crossover phenomena between various $l$-regimes; c) plot of the crossover temperature $T_{CO}$ vs $l$.

In a first step we must wonder if all the current terms of the previous $l$-series must be kept in the first term of Eq. (2.28) i.e., if the series must be truncated, for a given range of temperature $[T_{l_i,<}, T_{l_i,>}]$. As a result, for any $T$ belonging to $[T_{l_i,<}, T_{l_i,>}]$ we define the dominant term

$$u_{\max} = F_{L,L}\lambda_L(-\beta J)^2, \quad F_{L,L} = F(L,L,0), \quad L = l_{\max} \quad (2.29)$$

as well as the following ratio:

$$\frac{r_{l\pm 1}}{r_l} = \frac{F_{l\pm 1, l\pm 1}}{F_{l,l}}\left(\frac{\lambda_{l\pm 1}(-\beta J)}{\lambda_l(-\beta J)}\right)^2, \quad J = J_1 = J_2. \quad (2.30)$$

We have studied the thermal behavior of $r_{l+1}/r_l$ for various finite $l$-values. This work is reported in Fig. 2a. We observe that $\log_{10}(r_{l+1}/r_l)$ shows a decreasing linear behavior with respect to $k_BT/|J|S(S+1)$. We have zoomed Fig. 2a in the very low-temperature domain (Fig. 2b). If $r_{l+1}/r_l < 1$ $\log_{10}(r_{l+1}/r_l) < 0$ and $\log_{10}(r_{l+1}/r_l) > 0$ if $r_{l+1}/r_l > 1$. We can then point out a succession of *crossovers*, each crossover being characterized by a specific temperature called *crossover temperature* $T_{CO}$. $T_{CO}$ is the solution of the following equation:

$$r_l(T_{CO}) = r_{l+1}(T_{CO}) \quad (2.31)$$

i.e., owing to Eq. (2.30):

$$\frac{\lambda_{l+1}(|\tilde{J}|/k_BT_{CO})}{\lambda_l(|\tilde{J}|/k_BT_{CO})} = \left[\frac{F_{l,l}}{F_{l+1,l+1}}\right]^{1/2}, \quad |\tilde{J}| = JS(S+1). \quad (2.32)$$

For instance, for the reduced temperatures such as $k_BT/|\tilde{J}| \geq 0.255$, the value $l = 0$ is dominant i.e., $\lambda_0(-\beta J)$ represents the dominant term of the characteristic polynomial. All the other terms $\lambda_l(-\beta J)$ with $l > 0$ are subdominant. When $0.255 \geq k_BT/|\tilde{J}| \geq 0.043$ $l = 1$ is dominant so that $\lambda_1(-\beta J)$ is now the dominant term of the characteristic polynomial whereas $\lambda_0(-\beta J)$ has become the subdominant one as well as all the other terms $\lambda_l(-\beta J)$ with $l \neq 1$ etc... In that case the crossover temperature corresponding to the transition between the regimes respectively characterized by $l = 0$ and $l = 1$ is labeled $T_{CO_{0,1}}$. We have reported $k_BT_{CO}/|\tilde{J}|$ vs $l$ in Fig. 2c. As expected we observe that $T_{CO}$ rapidly decreases when $l$ increases. It means that, when the temperature tends to absolute zero, it appears a succession of closer and closer crossovers so that all the eigenvalues, characterized by an increasing $l$-value, successively play a role. But, when $T \approx 0$ K, all these eigenvalues intervene due to the fact that the crossover temperatures are closer and closer. *The discret eigenvalue spectrum tends to a continuum.* As a result we can say that $T = 0$ K *plays the role of* *critical temperature* $T_c$. This aspect will be more detailed later.

How interpreting this phenomena? In the 1$d$-case (infinite spin chain) we always have $\lambda_0(-\beta J)$ as dominant eigenvalue in the whole range of temperature, the integral $F_{0,0}$ being always equal to unity. In the 2$d$-case the situation is more complicated. The appearance of successive predominant eigenvalues is due to the presence of integrals $F_{l,l} \neq 1$, for any $l > 0$. A numerical fit shows that the ratio $F_{l,l} / F_{0,0}$ increases with $l$ *according to a logarithmic law*, more rapidly than the ratio $|\lambda_l(-\beta J)/\lambda_0(-\beta J)|^2$ which decreases with $l$, for a given temperature. The particular case $l \to +\infty$ will be examined in a forthcoming section.

Now, if taking into account the previous study Eq. (2.28) can be rewritten as:

$$Z_N(0) = (4\pi)^{8N^2}[u_{\max}(T)]^{4N^2}\{1 + S_1(N,T) + S_2(N,T)\},$$
$$T \in [T_{l_i,<}, T_{l_i,>}] \quad (2.33)$$

where $u_{\max}$ is given by Eq (2.21) with $l_{\max} = l_i = l_j = l$ and:

$$S_1(N,T) = \sum_{l=0, l \neq l_{\max}}^{+\infty} \left[\frac{u_{l,l}(T)}{u_{\max}(T)}\right]^{4N^2},$$

$$S_2(N,T) = \prod_{i=-(N-1)}^{N} \prod_{j=-(N-1)}^{N} \sum_{l_i=0}^{+\infty} \sum_{\substack{l_j=0, \\ l_j \neq l_i}}^{+\infty} \frac{u_{l_i,l_j}(T)}{u_{\max}(T)}. \quad (2.34)$$

In Appendix A we have studied $Z_N(0)$ in the thermodynamic limit, for temperatures $T > 0$ K, in the whole range $[0+\varepsilon, +\infty[$, with $\varepsilon << 1$. We show that: i) For a given range $[T_{l_i,<}, T_{l_i,>}]$ we have $Z_N(0) \sim (4\pi)^{8N^2} u_{\max}^{4N^2}$, with $L = l_{\max} = l$; ii) as the reasoning is valid for any $[T_{l_i,<}, T_{l_i,>}]$ we finally have

$$Z_N(0) = (4\pi)^{8N^2} \sum_{l=0}^{+\infty}{}^t \left[F_{l,l}\lambda_l(-\beta J)^2\right]^{4N^2}, \text{ as } N \to +\infty. \quad (2.35)$$

In the previous equation the special notation $\sum_{l=0}^{+\infty}{}^t$ recalls that the summation is truncated: In each temperature range $[T_{l_i,<}, T_{l_i,>}]$ the eigenvalue $\lambda_{l_i}(-\beta J_k)$ with $k = 1, 2$ is dominant. At the frontier $T = T_{l_i,<}$ $\lambda_{l_i}(-\beta J_k)$ and $\lambda_{l_i-1}(-\beta J_k)$ must be taken into account whereas if $T = T_{l_i,>}$ we have to consider $\lambda_{l_i}(-\beta J_k)$ and $\lambda_{l_i+1}(-\beta J_k)$. However the initial notation $\sum_{l=0}^{+\infty}$ can be also kept.

*As the expression of $Z_N(0)$ appears at the denominator of all the thermodynamic functions derived from $Z_N(0)$ such as the specific heat, the spin correlations, the correlation length and the static susceptibility, all the numerical properties derived from this study are also valid for these thermodynamic functions.*

### C. Thermal derivatives of the zero-field partition function of an infinite lattice

Once known the exact polynomial closed-form expression of the zero-field partition function $Z_N(0)$ as $N \to +\infty$, it then becomes possible to obtain the free energy $F$ and its thermal derivatives of main interest i.e., the internal energy $U$ and the specific heat $C_V$. From their respective definitions and the expression of $Z_N(0)$ given by Eq. (2.35) in the thermodynamic limit we have per lattice bond

$$F = -\frac{k_B T}{8N^2}\ln(Z_N(0)),$$

$$\frac{U}{8N^2} = -\frac{1}{2Z_N(0)}\sum_{l=0}^{+\infty}{}^t \sum_{i=1}^{2} |J_i| f_l(\beta|J_i|) z_l^{4N^2},$$

$$\frac{C_V}{8N^2} = \frac{1}{Z_N(0)}\sum_{l=0}^{+\infty}{}^t \left[\sum_{i=1}^{2} \frac{\beta|J_i|}{2} f_l(\beta|J_i|)\right]^2 z_l^{4N^2}$$

$$- \left[\frac{1}{Z_N(0)}\sum_{l=0}^{+\infty}{}^t \sum_{i=1}^{2} \frac{\beta|J_i|}{2} f_l(\beta|J_i|) z_l^{4N^2}\right]^2,$$

as $N \to +\infty$ (2.36)

with

$$f_l(\beta|J_i|) = \frac{1}{2l+1}\left[l\frac{\lambda_{l-1}(\beta|J_i|)}{\lambda_l(\beta|J_i|)} + (l+1)\frac{\lambda_{l+1}(\beta|J_i|)}{\lambda_l(\beta|J_i|)}\right],$$

$$Z_N(0) = (4\pi)^{8N^2} \sum_{l=0}^{+\infty}{}^t z_l^{4N^2}, z_l = F_{l,l}\lambda_l(\beta|J_1|)\lambda_l(\beta|J_2|),$$

as $N \to +\infty$. (2.37)

The special notation $\sum_{l=0}^{+\infty}{}^t$ has been defined after Eq. (2.35) and recalls that the $l$-summation is truncated.

Finally we define the Wilson ratio

$$R_W = \frac{k_B^2 \chi T}{G^2 C_V} \quad (2.38)$$

where the susceptibility $\chi$ is given by Eqs. (2.7) and (2.12).

### III. EXPRESSION OF THE SPIN CORRELATIONS, THE CORRELATION LENGTH AND THE SUSCEPTIBILITY

#### A. Generalities

Using the general definition of the spin-spin correlation given by Eq. (2.9) and expanding the exponential part on the infinite basis of spherical harmonics, we can write:

$$\begin{pmatrix} <S_{i,j}^z>, <S_{i+k,j+k'}^z> \\ <S_{i,j}^z.S_{i+k,j+k'}^z> \end{pmatrix} = \frac{(4\pi)^{8N^2}}{Z_N(0)} \sum_{l_{N,-N}=0}^{+\infty} \lambda_{l_{N,-N}}(-\beta J_1) \sum_{l'_{N,-N}=0}^{+\infty} \lambda_{l'_{N,-N}}(-\beta J_2) \times ...$$
$$\times \sum_{l_{-N,N-1}=0}^{+\infty} \lambda_{l_{-N,N-1}}(-\beta J_1) \sum_{m_{N,-N}=-l_{N,-N}}^{+l_{N,-N}} \sum_{m'_{N,-N}=-l'_{N,-N}}^{+l'_{N,-N}} ... \sum_{m_{-N,N-1}=-l_{-N,N-1}}^{+l_{-N,N-1}} \prod_{i=-N}^{N} \prod_{j=-N}^{N} F'_{i,j} \qquad (3.1)$$

where $F'_{i,j}$ is the following current integral

$$F'_{i,j} = \int d\mathbf{S}_{i,j} X_{i,j} Y_{l'_{i+1,j}, m'_{i+1,j}}(\mathbf{S}_{i,j}) Y_{l_{i,j-1}, m_{i,j-1}}(\mathbf{S}_{i,j}) \times$$
$$\times Y^*_{l_{i,j}, m_{i,j}}(\mathbf{S}_{i,j}) Y^*_{l'_{i,j}, m'_{i,j}}(\mathbf{S}_{i,j}), \qquad (3.2)$$

for site (i,j) (and a similar expression for site (i+k,j+k')). When $X_{i,j} = 1$, we have $F'_{i,j} = F_{i,j}$ (cf Eq. (2.17)). Thus, if calculating $<S_{i,j}^z>$ (or $<S_{i+k,j+k'}^z>$) we have a single integral $F'_{i,j}$ (or $F'_{i+k,j+k'}$) containing $\cos\theta_{i,j}$ (or $\cos\theta_{i+k,j+k'}$) whereas for $<S_{i,j}^z.S_{i+k,j+k'}^z>$ we have two integrals $F'_{i,j}$ and $F'_{i+k,j+k'}$ in the product of integrals appearing in Eq. (3.1).

**B. Calculation of the spin correlation. Consequences**

Now we wish to calculate the numerator of the spin correlation $<S_u^z>$. It is given by Eq. (3.1) in which we have $X_{k_1, k_2} = 1$ except at sites (i,j) or (i+k,j+k') where we use the following recursion relation:

$$\cos\theta_{i,j} Y_{l_{i,j}, m_{i,j}}(\mathbf{S}_{i,j}) = C_{l_{i,j}+1} Y_{l_{i,j}+1, m_{i,j}}(\mathbf{S}_{i,j})$$
$$+ C_{l_{i,j}-1} Y_{l_{i,j}-1, m_{i,j}}(\mathbf{S}_{i,j}). \qquad (3.3)$$

In the particular but important case on which we focus i.e., the thermodynamic limit ($N \to +\infty$), we have $m_{i,j} = 0$ and $m'_{i,j} = 0$ for any (i,j). Then $C_{l_{i,j}+1}$ and $C_{l_{i,j}-1}$ reduce to

$$C_{l_{i,j}+1} = \frac{l_{i,j}+1}{\sqrt{(2l_{i,j}+1)(2l_{i,j}+3)}},$$
$$C_{l_{i,j}-1} = \frac{l_{i,j}}{\sqrt{(2l_{i,j}+1)(2l_{i,j}-1)}}. \qquad (3.4)$$

In the particular case $l_{i,j} = 0$ as it occurs for the beginning of each $l$-series expansion, we have $C_1 = 1/\sqrt{3}$ and $C_{-1} = 0$. In the case of the calculation of the spin correlation, this transform can be equivalently applied to each of the four spherical harmonics appearing in Eq. (3.2). For instance, if we wish to calculate $<S_{i,j}^z>$, we directly apply Eq. (3.3) to $Y^*_{l_{i,j},0}(\mathbf{S}_{i,j})$. As a result the integral $F'_{i,j}$ defined in Eq. (3.2) can be written:

$$F'_{i,j} = C_{l_{i,j}+1} F_{l_{i,j}, l_{i,j}+1} + C_{l_{i,j}-1} F_{l_{i,j}, l_{i,j}-1} \qquad (3.5)$$

with:

$$F_{l_{i,j}, l_{i,j}+\varepsilon} = \int d\mathbf{S}_{i,j} Y_{l'_{i+1,j}, 0}(\mathbf{S}_{i,j}) Y_{l_{i,j-1}, 0}(\mathbf{S}_{i,j}) \times$$
$$\times Y^*_{l_{i,j}+\varepsilon, 0}(\mathbf{S}_{i,j}) Y^*_{l'_{i,j}, 0}(\mathbf{S}_{i,j}), \varepsilon = \pm 1, \qquad (3.6)$$

with $l_{i+1 j} = l_{i,j-1} = l_{i,j} = l'_{i,j} = l$ in the thermodynamic limit ($N \to +\infty$). We immediately retrieve the calculation of integrals appearing in that of the zero-field partition function [19]. As a result, if using Eq. (2.19), we can readily write:

$$F_{l_{i,j}, l+\varepsilon} = \frac{(2l+1)^2}{4\pi} \sum_{L_{i,j}=L_<}^{L_>} \frac{1}{2L_{i,j}+1} \left[ C_{l\ 0\ l\ 0}^{L_{i,j}\ 0} C_{l+\varepsilon\ 0\ l\ 0}^{L_{i,j}\ 0} \right]^2,$$
$$\varepsilon = \pm 1. \quad (3.7)$$

Thus, the non-vanishing condition of each current integral $F_{l_{i,j}, l_{i,j}+\varepsilon}$ is mainly due to that of the involved C.G. coefficients. We have previously seen that it allows one to write down *universal temperature-independent selection rules*. For integers $l$ the selection rule given by Eq. (2.25) is slightly modified. We now have $C_{l+\varepsilon\ 0\ l\ 0}^{L_{i,j}\ 0} \neq 0$ (with $\varepsilon = \pm 1$) and $C_{l\ 0\ l\ 0}^{L_{i,j},\ 0} \neq 0$ if respectively:

$$2l + \varepsilon + L_{i,j} = 2g_{i,j},\ 2l + L_{i,j} = 2g'_{i,j}. \qquad (3.8)$$

Reporting the $L$-value derived from the first equation i.e., $L = 2g - (2l + \varepsilon)$ in the second one, we must have $2g - \varepsilon = 2g'$. The unique solution is $\varepsilon = 0$ which is impossible in the present case because $\varepsilon = \pm 1$, exclusively. As a result $C_{l+\varepsilon\ 0\ l\ 0}^{L\ 0}$ and $C_{l\ 0\ l\ 0}^{L\ 0}$ do not vanish simultaneously but their product is always null. We immediately derive $F_{l,l+\varepsilon} = 0$ and $F'_{i,j} = 0$ so that:

$$<S^z_{i,j}> = 0, \quad <S^z_{i+k,j+k'}> = 0, \quad (3.9)$$

and:

$$\Gamma^z_{k,k'} = <S^z_{i,j}.S^z_{i+k,j+k'}>. \quad (3.10)$$

This result rigorously proves that the critical temperature is absolute zero i.e., $T_c = 0$ K.

### C. Calculation of spin-spin correlations

The z-z spin-spin correlation $<S^z_{i,j}.S^z_{i+k,j+k'}>$ has been defined in Eq. (3.1). At this step we recall that $(i,j)$, the reference site, can be chosen as $(0,0)$ due to the fact that all the lattice sites are equivalent. In addition we restrict the following study to $k > 0$ and $k' > 0$, without loss of generality.

#### C1. Calculation of the spin-spin correlation between first-nearest neighbors belonging to the same lattice line

In a first step we examine the case of the spin-spin correlation between first nearest-neighbors belonging to the same lattice line (or row), for instance $<S_{0,0}.S_{0,1}>$. The numerator of Eq. (3.1) can be calculated by the same method used for expressing the zero-field partition function $Z_N(0)$. Due to the imbricate character of integrals $F_{i,j}$ we recall that the process of integration can be achieved through three methods: i) Integrating from horizontal line $i = -N$ to $i = N$ between vertical lines $j = -N$ and $j = N$ or ii) vice versa; iii) integrating simultaneously from the four lattice lines $i = N$, $i = -N$, $j = -N$ and $j = N$ (confused on the torus) in the direction of the lattice heart characterized by site $(0,0)$.

In the thermodynamic limit ($N \rightarrow +\infty$) we have $m_{k,k'} = m'_{k,k'} = 0$ for any lattice site. In addition, we have seen in Appendix A that the $l$-polynomial expansion giving $Z_N(0)$ reduces to the dominant term for which $l_{k,k'} = l'_{k,k'} = l$ for all the horizontal and vertical bonds, notably for all the sites linked to the horizontal line $i$ of reference that we conventionally choose as $i = 0$. Thus, among the horizontal bonds of that line, a special care must be brought to the bond between sites $(0,0)$ and $(0,1)$ characterized by $l_{0,0}$: This is the *correlation domain*.

At these sites we respectively have $l'_{1,0} = l_{0,-1} = l'_{0,0} = l$ and $l'_{1,1} = l'_{0,1} = l_{0,1} = l$ due to the fact that these coefficients are involved in the highest-degree term of the characteristic polynomial associated with $Z_N(0)$: All the current integrals $F'_{k,k'}$ have been evaluated except $F'_{0,0}$ and $F'_{0,1}$. We have $F'_{k,k'} = F_{l,l}$ which is given by Eq. (3.7) in which $\varepsilon = 0$. In other words the integrand of current integrals $F'_{k,k'}$, with $(k,k') \ne (0,0)$ and $(k,k') \ne (0,1)$, does not have changed in contrast with the integrand of $F'_{0,0}$ and $F'_{0,1}$. As a result we have to determine the evolution of $l_{0,0}$ when passing from its value in the highest-degree term of $Z_N(0)$ i.e., $l_{0,0} = l$ to its new value in the numerator of Eq. (3.1) because the integrand of integral $F'_{0,0}$ has become $\cos\theta Y_{l_{0,0},0}(S_{0,0})[Y_{l,0}(S_{0,0})]^3$.

The decomposition law only intervenes at sites $(0,0)$ and $(0,1)$. Thus, at site $(0,0)$, for determining $l_{0,0}$ with respect to the other $l$-coefficients appearing in $F'_{0,0}$ i.e., $l'_{1,0} = l_{0,-1} = l'_{0,0} = l$, we first apply the decomposition law to the spherical harmonics $Y_{l,0}(S_{0,0})$. The corresponding result is

$$F'_{0,0} = C_{l+1}F_{l_{0,0},l+1} + C_{l-1}F_{l_{0,0},l-1} \quad (3.11)$$

with $F_{l_{0,0},l+\varepsilon}$ artificially written as:

$$F_{l_{0,0},l+\varepsilon} = \int dS_{0,0} [Y_{l,0}(S_{0,0})]^2 Y_{l_{0,0},0}(S_{0,0})Y_{l+\varepsilon,0}(S_{0,0}),$$
$$\varepsilon = \pm 1. \quad (3.12)$$

We immediately retrieve the calculation of integrals appearing in the zero-field partition function. In integral $F_{l_{0,0},l+\varepsilon}$ we begin by expressing the products of pairs of spherical harmonics as C.G. series (cf Eq. (2.18)). For instance we have

$$[Y_{l,0}(S_{0,0})]^2 = \frac{(2l+1)}{\sqrt{4\pi}} \sum_{L=0}^{2l} \frac{1}{\sqrt{2L+1}} \left[ C^{L\ 0}_{l\ 0\ l\ 0} \right]^2 Y_{L,0}(S_{0,0}),$$

$$Y_{l_{0,0},0}(S_{0,0})Y_{l+\varepsilon,0}(S_{0,0}) = \sum_{L'=|l_{0,0}-(l+\varepsilon)|}^{l_{0,0}+l+\varepsilon} \left[ \frac{(2(l+\varepsilon)+1)(2l_{0,0}+1)}{4\pi(2L'+1)} \right]^{1/2}$$

$$\times \left[ C^{L'\ 0}_{l_{0,0}\ 0\ l+\varepsilon\ 0} \right]^2 Y_{L',0}(S_{0,0}). \quad (3.13)$$

If inserting in Eq. (3.12) the non-vanishing condition of integral $F_{l_{0,0},l+\varepsilon}$ imposes $L = L'$ i.e., $l_{0,0} = l + \varepsilon$, $\varepsilon = \pm 1$. As a result all the lattice bonds are characterized by the integer $l$ whereas the unique bond of the correlation domain is characterized by $l \pm 1$. We have if $l_{0,0} = l + \varepsilon$

$$F_{l+\varepsilon,l+\varepsilon'} = \frac{(2l+1)[(2(l+\varepsilon)+1)(2(l+\varepsilon')+1)]^{1/2}}{4\pi} \times$$

$$\times \sum_{L=0}^{L_>} \frac{1}{2L+1} \left[ C^{L\ 0}_{l\ 0\ l\ 0} C^{L\ 0}_{l+\varepsilon\ 0\ l+\varepsilon'\ 0} \right]^2,$$

$$\varepsilon = \pm 1, \varepsilon' = \pm 1, \quad (3.14)$$

with $L_> = \min(2l, 2l + \varepsilon + \varepsilon')$. Thus, after applying at site $(0,0)$ the decomposition law given by Eqs. (3.3) and (3.4), each contribution can be written owing to the generic term $C_{l+\varepsilon}F_{l+\varepsilon,l+\varepsilon'}$.

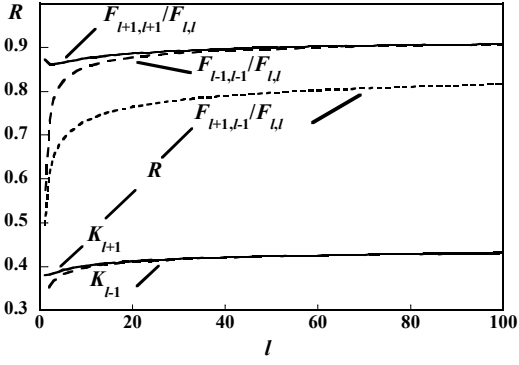

FIG. 3. Plots of the ratios $R = F_{l+1,l+1}/F_{l,l}$, $F_{l-1,l-1}/F_{l,l}$, and $F_{l+1,l-1}/F_{l,l}$ where integrals $F_{l+\varepsilon,l+\varepsilon'}$ ($\varepsilon = \pm 1$ and $\varepsilon' = \pm 1$) are defined by Eqs. (2.17), (2.26), (3.12) and (3.14) as well quantities $K_{l+1}$ and $K_{l-1}$ given by Eq. (3.16).

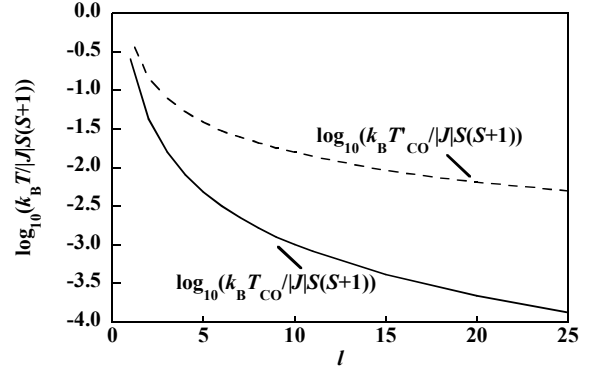

FIG. 4. Plots of $\log_{10}(k_B T_{CO}/|J|S(S+1))$ and $\log_{10}(k_B T'_{CO}/|J|S(S+1))$; the crossover temperatures $T_{CO}$ and $T'_{CO}$ are defined by the transcendental equations (2.32) and (3.17), respectively.

We have discarded the case in which we apply the decomposition law to $Y_{l_{0,0},0}(\mathbf{S}_{0,0})$ in Eq. (3.12). If doing so and imposing that the final value of integral $F'_{0,0}$ given by Eq. (3.12) must be similar i.e., independent of the application of the decomposition law, a work similar to the previous corresponding one leads to write $l_{0,0} = l$ and one of the coefficients $l'_{1,0}$, $l_{0,-1}$ or $l'_{0,0}$ is equal to $l + \varepsilon$, $\varepsilon = \pm 1$. This result is in contradiction with the fact that we must have $l'_{1,0} = l_{0,-1} = l'_{0,0} = l$. In addition the first nearest-neighbor integral $F'_{0,-1}$, $F'_{1,0}$ or $F'_{-1,0}$ vanishes. *As a result the decomposition law must uniquely be applied to a spherical harmonics $Y_{l,0}(\mathbf{S}_{0,0})$ for which l is known i.e., for a bond already characterized by an integer l.* If interpreting the geometrical nature of the corresponding correlation path it means that we are dealing with the beginning of a *loop which is forbidden*. In other words the unique correlation path is constituted by the bond between sites (0,0) and (0,1) to which the correlation domain is reduced. This point will be more detailed in a next subsection.

Finally we apply the decomposition law at site (0,1). We have to consider integral $F_{0,1}$. The work is similar to that one achieved for integral $F_{0,0}$ given by Eqs. (3.11) and (3.14) but the $l$-coefficients are now $l'_{1,1}$, $l'_{0,1}$ and $l_{0,1}$, with $l'_{1,1} = l'_{0,1} = l_{0,1} = l$ for reasons explained above. We also consider that $l_{0,0}$ is unknown whereas we have a single possibility for applying the decomposition law to spherical harmonics i.e., on $Y_{l'_{1,1},0}(\mathbf{S}_{0,1})$, $Y_{l_{0,1},0}(\mathbf{S}_{0,1})$ or $Y_{l'_{0,1},0}(\mathbf{S}_{0,1})$ (which are nothing but $Y_{l,0}(\mathbf{S}_{0,0})$). We verify that $l_{0,0} = l + \varepsilon$, $\varepsilon = \pm 1$, as independently found when calculating $F'_{0,0}$: The corresponding contribution is $F'_{0,1} = C_{l+1}F_{l+1,l+1} + C_{l-1}F_{l+1,l-1}$ if $l_{0,0} = l + 1$ and $F'_{0,1} = C_{l+1}F_{l-1,l+1} + C_{l-1}F_{l-1,l-1}$ if $l_{0,0} = l - 1$. As a result we can finally write

$$<S^z_{\bar{0},0}.S^z_{\bar{0},1}> = \frac{(4\pi)^{8N^2}}{Z_N(0)} \sum_{l=0}^{+\infty} {}^t\left[F_{l,l}\lambda_l(-\beta J_1)\lambda_l(-\beta J_2)\right]^{4N^2} \times$$
$$\times \left[K_{l+1}P_{1,l+1} + (1-\delta_{l,0})K_{l-1}P_{1,l-1}\right], \text{ as } N \to +\infty \quad (3.15)$$

where

$$K_{l+1} = C_{l+1}\left(C_{l+1}\frac{F_{l+1,l+1}}{F_{l,l}} + C_{l-1}\frac{F_{l+1,l-1}}{F_{l,l}}\right),$$

$$K_{l-1} = C_{l-1}\left(C_{l+1}\frac{F_{l-1,l+1}}{F_{l,l}} + C_{l-1}\frac{F_{l-1,l-1}}{F_{l,l}}\right),$$

$$P_{i,l+\varepsilon} = \frac{F_{l+\varepsilon,l+\varepsilon}}{F_{l,l}}\frac{\lambda_{l+\varepsilon}(-\beta J_i)}{\lambda_l(-\beta J_i)}, i = 1, 2, \varepsilon = \pm 1. \quad (3.16)$$

$\delta_{l,0}$ is the Kronecker symbol (due to the fact that $C_{l-1} = 0$ if $l = 0$). Integrals $F_{l,l}$ and $F_{l+\varepsilon,l+\varepsilon'}$ are respectively given by Eqs. (2.26) in which $m = m' = 0$ and (3.14), and the coefficient $C_{l+\varepsilon}$ is given by Eq. (3.4).

In Fig. 3 we have reported the various ratios of integrals $F_{l+1,l+1}/F_{l,l}$, $F_{l-1,l-1}/F_{l,l}$ and $F_{l+1,l-1}/F_{l,l}$. We note that all these ratios are lower than unity as well as the quantities $K_{l+1}$ and $K_{l-1}$ given by Eq. (3.16).

Now we have to examine the ratios $P_{1,l+1}$ and $P_{1,l-1}$ defined by Eq. (3.16). For physical reasons they must be lower (or equal) to unity in absolute value. They are composed of two ratios: $F_{l+\varepsilon,l+\varepsilon}/F_{l,l}$ and $\lambda_{l+\varepsilon}(-\beta J_i)/\lambda_l(-\beta J_i)$ with $\varepsilon = \pm 1$ or equivalently owing to their definition given by Eq. (2.15) $I_{l+\varepsilon+1/2}(-\beta J_i)/I_{l+1/2}(-\beta J_i)$. In absolute value and for a given relative temperature $\beta|J_i| = |J_i|/k_BT$ ($i = 1, 2$) we have $I_{l+3/2}(\beta|J_i|) < I_{l+1/2}(\beta|J_i|)$ for $\varepsilon = +1$ so that $|P_{i,l+1}| < 1$. However we always have $I_{l-1/2}(\beta|J_i|) > I_{l+1/2}(\beta|J_i|)$ for $\varepsilon = -1$. As a result and due to the fact that $F_{l-1,l-1}/F_{l,l} < 1$ we can have $|P_{i,l-1}| < 1$ or $|P_{i,l-1}| > 1$ which has no physical meaning. We proceed as for the thermal study of the current term of the $l$-polynomial expansion of $Z_N(0)$ where we have defined a crossover temperature $T_{CO}$ by Eq. (2.32) so that the eigenvalue $\lambda_{l_i}(\beta|J_i|)$ is dominant within the range $[T_{l_i,<}, T_{l_i,>}]$. Similarly, for studying the ratio $\lambda_{l-1}(\beta|J|)/\lambda_l(\beta|J|)$ appearing in $|P_{i,l-1}|$ given by Eq. (3.16) and due to the fact that $\max(|P_{i,l-1}|) = 1$, we define the new crossover temperature $T'_{CO}$ by

$$\frac{\lambda_{l-1}(|J|/k_B T'_{CO})}{\lambda_l(|J|/k_B T'_{CO})} = \frac{F_{l,l}}{F_{l-1,l-1}} \quad (3.17)$$

in the simplest case $J = J_1 = J_2$ without loss of generality. As for Eq. (2.32) we have numerically solved this transcendental equation and we have reported the corresponding results in Fig. 4.

We remark that, if comparing $T_{CO}$ and $T'_{CO}$, we always have $T'_{CO} > T_{CO}$. Then, if achieving the numerical study of $|P_{i,l-1}|$ we always have $|P_{i,l-1}| < 1$ if $T < T'_{CO}$. It means that, finally, $|P_{i,l-1}| < 1$ if $T < T_{CO} < T'_{CO}$. As a result, in the temperature range $[T_{l_i,<}, T_{l_i,>}]$ where the eigenvalue $\lambda_{l_i}(-\beta J)$ is dominant, we always have $|P_{i,l-1}| < 1$. This property remains true at the frontier $T = T_{l_i,<}$ where $\lambda_{l_i}(-\beta J)$ and $\lambda_{l_i-1}(-\beta J)$ must be taken into account whereas if $T = T_{l_i,>}$ we have to consider $\lambda_{l_i}(-\beta J)$ and $\lambda_{l_i+1}(-\beta J)$. The other eigenvalues are considered as neglected ones due to the fact that there are out of their temperature domain of predominance. Under these conditions we again justify the special notation $\sum_{l=0}^{+\infty}{}^t$ appearing in Eq. (3.15) which recalls that the $l$-summation is truncated.

Finally we note that it is easy to express $<S_{0,0}^z . S_{1,0}^z>$ by changing $J_1$ into $J_2$ in Eq. (3.15).

### C2. Calculation of the spin-spin correlation between spins belonging to the same lattice line

We first focus on the spin-spin correlation $<S_{0,0}^z . S_{0,2}^z>$ for illustrating the reasoning before its generalization to the spin-spin correlation $<S_{0,0}^z . S_{0,k'}^z>$. Due to the fact that we consider the thermodynamic limit ($N \to +\infty$) the characteristic polynomial is restricted to the highest-degree term for which all the bonds are characterized by integers $m = m' = 0$ and $l_{i,j}=l'_{i,j}=l$: $l_{0,-1} = l'_{0,0} = l'_{1,0} = l$ at site (0,0), $l'_{1,1} = l'_{0,1} = l$ at site (0,1) and $l'_{1,2} = l'_{0,2} = l_{0,2} = l$ at site (0,2) due to the fact that loops are forbidden. We then have to determine integers $l_{0,0}$ and $l_{0,1}$ characterizing the two horizontal bonds between sites (0,0) and (0,2) which now constitute the correlation domain.

At site (0,0) the treatment is similar to that one previously explained in Subsect. III.C1. The decomposition law leads to two contributions $C_{l+\varepsilon}F_{l+\varepsilon,l+\varepsilon}$ ($\varepsilon = \pm 1$) with $l_{0,0} = l + \varepsilon$ so that, for a fixed value of $\varepsilon$, the total contribution including the radial factor is $C_{l+\varepsilon}F_{l+\varepsilon,l+\varepsilon}\lambda_{l+\varepsilon}(-\beta J_1)$. At site (0,1), for a given $l_{0,0} = l + \varepsilon$, integral $F_{l+\varepsilon,l_{0,1}}$ can be expressed as integral $F_{l+\varepsilon,l_{0,0}}$, owing to Eqs. (3.12)-(3.14) where $l_{0,0}$ is replaced by $l_{0,1}$. The non-vanishing condition of this integral leads to $l_{0,1} = l_{0,0} = l + \varepsilon$ ($\varepsilon = \pm 1$) so that the corresponding contribution of site (0,1) is $F_{l+\varepsilon,l+\varepsilon}\lambda_{l+\varepsilon}(-\beta J_1)$. If taking into account the previous work of integration at site (0,0) the total contribution of sites (0,0) and (0,1) to the numerator of $<S_{0,0}^z . S_{0,2}^z>$ is $C_{l+\varepsilon}(F_{l+\varepsilon,l+\varepsilon}\lambda_{l+\varepsilon}(-\beta J_1))^2$.

Arriving at (0,2) the decomposition law is applied but all the $l$-coefficients have been already determined: The corresponding contribution is $C_{l+1}F_{l+1,l+1} + C_{l-1}F_{l+1,l-1}$ if $l_{0,1} = l + 1$ and $C_{l+1}F_{l-1,l+1} + C_{l-1}F_{l-1,l-1}$ if $l_{0,1} = l - 1$. As a result, for site (0,2), the two full contributions are respectively $C_{l\pm 1}[F_{l\pm 1,l\pm 1}\lambda_{l\pm 1}(-\beta J_1)]^2 (C_{l+1}F_{l\pm 1,l+1} + C_{l-1}F_{l\pm 1,l-1})$. Thus, as noted in the previous subsection, all the lattice bonds are characterized by integer $l$ whereas the two bonds of the correlation domain are characterized by integers $l_{0,0}$ and $l_{0,1}$ equal to $l \pm 1$.

If now considering the site $(0, k')$ the spin-spin correlation $<S_{0,0}^z . S_{0,k'}^z>$ can be easily derived from the previous calculation of $<S_{0,0}^z . S_{0,2}^z>$. At site (0,0) we have $l_{0,-1} = l'_{0,0} = l'_{1,0} = l$ and $l'_{1,K'} = l'_{0,K'} = l$ at sites (0,K') with $0 < K' < k'$, for which we have to determine $l_{0,K'}$. Finally $l'_{1,k'} = l'_{0,k'} = l_{0,k'} = l$ at site (0,k'). At site (0,0) the decomposition law is applied on integral $F_{0,0}$. The calculation is unchanged and the corresponding contribution is $C_{l+\varepsilon}F_{l+\varepsilon,l+\varepsilon}\lambda(-\beta J_1)$ (with $l_{0,0} = l \pm 1$). For all the sites (0,K') with $K' < k'$, there is no decomposition law and the work is similar to the one achieved at site (0,1): The non-vanishing condition of each encountered integral $F_{0,K'}$ leads to $l_{0,K'} = l + \varepsilon$, with $\varepsilon = \pm 1$, so that the full contribution of each site (0,K') is $F_{l+\varepsilon,l+\varepsilon}\lambda_{l+\varepsilon}(-\beta J_1)$. If taking into account all the contributions between site (0, 0) and site $(0,k'-1)$ the full contribution is $C_{l+\varepsilon}(F_{l+\varepsilon,l+\varepsilon}\lambda_{l+\varepsilon}(-\beta J_1))^{k'}$. Arriving at site (0,k') the decomposition law is applied on integral $F'_{0,k'}$ but all the $l$-coefficients have been already determined: The corresponding contribution is $C_{l+1}F_{l+1,l+1} + C_{l-1}F_{l+1,l-1}$ if $\varepsilon = +1$ and $C_{l+1}F_{l-1,l+1} + C_{l-1}F_{l-1,l-1}$ if $\varepsilon = -1$. Finally all the lattice bonds are characterized by integer $l$ whereas the $k'$ bonds of the correlation domain are characterized by integers $l_{0,0}$, $l_{0,1}$, ..., $l_{0,k'}$ all equal to $l \pm 1$.

As a result the spin-spin correlation $<S_{0,0}^z . S_{0,k'}^z>$ can be finally written

$$<S_{0,0}^z . S_{0,k'}^z> = \frac{(4\pi)^{8N^2}}{Z_N(0)} \sum_{l=0}^{+\infty}{}^t [F_{l,l}\lambda_l(-\beta J_1)\lambda_l(-\beta J_2)]^{4N^2} \times$$
$$\times [K_{l+1}(P_{1,l+1})^{k'} + (1-\delta_{l,0})K_{l-1}(P_{1,l-1})^{k'}],$$
$$k' > 0, \text{ as } N \to +\infty \quad (3.18)$$

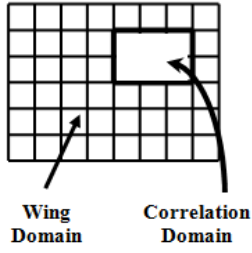
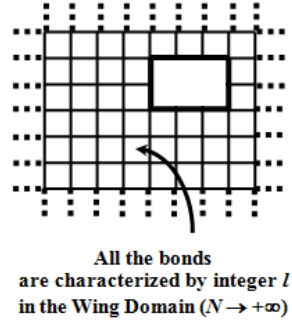

FIG. 5. a) Wing domain and correlation domain for a finite lattice showing edges; b) case of an infinite lattice ($N \to +\infty$); all the bonds of the wing domain are characterized by integer $l$.

where the coefficients $K_{l+1}$ and $K_{l-1}$ are given by Eq. (3.16). The special notation $\sum_{l=0}^{+\infty} {}^t$ recalls that the $l$-summation is truncated. $<S_{0,0}^z \cdot S_{k,0}^z>$ can be obtained by exchanging $J_1$ and $J_2$ i.e., $P_{1,l+\varepsilon}$ against $P_{2,l+\varepsilon}$ in Eq. (3.18).

### C3. Calculation of the spin-spin correlation between any couple of lattice sites

Now we focus on the general case i.e., the calculation of the spin-spin correlation $<S_{0,0}^z \cdot S_{k,k'}^z>$ between any couple of lattice sites (0,0) and ($k,k'$).

We have previously seen that, due to the imbricate character of site integrals $F_{i,j}$ the process of integration can be achieved through three methods: i) Integrating from horizontal edge line $i = N$ to $i = -N$ (confused with $i = N$) between vertical edge line $j = -N$ and $j = N$ (confused with $j = -N$) or ii) vice versa; iii) integrating simultaneously from the four lattice edges $i = N$, $i = -N$, $j = -N$ and $j = N$ in the direction of the lattice heart characterized by site (0,0). For sake of simplicity we choose case iii for the integration process.

As we exclusively focus on the case of infinite lattices the polynomial expression of the numerator of Eq. (3.1) is such as all the integrals $F_{i,j}$ not involved in correlation paths are characterized by a collection of integers $l'_{i+1,j}$, $l_{i,j-1}$, $l'_{i+1,j}$ and $l_{i,j}$ which are all equal to $l$. This part of the lattice constitutes the *wing domain*. The remaining part which contains all the possible correlation paths is the *correlation domain*. This must be true for a finite lattice showing edges (Fig. 5a) or an infinite lattice (thermodynamic limit, Fig. 5b). Now we have to answer to the following question: *How to define the frontier between the wing domain and the correlation one?*

In Subsect. III.C1 we have seen that, for line $i = 0$, the decomposition law given by Eqs. (3.3)-(3.5) must exclusively concern a spherical harmonics $Y_{l,0}(S_{0,j})$. In that case too, at site (0,1), if applying the decomposition law to $Y_{l_{0,1},0}(S_{0,1})$ we derive $l'_{0,1} = l + \varepsilon$ (with $\varepsilon = \pm 1$, Fig. 6a) or $l'_{1,1} = l + \varepsilon$ (Fig. 6b). Integral $F_{-1,1}$ or $F_{1,1}$ vanishes (*cf* Eqs. (3.5)-(3.8)). It means that the correlation path on line $i = 0$ showing *a loop outside* (Fig. 6a) or *inside* (Fig. 6b) *the correlation domain is forbidden*. A similar reasoning can be achieved for the horizontal line $i = k$ (Fig. 6a and Fig. 6b) or the vertical lines $j = 0$ and $j = k'$ (Fig. 6c). This allows one to conclude that the correlation domain between sites (0,0) and ($k,k'$) is a rectangle built on these sites and the other sites ($k$,0) and (0,$k'$). This rectangle contains all the correlation paths. All the previous results are summarized in the following theorem:

*Theorem 1*:

*For calculating the numerator of the spin-spin correlation $<S_{i,j}^z \cdot S_{i+k,j+k'}^z>$, it is necessary to take into account two domains: A correlation domain which is a rectangle of vertices (i j), (i,j+k'), (i+k,j+k') and (i+k,j) within which all the correlation paths are confined, and a remaining domain called wing domain. In both domains, for an infinite lattice, we have m = 0. All the bonds of the wing domain are characterized by the same coefficient l, including the bonds linked to the correlation domain.*

At site (0,0), if wishing to calculate the correlation with any lattice site, the decomposition law leads to two types of correlation path respectively labeled cases i and ii (see Fig. 7a) because there are two possible choices: $l_{0,0} = l + \varepsilon$ and $l'_{0,0} = l$ (case i) or $l'_{0,0} = l + \varepsilon$ and $l_{0,0} = l$ (case ii).

i) Let us consider the correlation path of case i. This path starts at site (0,0), follows the horizontal line $i = 0$ up to site (0,$k'$) and, from this site, finally runs along the vertical line $j = k'$ for terminating at site ($k,k'$). The work of integration in the wing domain is done so that the lattice is simultaneously swept for the horizontal lines starting at $i = -N$ and ending at $i = -1$, line $i = N$ down to $i = k + 1$, the vertical lines $j = -N$ to $j = -1$, and $j = N$ to $j = k' + 1$. All the horizontal bonds of the correlation domain not involved in the correlation path are characterized by integer $l$ (due to the fact that we deal with the highest-degree term) whereas those belonging to the correlation path show the integer $l + \varepsilon$ ($\varepsilon = \pm 1$). We have a similar reasoning for the vertical lines $j$. In addition, when the wing domain has been fully swept (i.e., the integration process being over), the correlation domain can be also swept from line $i = k$ to line $i = 0$, and line $j = 0$ to $j = k'$ (or vice versa). It finally means that, when the integration process has been achieved for a whole vertical line $j$ from $j = 0$ to $j = k'$ but also for a whole vertical line $i$ from $i = k$ to $i = 0$, it is impossible to go backwards. In other words, *loops inside the correlation domain are forbidden* (see Fig. 6d).

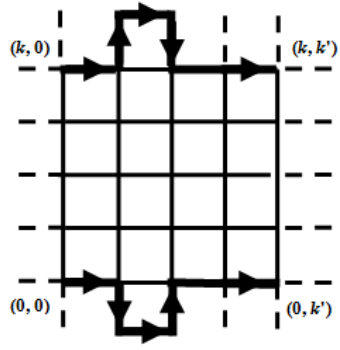
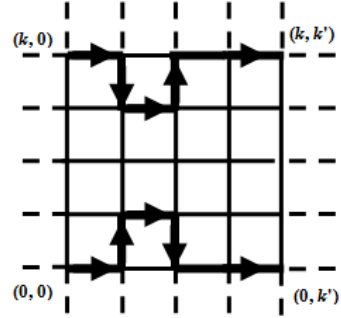

FIG. 6a    FIG. 6b

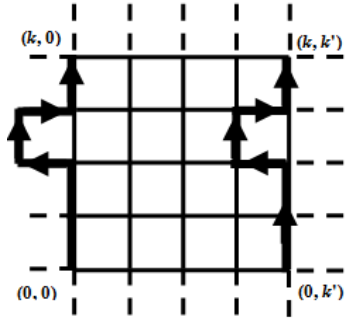
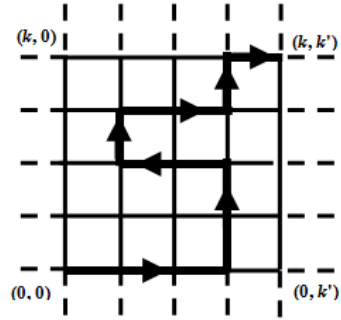

FIG. 6c    FIG. 6d

FIG. 6. a), b), c) Examples of forbidden correlation paths (loops) at the frontier between the wing domain and the correlation domain, for an infinite lattice ($N \rightarrow +\infty$); the correlation path is characterized by a thick line and each bond of the path by integer $l + \varepsilon$ ($\varepsilon = \pm 1$); the other bonds which do not belong to correlation paths are characterized by a thin line and integer $l$; d) another example of forbidden correlation path showing a loop inside the correlation domain.

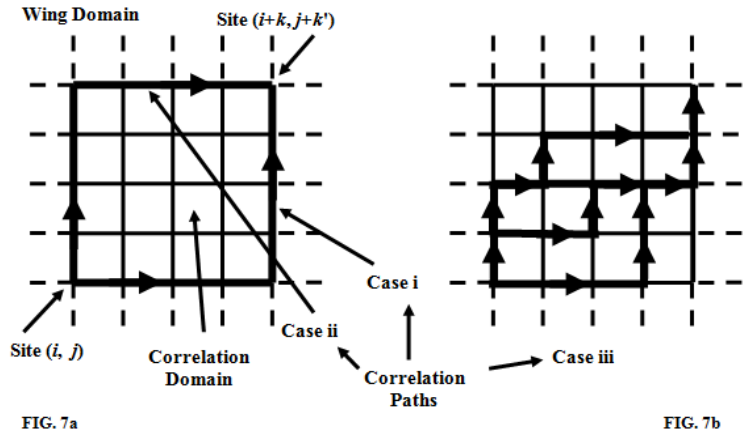

FIG. 7a    FIG. 7b

FIG.7. a) Correlation paths along the frontier between the correlation and the wing domains (cases i and ii); examples of correlation paths inside the correlation domain (case iii); all these paths are characterized by the shortest length between sites $(i, j)$ and $(i + k, j + k')$ and are equivalent i.e., they show the same weight.

Thus, in case i, all the vertical bonds linked to the horizontal line $i = 0$ as well as all the horizontal bonds linked to the vertical line $j = k'$ are characterized by integer $l$. For line $i = 0$, the calculation of the contribution to the spin-spin correlation $<S_{0,0}^z.S_{k,k'}^z>$ is strictly the same one as that one described for evaluating $<S_{0,0}^z.S_{0,k'}^z>$. At site (0,0) we have $l_{0,0} = l + \varepsilon$ and $l'_{0,0} = l$ and the full contribution is $C_{l+\varepsilon}F_{l+\varepsilon,l+\varepsilon}\lambda_{l+\varepsilon}(-\beta J_1)$, ($\varepsilon = \pm1$). For all the remaining sites $K'$ of line $i = 0$ such as $0 < K' < k'$, each bond brings the contribution $F_{l+\varepsilon,l+\varepsilon}\lambda(-\beta J_1)$ to the correlation so that, for all the sites between (0,0) and (0,$k'$ –1), we have the contribution $C_{l+\varepsilon}(F_{l+\varepsilon,l+\varepsilon}\lambda_{l+\varepsilon}(-\beta J_1))^{k'}$. At site (0,$k'$) the integration work is slightly different due to the fact that we have $l_{0,k'} = l'_{0,k'} = l$ but, now, we have to determine $l'_{1,k'}$. The non-vanishing condition of integral $F_{0,k'}$ given by Eq. (3.12) in which $l_{0,0}$ is replaced by $l'_{1,k'}$ leads to $l'_{1,k'} = l + \varepsilon$ ($\varepsilon = \pm1$). Thus, the corresponding contribution is $F_{l+\varepsilon,l+\varepsilon}\lambda_{l+\varepsilon}(-\beta J_2)$. A similar work of integration done on the horizontal line $i = 0$ can be achieved for all the sites ($K,k'$) belonging to the vertical line $j = k'$, such as $0 < K < k$: Due to the integration achieved for all the remaining bonds out of the correlation path we have $l_{K,k'-1} = l_{K,k'} = l$; the non-vanishing condition of integral $F_{K,k'}$ leads to $l'_{K,k'} = l'_{k-1,k'} = l + \varepsilon$ ($\varepsilon = \pm1$). The corresponding contribution of the correlation path bonds is $(F_{l+\varepsilon,l+\varepsilon}\lambda_{l+\varepsilon}(-\beta J_2))^k$. At site ($k,k'$) the decomposition law is applied on integral $F_{k,k'}$: All the involved integers $l_{k-1,k'}$, $l'_{k+1,k'}$ and $l_{k,k'}$ have been already determined due to the integration in the wing domain ($l'_{k+1,k'}$ and $l_{k,k'}$) or along the correlation path ($l_{k-1,k'}$); they are all equal to $l$ and $l'_{k,k'} = l + \varepsilon$ ($\varepsilon = \pm1$). The result of the integration work is then similar than in Subsec. III.C1 (site (0,1)) or in Subsec. III.C2 (site (0,$k'$)) and gives the numerical factor $C_{l+1}F_{l+\varepsilon,l+1} + C_{l-1}F_{l+\varepsilon,l-1}$ ($\varepsilon = \pm1$). As a result the final contribution of all the sites of the correlation path between sites (0,0) and ($k,k'$) is
$C_{l+\varepsilon}(C_{l+1}F_{l+\varepsilon,l+1} + C_{l-1}F_{l+\varepsilon,l-1})(F_{l+\varepsilon,l+\varepsilon}\lambda_{l+\varepsilon}(-\beta J_1))^{k'} \times$
$\times (F_{l+\varepsilon,l+\varepsilon}\lambda_{l+\varepsilon}(-\beta J_2))^k$ ($\varepsilon = \pm1$) (case i).

ii) Let us consider the correlation path of case ii. This path starts at site (0, 0), follows the vertical line $j = 0$ up to site ($k$,0) and, from this site, finally runs along the horizontal line $i = k$ for terminating at site ($k,k'$). At site (0,0) we have the other possibility $l'_{0,0} = l + \varepsilon$ and $l_{0,0} = l$ and the corresponding contribution is now $C_{l+\varepsilon}F_{l+\varepsilon,l+\varepsilon}\lambda_{l+\varepsilon}(-\beta J_2)$. For all the remaining sites $K$ of line $j = 0$ such as $0 < K < k$, each bond brings the contribution $F_{l+\varepsilon,l+\varepsilon}\lambda_{l+\varepsilon}(-\beta J_2)$ to the correlation path so that for all the sites between (0,0) and ($k$ –1,0), we have the following contribution $C_{l+\varepsilon}(F_{l+\varepsilon,l+\varepsilon}\lambda_{l+\varepsilon}(-\beta J_2))^k$. At site ($k$,0) the integration work is slightly different. The integers $l_{k-1,0}$ and $l'_{k+1,0}$ have been already determined due to the integration in the wing domain ($l_{k-1,0} = l'_{k+1,0} = l$) and $l'_{k,0} = l + \varepsilon$ ($\varepsilon = \pm1$) due to the integration on the correlation path at site ($k$–1,0) but, now, $l_{k,0} \neq l$. The non-vanishing condition of integral $F_{k,0}$ given by Eq. (3.12) in which $l_{0,0}$ is replaced by $l_{k,0}$ leads to $l_{k,0} = l + \varepsilon$ ($\varepsilon = \pm1$). A similar work of integration can be achieved for all the sites ($k,K'$) belonging to the horizontal line $i = k$, such as $0 < K' < k'$: Their full contribution is $(F_{l+\varepsilon,l+\varepsilon}\lambda_{l+\varepsilon}(-\beta J_1))^{k'}$. At site ($k,k'$) the decomposition law is applied on integral $F_{k,k'}$: All the involved integers $l'_{k,k'}$, $l'_{k+1,k'}$ and $l_{k,k'}$ have been already determined and are all equal to $l$, $l_{k-1,k'} = l + \varepsilon$ ($\varepsilon = \pm1$) due to integration at site ($k$–1,$k'$). The result of the integration work gives the numerical factor $C_{l+1}F_{l+\varepsilon,l+1} + C_{l-1}F_{l+\varepsilon,l-1}$ ($\varepsilon = \pm1$) like in case i. Finally the contribution of all the sites of the correlation path is
$C_{l+\varepsilon}(C_{l+1}F_{l+\varepsilon,l+1} + C_{l-1}F_{l+\varepsilon,l-1})(F_{l+\varepsilon,l+\varepsilon}\lambda_{l+\varepsilon}(-\beta J_1))^{k'} \times$
$\times (F_{l+\varepsilon,l+\varepsilon}\lambda_{l+\varepsilon}(-\beta J_2))^k$ ( case ii, $\varepsilon = \pm1$).

At this step two remarks must be made: (i) The correlation paths of cases i and ii contain the same number of horizontal bonds characterized by $J_1$ and vertical bonds characterized by $J_2$; their respective contributions to the corresponding spin-spin correlation are equal which means that the spin-spin correlations are equal as expected; (ii) *these correlation paths correspond to the shortest possible path between sites* (0,0) *and* ($k,k'$). This constitutes an illustration of the principle of least action (Maupertuis' principle for classical spins).

iii) Case iii is a mix of cases i and ii (see Fig. 7b). For instance one can begin as in case i: The correlation path is composed of horizontal bonds up to the current site (0,$K'$) of line $i= 0$. At this site the integrand of $F_{0,K'}$ given by Eq. (2.17) is $Y_{l'_{1,K'},0}(S_{0,K'})Y_{l+\varepsilon,0}(S_{0,K'})Y_{l_{0,K'},0}(S_{0,K'})Y_{l,0}(S_{0,K'})$ due to the integration process in the wing domain which imposes $l'_{0,K'} = l$ and the integration on site (0,$K'$–1) which gives $l_{0,K'-1} = l + \varepsilon$ ($\varepsilon = \pm1$). As a result it remains two possibilities corresponding for the choice of integers $l'_{1,K'}$ and $l_{0,K'}$: (i) The non-vanishing condition of integral $F_{0,K'}$ given by Eq. (3.14) allows to write $l'_{1,K'} = l + \varepsilon$ ($\varepsilon = \pm1$) and $l_{0,K'} = l$ due to the fact the polynomial giving the numerator of $<S_{0,0}^z.S_{k,k'}^z>$ (*cf* Eq. (3.1)) is restricted to its higher-degree term (case iii1); (ii) a similar reasoning can lead to choose $l_{0,K'} = l + \varepsilon$ ($\varepsilon = \pm1$) and $l'_{1,K'} = l$ (case iii2). In case iii1 the correlation path follows a vertical bond whereas in case iii2 we deal with a horizontal bond. This situation can be finally encountered for each lattice site and is illustrated by the correlation paths of Fig. 7b.

When comparing with the correlation paths i and ii we derive that: (i) The correlation paths of cases i, ii and iii contain the same number of horizontal bonds (characterized by $J_1$) and vertical bonds (characterized by $J_2$); all the combinations involving horizontal and vertical bonds allowing to write the spin-spin correlation are equal which means that the spin-spin correlations show a unique expression, as expected for this kind of lattice; (ii) *these correlation paths correspond to the shortest possible paths between sites* (0,0) *and* ($k,k'$). It is then easy to determine the total number of these paths: It is simply $\binom{k+k'}{k'}$; thus all the paths are equivalent and are characterized by the weight $\binom{k+k'}{k'}^{-1}$.

*Theorem 2:*

*All the correlation paths show the same length inside the correlation domain. These paths are the shortest possible ones in agreement with Maupertuis' principle. They respectively involve the same number of horizontal and vertical bonds than the horizontal and vertical sides of the correlation rectangle, for a 2d infinite square lattice.*

As a result the spin-spin correlation $<S^z_{0,0}.S^z_{k,k'}>$ can be written:

$$<S^z_{0,0}.S^z_{k,k'}> = \frac{(4\pi)^{8N^2}}{Z_N(0)} \sum_{l=0}^{+\infty} {}^t[F_{l,l}\lambda_l(-\beta J_1)\lambda_l(-\beta J_2)]^{4N^2} \times$$
$$\times [K_{l+1}(P_{1,l+1})^{k'}(P_{2,l+1})^k$$
$$+ (1-\delta_{l,0})K_{l-1}(P_{1,l-1})^{k'}(P_{2,l-1})^k],$$

$$k' > 0, \text{ as } N \to +\infty \quad (3.19)$$

where the coefficients $K_{l+1}$ and $K_{l-1}$ and the ratio $P_{i,l+\varepsilon}$ ($i = 1, 2$) are given by Eq. (3.16) and the integrals $F_{l+\varepsilon,l+\varepsilon'}$ by Eq. (3.14).

### D. Expression of the correlation length and the susceptibility

The correlation length can be derived owing to Eq. (2.6) so that

$$\xi = \left[ \frac{\sum_{l=0}^{+\infty} {}^tz_l^{4N^2}[N_{l+1} + (1-\delta_{l,0})N_{l-1}]}{\sum_{l=0}^{+\infty} {}^tz_l^{4N^2}[D_{l+1} + (1-\delta_{l,0})D_{l-1}]} \right]^{1/2}, \text{ as } N \to +\infty \quad (3.20)$$

with

$$D_{l+\varepsilon} = K_{l+\varepsilon}\frac{1}{(1-|P_{1,l+\varepsilon}|)(1-|P_{2,l+\varepsilon}|)},$$

$$N_{l+\varepsilon} = D_{l+\varepsilon}\left[\frac{|P_{1,l+\varepsilon}|(1+|P_{1,l+\varepsilon}|)}{(1-|P_{1,l+\varepsilon}|)^2} + \frac{|P_{2,l+\varepsilon}|(1+|P_{2,l+\varepsilon}|)}{(1-|P_{2,l+\varepsilon}|)^2}\right] \quad (3.21)$$

where $z_l$ is defined in Eq. (2.37); $P_{i,l+\varepsilon}$ ($i = 1, 2$) and the coefficients $K_{l+\varepsilon}$ are given by Eq. (3.16) and the integrals $F_{l+\varepsilon,l+\varepsilon'}$ by Eq. (3.14).

The total susceptibility per lattice site $(i, j)$ $\chi_{i,j}$ is defined by Eq. (2.7). Due to the isotropic character of spin couplings we have previously seen that one can also define a susceptibility per axis $\chi_{i,j} = \chi_{i,j}/3$ (*cf* Eq. (2.11)). As a result, for a practical purpose, we can consider the static susceptibility per square unit cell and averaged per lattice site which can be written as:

$$\chi = \frac{1}{4}(\chi_{0,0} + \chi_{0,1} + \chi_{1,0} + \chi_{1,1}) \quad (3.22)$$

where the susceptibility per site, referred per lattice axis, is

$$\chi_{k,k'} = \frac{1}{Z_N(0)} \sum_{l=0}^{+\infty} {}^t[F_{l,l}\lambda_l(-\beta J_1)\lambda_l(-\beta J_2)]^{4N^2}[\chi_{l+1} + (1-\delta_{l,0})\chi_{l-1}], \text{ as } N \to +\infty \ (k = 0 \text{ or } 1, k' = 0 \text{ or } 1) \quad (3.23)$$

with:

$$\chi_{l+\varepsilon} = \frac{\beta}{2}K_{l+\varepsilon}\frac{(G^2 + G'^2)W_{1,l+\varepsilon} + 4GG'W_{2,l+\varepsilon}}{W_{3,l+\varepsilon}}, G \neq G', \ \chi_{l+\varepsilon} = \beta G^2 K_{l+\varepsilon}\frac{(1+P_{1,l+\varepsilon})(1+P_{2,l+\varepsilon})}{(1-P_{1,l+\varepsilon})(1-P_{2,l+\varepsilon})}, G = G',$$
$$\varepsilon = \pm 1, \text{ as } N \to +\infty, \quad (3.24)$$

and

$$W_{1,l+\varepsilon} = [1+(P_{1,l+\varepsilon})^2][1+(P_{2,l+\varepsilon})^2] + 4P_{1,l+\varepsilon}P_{2,l+\varepsilon}, \ W_{2,l+\varepsilon} = P_{1,l+\varepsilon}(1+(P_{2,l+\varepsilon})^2) + P_{2,l+\varepsilon}(1+(P_{1,l+\varepsilon})^2),$$
$$W_{3,l+\varepsilon} = [1-(P_{1,l+\varepsilon})^2][1-(P_{2,l+\varepsilon})^2] \quad (3.25)$$

where the coefficients $K_{l+1}$ and $K_{l-1}$ and the ratio $P_{i,l+\varepsilon}$ ($i = 1, 2$) are given by Eq. (3.16)

### IV. LOW-TEMPERATURE STUDY

In this section we examine the low-temperature behaviors of the free energy and specific heat densities on the one hand and, on the other one, the correlation length and the static susceptibility density. These latter thermodynamic functions are respectively characterized by a denominator of the type $1-|P_{i,l+\varepsilon}|$ or $1-(P_{i,l+\varepsilon})^2$, with $i = 1, 2$, where $P_{i,l+\varepsilon}$ is given by Eq. (3.16). As a result the study of this parameter (i.e., the study of ratios $\lambda_{l+\varepsilon}(-\beta|J_i|)/\lambda_l(-\beta|J_i|)$ and $F_{l,l+\varepsilon}/F_{l,l}$) is crucial.

### A. Preliminaries

We first examine the ratio $F_{l+\varepsilon,l+\varepsilon}/F_{l,l}$ ($\varepsilon = \pm 1$) where integrals $F_{l,l}$ and $F_{l+\varepsilon,l+\varepsilon}$ are respectively given by Eqs. (2.26), and (3.14). In Fig. 3 each of these ratios has been reported. They are always lower than unity and tends to unity in the infinite $l$-limit. In addition, in Figs. 2a and 2b, we can see that, in the low-temperature limit, the eigenvalues $\lambda_l(-\beta|J_i|)$ characterized by an infinite $l$ become dominant. In that case it is possible to obtain a closed-form expression of this ratio. If expressing the spherical harmonics involved in the definition of integral $F_{l,l}$ given by Eq. (2.17) in which $m = 0$ in the infinite $l$-limit [23]

$$Y_{l,0}(\theta,\varphi) \approx \frac{1}{\pi\sqrt{\sin\theta}}\left\{\left(1-\frac{3}{8l}\right)\cos\left((2l+1)\frac{\theta}{2}-\frac{\pi}{4}\right)\right.$$
$$\left. -\frac{1}{8l\sin\theta}\cos\left((2l+3)\frac{\theta}{2}-\frac{3\pi}{4}\right)\right\} + O\left(\frac{1}{l^2}\right), \text{ as } l \to +\infty,$$

$$\varepsilon' \le \theta \le \pi - \varepsilon', \ 0 < \varepsilon' << 1/l, \ 0 \le \varphi \le 2\pi. \quad (4.1)$$

we have the exact asymptotic result:

$$\frac{F_{l+\varepsilon,l+\varepsilon}}{F_{l,l}} \to 1 + O\left(\frac{1}{l^2}\right), \ \varepsilon = \pm 1, \text{ as } l \to +\infty. \quad (4.2)$$

It means that the ratios $P_{i,l+\varepsilon}$ defined by Eq.(3.16) reduce to the ratio of eigenvalues $\lambda_{l+\varepsilon}(-\beta|J_i|)/\lambda_l(-\beta|J_i|)$

$$P_{i,l\pm 1} \approx \frac{\lambda_{l\pm 1}(-\beta J_i)}{\lambda_l(-\beta J_i)}, \text{ as } T \to 0 \quad (4.3)$$

i.e., due to Eq. (2.15), $I_{l+\varepsilon+1/2}(-\beta|J_i|)/I_{l+1/2}(-\beta|J_i|)$, with $\varepsilon = \pm 1$ ($i = 1, 2$) and presently $I_{l+1/2}(-\beta|J_i|) \approx I_l(-\beta|J_i|)$.

For simplifying the discussion we restrict to the case $J_1 = J_2 = J$ without loss of generality. Intuitively, in the low-temperature limit, we must consider the three cases $\beta|J| >> l$, $\beta|J| \sim l$ and $\beta|J| << l$. The behavior of Bessel function $I_l(-\beta|J|)$ as $l \to +\infty$ and $\beta|J| \to +\infty$ has been established by Olver [24]. In a previous paper [19] we have extended this work for a large order $l$ (but not necessarily infinite) and for any real argument $\beta|J|$ varying from a finite value to infinity. Here we briefly summarize the key results. The study of the Bessel differential equation in the large $l$-limit necessitates the introduction of the dimensionless auxiliary variables:

$$\zeta = -\frac{J}{|J|}\left[\sqrt{1+z^2} + \ln\left(\frac{|z|}{1+\sqrt{1+z^2}}\right)\right], \ |z| = \frac{\beta|J|}{l}. \quad (4.4)$$

The numerical study of $|\zeta|$ is reported in Fig. 8. As expected we observe that there are two branches. $|\zeta|$ vanishes for a numerical value of $|z_0|^{-1}$ very closed to $\pi/2$ so that there are 3 domains which will be physically interpreted in next subsection. Let be $T_0$ the corresponding temperature. We set:

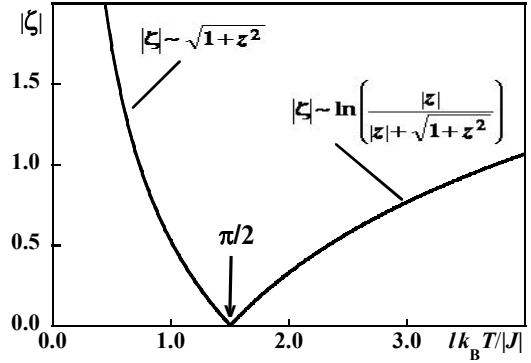

FIG. 8. Thermal variations of $|\zeta|$ for various values of $lk_BT/|J| = 1/|z|$.

$$l\frac{k_BT_0}{|J|} = \frac{\pi}{2}. \quad (4.5)$$

In the formalism of renormalization group $T_0$ is called a *fixed point*. In the present 2d case we have $l \to +\infty$. As a result we derive that $T_0 \to T_c = 0$ K as $l \to +\infty$ so that the critical temperature can be seen as a fixed point. In other words it means that all the thermodynamic functions can be expanded as series of current term $|T - T_0|$ near $T_0 \approx T_c = 0$ K, in the infinite $l$-limit.

At this step we must recall that the spin modulus $S(S+1) \sim S^2$, as $S \to +\infty$, is absorbed in the exchange energy $J$. Chakravarty et al. [6] as well as Chubukov et al. [10] have written that the action $S/\hbar$ (which allows one to calculate the partition function) is proportional to $J/2$. In addition these authors have considered the spin density $S/a$ where $a$ is the lattice spacing. In our case the lattice spacing between two similar Landé factors $G$ or $G'$ is $2a$. As a result, the left member of Eq. (4.5) can be written as $lk_BT_0/(|J|/2)(S/2a)^2$ so that the right member $\pi/2$ becomes $4\pi$. We must keep in mind this remark because it will be very useful later.

In Fig. 8 we observe that, below $T_c$, we have $|z| >> 1$ (i.e., $\beta|J| >> l$) and $|\zeta| \sim (1 + z^2)^{1/2}$ (i.e., $|\zeta| \sim |z|$). Above $T_c$, we have $|z| << 1$ (i.e., $\beta|J| << l$) and $|\zeta| \sim 1 + \ln(|z|)/(1 + (1 + z^2)^{1/2})$ (i.e., $|\zeta| \sim 1 + \ln(|z|/2)$). Finally, if $T \approx T_0 = T_c$, $|z| \approx z_0$ i.e., $\beta|J| \approx l$ and $|\zeta| \approx 0$.

For convenience we now introduce the coupling constant $g$ at temperature $T$ as well as its reduced value $g^*$:

$$g = \frac{k_BT}{|J|}, \ g^* = \frac{T}{T_c}. \quad (4.6)$$

$g$ measures the strength of spin fluctuations. $g^*$ is a universal parameter and is $l$-independent. At the critical point $T_0 = T_c$ we have $g^* = 1$. Owing to Eq. (4.4), the critical coupling $g_c$ can be written as:

$$g_c = \frac{k_BT_c}{|J|}, \ g_c = \frac{\pi}{2l}. \quad (4.7)$$

Chubukov *et al.* have found that, at the critical temperature $T_c$, the critical coupling is $g_c = 4\pi/\Lambda$ where $\Lambda = 2\pi/a$ is a relativistic cutoff parameter (*a* being the lattice spacing) [10]. Haldane has evaluated $g_c$ in the case of a classical spin lattice [11,12]. He proposed $g_c = 2\sqrt{d}a/S$ or equivalently $g_c = 2a/S$ if referring to the vertical rows or horizontal lines of the 2*d*-lattice characterized by the same exchange energy $J = J_1 = J_2$. In our case $S = 1$ so that $g_c = 2a$ or

$$g_c^* = \frac{4\pi}{\Lambda}, \quad \Lambda = \frac{2\pi}{a}. \tag{4.8}$$

We introduce: i) The thermal de Broglie wavelength $\lambda_{DB}$, ii) the low-temperature spin wave celerity $c = 2\sqrt{2}|J|Sa/\hbar$ along the diagonal of the lattice (i.e., $c_u = 2|J|a/\hbar$ along the vertical rows ($u = y$) or horizontal lines ($u = x$) of the lattice characterized by the same exchange energy $J = J_1 = J_2$ and spacing *a*, with $\sqrt{S(S+1)} = 1$) and iii) the slab thickness $L_\tau$ of the *D*-space-time ($D = 3$):

$$\lambda_{DB} = 2\pi \frac{\hbar c}{k_B T}, \quad L_\tau = \frac{\hbar c}{k_B T}. \tag{4.9}$$

In other words we have $L_\tau = \lambda_{DB}/2\pi$. By definition we must have

$$\lambda_{DB} \gg a \tag{4.10}$$

i.e., $\Lambda = 2\pi/a \gg 2\pi/\lambda_{DB}$ or equivalently

$$\Lambda \gg L_\tau^{-1}. \tag{4.11}$$

At the critical point $T_c = 0$ K $\lambda_{DB} \to +\infty$ (as well as $L_\tau$). It means that spins are strongly correlated. For finite temperatures $\lambda_{DB}$ and $L_\tau$ become finite. The adequate tool for estimating the correlation between any couple of spins is the correlation length $\xi$. As a result $\lambda_{DB}$ (or $L_\tau$) appears as the good unit length for measuring $\xi$.

Under these conditions we generalize the application of the cutoff parameter $\Lambda$. $|z|$ defined by the second of Eq. (4.4) can be rewritten if using Eq. (4.6)

$$|z^*| = \frac{\beta|J|}{\Lambda} = \frac{z_c^*}{g^*}, \quad z_c^* = \frac{1}{4\pi} \tag{4.12}$$

and $|z^*|\Lambda = |z|l = \beta|J|$ finally appears as

$$|z^*|\Lambda = \beta|J| = \frac{L_\tau}{2a\sqrt{2}} \tag{4.13}$$

if using the relation $\hbar c = 2\sqrt{2}|J|Sa$ (with $S = 1$). Thus, if considering the correlation length as a scaling parameter near the critical point $T_c = 0$ K, its measure $\xi$ along the diagonal of the lattice (if $J = J_1 = J_2$) characterized by a spacing *a* (i.e., in the $\Lambda$-scale) is $\xi^* = \xi(2a\sqrt{2})$ (or $\xi^*\Lambda = 4\pi\sqrt{2}\xi$) and $\xi_\tau^* = \xi L_\tau$ along the slab thickness of the *D*-space-time (i.e., the *i*$\tau$-axis), due to *scale invariance*. These respective notations can be generalized to any physical parameter. We have the dimensionless relations near $T_c = 0$ K and for a fixed value of $T$

$$\xi = \frac{\xi^*}{2a\sqrt{2}} = \frac{\xi_\tau^*}{L_\tau}, \quad \frac{\xi_\tau^*}{\xi^*} = \frac{L_\tau}{2a\sqrt{2}} = \beta|J|, \text{ as } T \to 0. \tag{4.14}$$

In Appendix B we have established the *l*-polynomial expansion of the ratio $P_{l\pm 1} \approx \lambda_{l\pm 1}(lz)/\lambda_l(lz)$ (notably containing the dimensionless variable $l|\zeta|$), near $T_c = 0$ K. The argument $lz$ is replaced by $z^*\Lambda$ and we have shown that $\lambda_{l\pm 1}(lz)/\lambda_l(lz) \sim \lambda_{\Lambda\pm 1}(z^*\Lambda)/\lambda_\Lambda(z^*\Lambda)$ i.e., $P_{l\pm 1} \approx P_{\Lambda\pm 1}$, as $T \to T_c = 0$. Simultaneously, $l|\zeta|$ must be replaced by $|\zeta^*|\Lambda$ whose expression has been established in Appendix B. We have

$$P_{\Lambda\pm 1} \approx \frac{\lambda_{\Lambda\pm 1}(z^*\Lambda)}{\lambda_\Lambda(z^*\Lambda)} = -\frac{J}{|J|}\left\{\mp\left(\frac{1}{|z^*|} + \frac{1}{|z^*|\Lambda}\right) + \frac{I'_\Lambda(|z^*|\Lambda)}{I_\Lambda(|z^*|\Lambda)}\right\}, \text{ as } T \to 0, \quad I'_\Lambda(|z^*|\Lambda) = \frac{dI_\Lambda(|z^*|\Lambda)}{d(|z^*|\Lambda)},$$

$$\frac{I'_\Lambda(|z^*|\Lambda)}{I_\Lambda(|z^*|\Lambda)} \approx \frac{1}{u^*|z^*|}\left[1 - \frac{u^*}{\Lambda} - \frac{u^{*2}}{2\Lambda^2} - 2\exp(-|\zeta^*|\Lambda)\left(1 - \frac{u^*}{2\Lambda} - \frac{3u^{*2}}{8\Lambda^2}\right) + O(\Lambda^{-3})\right], \text{ as } T \to 0, \tag{4.15}$$

where the parameter $u^*$ is given by:

$$u^* = \frac{1}{\sqrt{1+z^{*2}}}. \tag{4.16}$$

As a result, in the low-temperature domain, it becomes possible to expand the ratios $\lambda_{\Lambda\pm 1}(z^*\Lambda)/\lambda_\Lambda(z^*\Lambda)$ vs a general parameter that we now need to define.

Chakravarty *et al.* [6] have introduced the physical parameters $\rho_s$ and $\Delta$ defined as:

$$\rho_s = |J|(1 - g^*), \quad \Delta = |J|(g^* - 1). \tag{4.17}$$

In the 2*d*-case $\rho_s$ and $\Delta$ have the dimension of an energy $JS^2$ (in our case *J*). $\rho_s$ is the *spin stiffness* of the ordered ground state (Néel state for an antiferromagnet) and $\Delta$ is the *T=0-energy gap between the ground state and the first excited state*. In the framework of the classical spin approximation the spectrum is quasi continuous. In our case it means that $\Delta$ is very small. The writing of $\rho_s$ and $\Delta$ is motivated by the fact that these parameters must obey Josephson's scaling law near $T_c$ [26]:

$$\rho_s \approx (g_c - g)^{(D-2)\nu}, \Delta \approx (g - g_c)^{(D-2)\nu} \qquad (4.18)$$

where $D = d + 1$ is the space-time dimension (here $D = 2 + 1$) and $\nu$ is the usual correlation length critical exponent ($\nu = 1$ as evaluated by Chakravarty et al. [6]). At the critical point $T_c = 0$ K $g^* = 1$: $\rho_s$ and $\Delta$ vanish and, near critically, we have $\rho_s \ll |J|$ and $\Delta \ll |J|$ where $|J|$ finally appears as the bare value of $\rho_s$ and $\Delta$ i.e., their value at 0 K. For all the previous reasons we are led to introduce the following parameters:

$$\frac{\rho_s}{k_B T} = \frac{1}{g} - \frac{1}{g_c} \ (T < T_c), \frac{\Delta}{k_B T} = 4\pi\left(\frac{1}{g_c} - \frac{1}{g}\right) \ (T > T_c)$$
(4.19)

where the factor $4\pi$ appears in $\Delta$ for notational convenience (the correspondence with the notation of Chakravarty et al. is assumed in Eq. (4.18) by multiplying $\Delta$ by the factor $4\pi$).

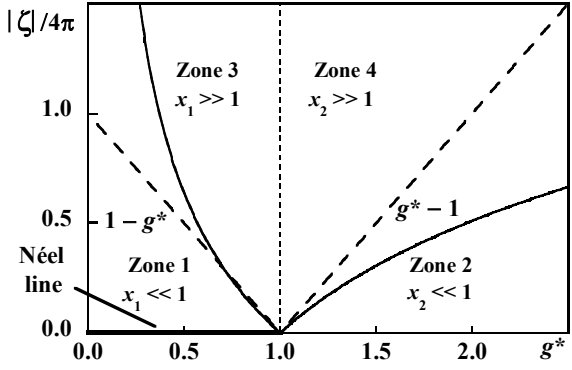
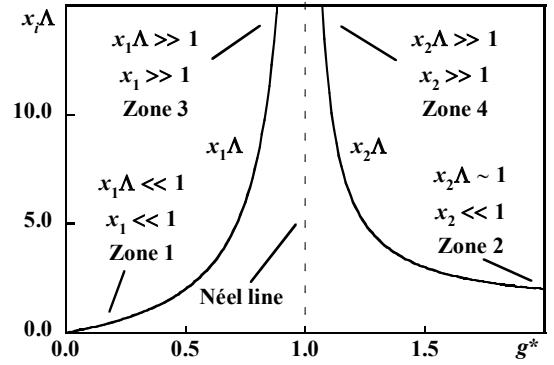

FIG. 9. Thermal variations of $|\zeta|/4\pi$ vs $g^*$ respectively defined by Eqs. (4.4) and (4.6) and domains of predominance vs dimensionless parameters $x_1$ and $x_2$ defined by Eq. (4.21); $x_1$ and $x_2$ control the scaling properties of the magnetic system.

FIG. 10. Plot of $x_1\Lambda$ and $x_2\Lambda$ defined by Eq. (4.22) vs $g^*= T/T_c$ and domains of predominance derived from the asymptotic behaviors of $x_1\Lambda$ and $x_2\Lambda$.

The expression giving $g^*$ must be handled with care because we have $T_0 \to T_c = 0$ K as $\Lambda \to +\infty$. Thus, when $g^* < 1$, it means that this ratio remains finite by always imposing $T < T_0$ when $T_0 \to T_c$ i.e., owing to Eqs. (4.6) and (4.7) which can be rewritten as $|z^*|/z_c^* = 1/g^*$, $\beta|J| > \Lambda$. When $g^* > 1$ we always have $T > T_0$ and $\beta|J| < \Lambda$. Finally $\beta|J| \sim \Lambda$ means that we have $T \sim T_0 = T_c$ in the vanishing temperature limit.

Now we express all the scaling parameters previously encountered. In a first step, we expand $|\zeta|$ in the low-temperature limit. We have the following asymptotic behaviors:

$$\frac{|\zeta|}{4\pi} \approx |\zeta_F|_<, |\zeta_F|_< = 1 - g^*, T < T_c,$$

$$\frac{|\zeta|}{4\pi} \approx |\zeta_F|_>, |\zeta_F|_> = g^* - 1, T > T_c, \qquad (4.20)$$

so that the thermal study of $|\zeta|$ is reduced to two domains: $g^* < 1$ i.e., $g < g_c$ ($T < T_c$) and $g^* > 1$ i.e., $g > g_c$ ($T > T_c$). Each of these domains can be divided itself into two subdomains according to as $|\zeta| > |\zeta_F|$ or $|\zeta| < |\zeta_F|$. In other words, in Fig. 9, if considering a vertical line $g^* = K$ where $K$ is any positive constant, with $g^* < 1$ or $g^* > 1$, as well as two different temperatures $T_1$ and $T_2$ such as $T_1 > T_2$ for instance, we have $T_1\Lambda_1 = T_2\Lambda_2$ due to the second of Eq. (4.4) so that $\Lambda_1 < \Lambda_2$ with $\Lambda_1 \gg 1$ and $\Lambda_2 \gg 1$. If examining Eq. (4.20) near $T_0$ a small variation $dg^* < 0$ if $T < T_0$ or $dg^* > 0$ if $T > T_0$ when $T_0 \to T_c = 0$ K always implies $d|\zeta| > 0$. Physically it means that we must choose a parameter proportional to $(1 - g^*)^{-a}$ if $T < T_0$ and another one proportional to $(g^* - 1)^{-a}$ if $T > T_0$ with $a \geq 1$. In the limit $T \to T_0 \to T_c = 0$ K (i.e., $g \to g_c$ or $g^* \to 1$) these parameters must become infinite. If examining Eq. (4.19) $(\rho_s/k_B T)^{-1}$ if $T < T_0$ and $(\Delta/k_B T)^{-1}$ if $T > T_0$ are the good candidates.

As a result we are led to introduce the following parameters:

$$x_1 = \frac{k_B T}{2\pi\rho_s}, x_2 = \frac{k_B T}{\Delta} \qquad (4.21)$$

where the factor $2\pi = g_c\Lambda/2$ also appears for notational convenience (as noted for $\Delta$ after Eq. (4.19)). Thus, if using the definition of $\rho_s$ and $\Delta$ (cf Eq. (4.18)), we verify that $x_1$ and $x_2$ are functions which conveniently vary with $g^*$. As $\rho_s$ and $\Delta$ vanish at $T_0 = T_c$, $x_1$ and $x_2$ become infinite at this fixed point. Finally $x_1$ and $x_2$ can be also written as:

$$x_1 = \frac{2g^*}{1 - g^*}\Lambda^{-1}, x_2 = \frac{g^*}{g^* - 1}\Lambda^{-1}. \qquad (4.22)$$

When choosing different scales $\Lambda$ near $T_c = 0$ K i.e., through a dilation or a translation operation, it is easy to show that the corresponding expressions of $x_1$ and $x_2$ tend towards the same universal limit. Thus $x_1$ and $x_2$ are scaling parameters as well as $|z^*|/z_c^*$ and $|\zeta^*|\Lambda$ (see Appendix B). $x_1$ and $x_2$ appear as functions of the universal parameter $g^*$. From a physical point of view and as noted by Chakravarty et al. [6] as well as by Chubukov et al. [10], $x_1$ and $x_2$ control the scaling properties of the magnetic system. $x_1\Lambda$ and $x_2\Lambda$ have been plotted in Fig. 10.

In summary, the comparison between Figs. 9 and 10 allows one to write that for $g^* < 1$ ($T < T_c$) we have two possibilities: if $|\zeta| > |\zeta_F|_<$ then $x_1 > 1$ and if $|\zeta| < |\zeta_F|_<$ then $x_1 < 1$. Similarly, for $g^* > 1$ ($T > T_c$) we have: $x_2 > 1$ if $|\zeta| > |\zeta_F|_>$ and $x_2 < 1$ if $|\zeta| < |\zeta_F|_>$.

There is an analytical continuity between $x_1$ and $x_2$ while passing through $T_0 = T_c$. We then guess that there are only 3 domains of predominance: $x_1 \ll 1$ ($T < T_c$ and $|\zeta| < |\zeta_F|_<$, Zone 1) i.e., $\rho_s \gg k_B T$, $x_2 \ll 1$ ($T > T_c$ and $|\zeta| < |\zeta_F|_>$, Zone 2) i.e., $\Delta \gg k_B T$; finally $x_1 \gg 1$ ($T < T_c$ and $|\zeta| > |\zeta_F|_<$, Zone 3) i.e., $\rho_s \ll k_B T$ and $x_2 \gg 1$ ($T > T_c$ and $|\zeta| > |\zeta_F|_>$, Zone 4) i.e., $\Delta \ll k_B T$. In this latter domain, along the line $T = T_c$ ($g^* = 1$), we directly reach the *Néel line* (see Fig. 10). Each of these domains previously described corresponds to a particular magnetic regime. The physical meaning of each regime is going to arise from the low-temperature study of the ratio $\lambda_{\Lambda+1}(z^*\Lambda)/\lambda_\Lambda(z^*\Lambda)$ vs $x_1$ or $x_2$.

However it remains a last step i.e., expressing $|\zeta^*|\Lambda$ as a scaling parameter vs $x_1$ or $x_2$. In Appendix B we have rigorously shown that

$$|\zeta^*|\Lambda \approx 2\,\mathrm{argsh}\left(\frac{\exp(-1/x_1)}{2}\right), \text{ (Zones 1 and 3), (4.23)}$$

$$|\zeta^*|\Lambda \approx 2\,\mathrm{argsh}\left(\frac{\exp(1/2x_2)}{2}\right), \text{ (Zones 2 and 4). (4.24)}$$

The corresponding behaviors are reported in Fig. 11. As a result behaviors can be derived for the four zones of the magnetic diagram given by Fig. 9. In Zone 1 ($x_1\Lambda \ll 1$, $x_1 \ll 1$) we have

$$|\zeta^*|\Lambda \approx \exp(-1/x_1), \, x_1 \ll 1 \text{ (Zone 1).} \quad (4.25)$$

In Zone 2 ($x_2\Lambda \sim 1$, $x_2 \ll 1$)

$$|\zeta^*|\Lambda \approx \frac{1}{x_2} + 2\exp(-1/x_2), \, x_2 \ll 1 \text{ (Zone 2). (4.26)}$$

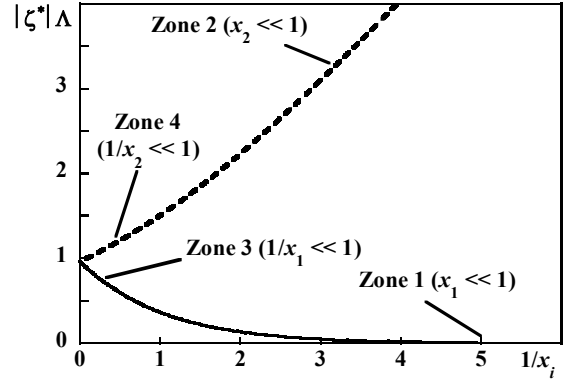

FIG. 11. Plot of $|\zeta^*|\Lambda$ vs the scaling parameters $x_1$ and $x_2$ defined by Eqs. (4.21) and (4.22).

In Zones 3 ($x_1\Lambda \gg 1$, $x_1 \gg 1$) and 4 ($x_2\Lambda \gg 1$, $x_2 \gg 1$) we have

$$|\zeta^*|\Lambda \approx 2\ln\left(\frac{1+\sqrt{5}}{2}\right) - \frac{2}{\sqrt{5}x_1}, \, x_1 \gg 1 \text{ (Zone 3), (4.27)}$$

$$|\zeta^*|\Lambda \approx 2\ln\left(\frac{1+\sqrt{5}}{2}\right) + \frac{1}{\sqrt{5}x_2}, \, x_2 \gg 1 \text{ (Zone 4).(4.28)}$$

At the common frontier between Zones 3 and 4, when directly reaching $T_c$, $x_1$ and $x_2$ become infinite (cf Fig. 10) and $|\zeta^*|\Lambda$ shows the common limit:

$$|\zeta_c^*|\Lambda \approx 2\ln\left(\frac{1+\sqrt{5}}{2}\right), \, x_1 \to +\infty, x_2 \to +\infty. \quad (4.29)$$

The ratio $\alpha = (1+\sqrt{5})/2$ is the golden mean. As a result, starting from a closed expression of $|\zeta|$ given by Eq. (4.4) (see Eq. (B.1)) we directly obtained for $|\zeta^*|\Lambda$ the result of Chubukov et al. derived from a renormalization technique [10]. Adopting their notation we define:

$$|\zeta^*|\Lambda \approx X_i(x_i), \, i = 1, 2. \quad (4.30)$$

Using all the previous results detailed in Appendix C the ratio $I'_{\Lambda\pm1}(z^*\Lambda)/I_\Lambda(z^*\Lambda)$ appearing in Eq. (4.15) and allowing to define the other ratio $P_{\Lambda\pm1} \approx \lambda_{\Lambda\pm1}(z^*\Lambda)/\lambda_\Lambda(z^*\Lambda)$, as $T \to 0$ can be expressed. This latter ratio appears in the closed-form expressions of the correlation length $\xi$ (cf Eqs. (3.20), (3.21)) and the susceptibility $\chi$ (cf Eq. (3.22)-(3.25)). Its low-temperature behaviors are then crucial for understanding the corresponding ones of $\xi$ and $\chi$. In Appendix C we have found

$$P_{\Lambda \pm 1} \approx -\frac{J}{|J|}\left[1 - \frac{8\pi}{e}|\zeta^*|\Lambda\left(1 - \frac{x_1}{2}\right) + ...\right],$$

as $T \to 0$, Zone 1 ($x_1\Lambda \ll 1$, $x_1 \ll 1$),

$$P_{\Lambda \pm 1} \approx -\frac{J}{|J|}\frac{x_2}{z_c^*}\left[\mp\frac{1}{x_2} + (|\zeta^*|\Lambda - 1) + ...\right], \text{ as } T \to 0,$$

Zone 2 ($x_2\Lambda' \sim 1$, $x_2 \ll 1$),

$$P_{\Lambda \pm 1} \approx -\frac{J}{|J|}\frac{1}{z_c^*}\left[\mp 1 + 2|\zeta^*|\Lambda - 1 + ...\right], \text{ as } T \to 0,$$

Zone 3 ($x_1\Lambda \gg 1$, $x_1 \gg 1$),
Zone 4 ($x_2\Lambda \gg 1$, $x_2 \gg 1$), as $T \to 0$, (4.31)

where $|\zeta^*|\Lambda$ is respectively given by Eqs. (4.23)-(4.29).

### B. Spin-spin correlations and correlation length

If considering the general $\Lambda$-expression of the spin-spin correlation given by Eq. (3.19) due to the fact that, near $T_c = 0$ K, $l$ is replaced by $\Lambda \to +\infty$, we have $C_{l\pm 1} \sim C_{\Lambda \pm 1} \to 1/2$ and $K_{l\pm 1} \sim K_{\Lambda \pm 1} \to 1/2$ due to Eqs. (3.4) and (3.16). As a result we can write the spin-spin correlation:

$$<\mathbf{S}_{00}\cdot\mathbf{S}_{k,k'}> \approx \frac{1}{2}\left[P_{\Lambda+1}^{k+k'} + P_{\Lambda-1}^{k+k'}\right], \text{ as } T \to 0 \quad (4.32a)$$

where $P_{\Lambda \pm 1}$ given by Eq. (4.31) reduces to

$$P_{\Lambda \pm 1} \approx -\frac{J}{|J|}\frac{\lambda_{\Lambda \pm 1}(|z^*|\Lambda)}{\lambda_{\Lambda}(|z^*|\Lambda)} \quad (4.32b)$$

if $J_1 = J_2 = J$ because $F_{\Lambda \pm 1, l\Lambda \pm 1}/F_{\Lambda,\Lambda} \to 1$ as $\Lambda \to +\infty$.

In Appendix C we have expressed the spin-spin correlation in this case, near $T_c = 0$ K, in the four zones of the magnetic phase diagram. In Appendix D we have shown that each expression (except for Zone 1) has to be renormalized.

For $k$ and $k'$ finite but not too large the spin-spin correlation given by Eqs. (D3)-(D6) can be reduced to the first order so that its renormalized expression can be written under the generic form for the nearest neighbors

$$\left|<\widetilde{\mathbf{S}_{00}\cdot\mathbf{S}_{k,k'}}>\right| \approx 1 - (k+k')f(x_i) \, (i=1 \text{ or } 2),$$

as $T \to 0$, (4.33)

on condition that $(k+k')f(x_i) \ll 1$, with

$$f(x_1) = \frac{8\pi}{e}|\zeta^*|\Lambda\left(1 - \frac{x_1}{2}\right), \text{ Zone 1 } (x_1\Lambda \ll 1, x_1 \ll 1),$$

$$f(x_i) = 2|\zeta^*|\Lambda, i = 1, \text{ Zone 3 } (x_1\Lambda \gg 1, x_1 \gg 1),$$

$$i = 2, \text{ Zone 4 } (x_2\Lambda \gg 1, x_2 \gg 1). \quad (4.34a)$$

Indeed, in Zone 1 and 3, we always have $|\zeta^*|\Lambda < 1$ near $T_c = 0$ K. In Zone 4 we must have $1/x_2 < \sqrt{5}(1-C) = 0.084$. Zone 2 for which

$$f(x_2) = |\zeta^*|\Lambda \gg 1, x_2\Lambda \sim 1, x_2 \ll 1 \quad (4.34b)$$

is a special case discussed below (no long-range order). In Appendix D we have shown that the Ornstein-Zernike law is fulfilled when $k$ and $k' \to +\infty$

$$\left|<\widetilde{\mathbf{S}_{00}\cdot\mathbf{S}_{k,k'}}>\right| \approx \exp\left(-\frac{k+k'}{\sqrt{2}}\frac{1}{\xi\sqrt{2}}\right), k+k' \to +\infty,$$

as $T \to 0$, (4.35)

where $\xi$ represents the universal measure of the correlation length along the lattice diagonal, near $T_c = 0$ K, with $\xi_x = \xi_y$ and $\xi = \xi_x\sqrt{2}$ if $J = J_1 = J_2$. We find

$$\xi_x \approx \frac{2}{f(x_i)} = (|\zeta^*|\Lambda)^{-1}, \text{ as } T \to 0. \quad (4.36)$$

Then, if using the definition of the correlation length given by Eqs. (2.6) and (2.8) and owing to Eq. (4.14), we can write

$$\xi_x = \frac{\xi_x^*}{2a} \approx \frac{\xi_\tau^*}{\hbar c/k_B T} = \frac{2}{f(x_i)} = (|\zeta^*|\Lambda)^{-1} = X_i(x_i)^{-1}, i = 1,2,$$

as $T \to 0$ (4.37)

where $2a$ represents the distance between two Landé factors $G$ and $L_\tau = \hbar c / k_B T$ is the de Broglie length; $f(x_i)$ is given by Eq. (4.34). It means that we can immediately derive the low-temperature correlation length $\xi$.

Recalling that $x_1 = k_B T/2\pi\rho_s$ where $\rho_s$ is the spin stiffness, the lattice spacing $a$ is such as $a = \hbar c/2|J|$ with $\rho_s \approx |J|$ as $g^* = T/T_c$ vanishes with $T$ near $T_c = 0$ K. For a lattice of double spacing $2a$ we finally derive in Zone 1:

$$\xi_x^* = \frac{e}{8}\frac{\hbar c}{2\pi\rho_s}\exp\left(\frac{2\pi\rho_s}{k_B T}\right)\left(1 + \frac{k_B T}{4\pi\rho_s}\right), x_1 \ll 1 \text{ (Zone 1). (4.38)}$$

We exactly retrieve the result first obtained by Hasenfratz and Niedermayer [27] and confirmed by Chubukov *et al.* [10]. This characterizes the *Renormalized Classical Regime* (RCR) for which $\rho_s \gg k_B T$: The divergence of $\xi$ describes a *long-range order* when $T$ approaches $T_c = 0$ K. Spins are aligned ($J < 0$) or antialigned ($J > 0$) inside quasi rigid quasi independent Kadanoff square blocks of side $\xi$ if $J = J_1 = J_2$.

In Zone 2 ($\Lambda x_2 \sim 1$, $x_2 \ll 1$) where $x_2 = k_B T/\Delta$ we have

$$\xi_\tau^* \approx \frac{\hbar c}{\Delta}, x_2 \ll 1 \text{ (Zone 2)} \quad (4.39)$$

We deal with the *Quantum Disordered Regime* (QDR) characterized by $\Delta \gg k_B T$. Owing to Eq. (4.21) we have $\Delta = k_B T/x_2$ so that $\xi_\tau^* \approx L_\tau x_2 \ll L_\tau$ as $x_2 \ll 1$. Equivalently, due to Eqs. (4.14) and (4.39) we have $\xi_x^* \approx 2ax_2 \ll 2a$: We then deal with a *short-range order*. The magnetic structure is ma-de of spin dimers or aggregates of spin dimers organized in Kadanoff blocks of small size $\xi_\tau$ that we can assimilate to blobs weakly interacting between each others. We deal with a *spin-fluid*. The detailed study is out of the scope of the present article. From a formal point of view it is often phrased in term of Resonating Valence Bonds (RVB) between pairs of quantum spins (considered here in the classical spin approximation) [28]-[30].

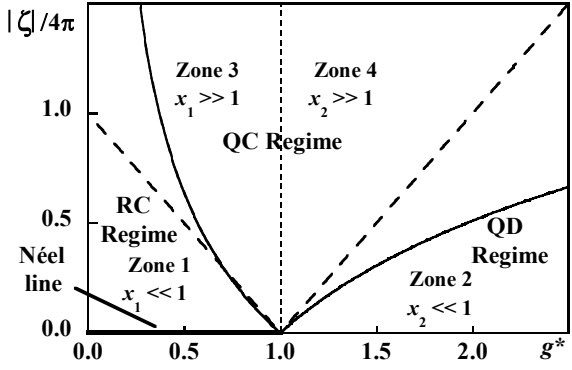

FIG. 12. Magnetic regime for each domain of predominance of $|\zeta|/4\pi$ vs $g^*$ respectively defined by Eqs. (4.4) and (4.6); the abbreviations stand for Renormalized Classical (RC), Quantum Critical (QC) and Quantum Disordered (QD) regimes.

In Zones 3 ($x_1 \gg 1$) and 4 ($x_2 \gg 1$) we have

$$\xi_\tau^* \approx C^{-1} \frac{\hbar c}{k_B T}\left(1 + \frac{2}{C\sqrt{5}x_1}\right), \; x_1 \gg 1 \text{ (Zone 3)},$$

$$\xi_\tau^* \approx C^{-1} \frac{\hbar c}{k_B T}\left(1 - \frac{1}{C\sqrt{5}x_2}\right), \; x_2 \gg 1 \text{ (Zone 4)} \quad (4.40)$$

where we have set:

$$C = 2\ln\left(\frac{1+\sqrt{5}}{2}\right) = 0.962\ 424, \; C^{-1} = 1.039\ 043. \quad (4.41)$$

We now deal with the *Quantum Critical Regime* (QCR). In Zone 3 we have $\rho_s \ll k_B T$ whereas in Zone 4 $\Delta \ll k_B T$. The divergence of $\xi$ also describes a *long-range order* when $T$ approaches $T_c = 0$ K. Spins are aligned ($J < 0$) or antialigned ($J > 0$) inside quasi rigid quasi independent Kadanoff square blocks of side $\xi$ if $J = J_1 = J_2$. But, if comparing with the *Renormalized Classical Regime* (RCR) and the *Quantum Disordered* one (QDR), for the same temperature $T$, we have $\xi_{QDR} < \xi_{QCR} < \xi_{RCR}$. As a result Kadanoff blocks show a smaller size. However we shall see while studying the specific heat $C_V$ that the exact knowledge of this magnetic structure remains problematic.

As a result each behavior of the correlation length characterizes a magnetic regime. All the predominance domains of these regimes are summarized in Fig. 12.

At the frontier between Zones 3 ($\rho_s \ll k_B T$) and 4 ($\Delta \ll k_B T$) i.e., along the vertical line reaching the Néel line at $T_c$, $x_1$ and $x_2$ become infinite so that:

$$\xi_\tau^* \approx C^{-1}\frac{\hbar c}{k_B T}, \; T = T_c. \quad (4.42)$$

i.e., $\xi_\tau^* \approx L_\tau$ as $C^{-1}$ is close to unity, as predicted by the renormalization group analysis [6,10]. As a result $\xi_\tau^*$ diverges according to a $T^{-1}$-law i.e., the critical exponent is:

$$\nu = 1 \quad (4.43)$$

in the $D$-space-time.

Now we have to estimate the critical exponent $\eta$ characterizing the decay law of the spin-spin correlation near the critical point $T_c = 0$ K. If $r \gg a$ it has been proposed by Chubukov et al. that $|\langle S_{00} \cdot S_{0,r}\rangle| \sim \exp(-r/\xi)/r^{(D-2+\eta)}$ (law 1) where $D = d + 1$ is the space-time dimension [10]. Through a renormalization group approach they found $\eta = 0$, with $D = 3$. But there is also another decay law involving the dimension $d$ of the crystallographic space $|\langle S_{00} \cdot S_{0,r}\rangle| \sim \exp(-r/\xi)/r^{(d-2+\eta)}$ (law 2). Comparing with the decay law given by Eq. (4.35) obtained if $J_1 = J_2 = J$ we can write along the lattice diagonal $|\langle S_{00} \cdot S_{k,k'}\rangle| \approx 1 - r/\tilde\xi$ with $r = (k+k')/\sqrt{2} \to +\infty$, $\tilde\xi = \xi_x\sqrt{2}$ and $r/\tilde\xi \ll 1$ so that

$$\eta = -1 \; (D = 3, \text{law 1}), \; \eta = 0 \; (d = 2, \text{law 2}). \quad (4.44)$$

Near $T_c = 0$ K the susceptibility $\chi$ behaves as $T^{-\gamma}$ with $\gamma = \nu(2-\eta)$ which does not directly depends on the dimension $d$ or $D$. As a result the exact knowledge of the coefficient $\gamma$ is going to select the good choice between law 1 and law 2.

Let us recall that, for a spin chain ($D = 2$) characterized by a single exchange energy $J$ between first-nearest neighbors, the correlation length $\xi$ behaves as $\beta|J|$ and the susceptibility $\chi$ is such as $\beta\xi(G \pm G')^2$ for a ferromagnet or a noncompensated antiferromagnet, near $T_c = 0$ K, thus leading to $\nu = 1$, $\gamma = 2$. The universal relation $\gamma = \nu(2 - \eta)$ leads to $\eta = 0$. As $|\langle S_{00} \cdot S_{0,r}\rangle| \sim 1 - r/\xi$ (with $r \gg a$ and $r/\xi \ll 1$) it means that this result validates the decay law $\exp(-r/\xi)/r^{(D-2+\eta)}$. As a result the study of the susceptiblity near $T_c = 0$ K when $D = 3$ must bring a definitive response. As the decay law is universal we expect that law 1 i.e., $|\langle S_{00} \cdot S_{0,r}\rangle| \sim r^{-(D-2+\eta)}$ is fulfilled instead of law 2 where $d = D - 1$ is involved.

The low-temperature expression of the correlation length $\xi$ can be also directly retrieved from the basic definition given by Eqs. (2.6), (3.20) and (3.21) which involves spin-spin correlations. In the low-temperature limit, these correlations have been expressed through Eqs. (4.32)-(4.34). If considering the general $\Lambda$-expression of the correlation length given by Eqs. (3.20) and (3.21), near $T_c = 0$ K, $l$ is replaced by $\Lambda \to +\infty$, $C_{l\pm 1} \sim C_{\Lambda\pm 1} \to 1/2$ and $K_{l\pm 1} \sim K_{\Lambda\pm 1} \to 1/2$ due to Eqs. (3.4) and (3.16). As a result, if calculating $\xi$ from the basic definition given by Eqs. (4.36) and (4.37) we retrieve $\xi = \xi_x \sqrt{2}$ with $\xi_x \approx 2/f(x_i) = (|\zeta^*|\Lambda)^{-1}$, $i = 1,2$. Owing to Eq. (4.33) we can write the correlation length as

$$\xi = \xi_x \sqrt{2} \approx \frac{2\sqrt{2}}{1 \mp |<\widetilde{\boldsymbol{S}_{00}.\boldsymbol{S}_{01}}>|}, \text{ as } T \to 0, \quad (4.45)$$

where $|<\widetilde{\boldsymbol{S}_{00}.\boldsymbol{S}_{01}}>|$ is the renormalized spin-spin correlation between first-nearest neighbors (0,0) and (0,1) for instance, with $|<\widetilde{\boldsymbol{S}_{00}.\boldsymbol{S}_{01}}>| = |<\widetilde{\boldsymbol{S}_{00}.\boldsymbol{S}_{10}}>|$ as $J = J_1 = J_2$. In Eq. (4.45) the sign − appearing in the denominator concerns Zones 1, 3 and 4 characterized by a long-range order whereas the sign + is for Zone 2 characterized by a short-range order. As a result, if using Eqs. (4.32)-(4.34), we exactly retrieve, for each zone of the magnetic phase diagram, the same corresponding low-temperature expression of the correlation length given by Eqs. (4.37)-(4.41).

Finally there is a third method for finding the low-temperature behavior of the correlation length. In Eq. (2.30) we have defined the convergence ratio $r_{\Lambda+1}/r_\Lambda$ of the characteristic polynomial associated with the zero-field partition function. Near the critical point we have $r_{\Lambda+1}/r_\Lambda \sim r_{\Lambda-1}/r_\Lambda \sim |<\boldsymbol{S}_{00}.\boldsymbol{S}_{0,1}>|^2$ so that $\xi_x$ behaves as $1/(1-\sqrt{r_{\Lambda+1}/r_\Lambda})$. This relation is exclusively valid when $T_c = 0$ K because the correlation function $\Gamma_{k,k'}$ reduces to $<\boldsymbol{S}_{i,j}.\boldsymbol{S}_{i+k,j+k'}>$. When $T_c \neq 0$ K this is no more valid because $<\boldsymbol{S}_{i,j}>$ and $<\boldsymbol{S}_{i+k,j+k'}>$ appearing in the general definition of $\Gamma_{k,k'}$ given by Eq. (2.8) do not vanish.

Finally, in the general case $J_1 \neq J_2$, using the definition of the correlation length given by Eqs. (2.6) and (2.8), it is easy to show

$$\xi_x \approx \frac{2}{f_1(x_i)}, \xi_y \approx \frac{2}{f_2(x_i)}, \xi = \sqrt{\xi_x^2 + \xi_y^2}, \text{ as } T \to 0, \quad (4.46)$$

where $f_k(x_i)$ is given by Eq. (4.34) in which $J$ is replaced by $J_k$ ($k = 1$ for the $x$-axis and 2 for the $y$-axis) in $x_i$ ($i = 1, 2$).

### C. Free energy and specific heat densities

#### C1. Free energy density

The free energy density for the two-dimensional lattice is defined as

$$\mathcal{F}(T) = -\frac{k_B T}{S} \ln Z_N(0) \quad (4.47)$$

where $Z_N(0)$ is the zero-field partition function given by Eq. (2.35) and $S$ the lattice surface. In the vicinity of the critical point we have seen that the lattice is composed of independent blocks of spins. When $J = J_1 = J_2$ we deal with square blocks of side length $\xi$ where $\xi$ is the correlation length (cf Eq. (4.42)). Thus we have $S = \xi^2$ i.e., $S = C^{-2}(\hbar c/k_B T)^2$ where $C$ is given by Eq. (4.41). The free energy density can be rewritten

$$\mathcal{F}(T) = -\hbar c C^{-1} \xi^{-D} \ln Z_N(0), D = 3. \quad (4.48)$$

As a result, near $T_c = 0$ K, $\mathcal{F}(T)$ has the form $\hbar c \Upsilon \xi^{-D}$ where $\Upsilon$ is a universal number defined by

$$\Upsilon = C^{-1} = 1.039\ 043 \quad (4.49)$$

so that the hyperscaling hypothesis is verified [31]. Finally $\xi^D$ represents the volume of the slab of space-time which is infinite in $D-1$ dimensions but of finite length $L_\tau$ in the remaining direction along the $i\tau$-axis. In our case $D = 3$.

As a result we can define the free energy density per lattice bond:

$$\frac{\mathcal{F}(T)}{8N^2} = -\frac{\hbar c C^2}{8N^2}\left(\frac{k_B T}{\hbar c}\right)^3 \ln Z_N(0). \quad (4.50)$$

In Appendix E we have established the expression of the free energy density near the critical point $T_c = 0$ K through a saddle-point method. In the thermodynamic limit we find

$$\frac{\mathcal{F}(T)}{8N^2} = \hbar c C \left[ \frac{1}{L_\tau} \int \frac{d^2 k}{(2\pi)^2} \ln(1 - \exp(-u_k)) \right. $$
$$+ \frac{1}{2} \int \frac{d^3(P/\hbar)}{(2\pi)^3} \ln\left(\frac{(P/\hbar)^2 + (m_0 c/\hbar)^2}{\Lambda^{-2}}\right)$$
$$\left. - \frac{(m_0 c/\hbar)^2 - 1}{2g} \right], \text{ as } N \to +\infty \quad (4.51)$$

with

$$\frac{d^3(P/\hbar)}{(2\pi)^3} = \frac{d^2 k}{(2\pi)^2} \frac{d(\omega/c)}{2\pi}, \quad (4.52)$$

$$u_k = \frac{L_\tau}{\hbar c}\sqrt{(\hbar k)^2 + (m_0 c^2)^2}, u_k = \frac{\varepsilon_k}{k_B T}, \varepsilon_k = \sqrt{(\hbar k)^2 + (m_0 c^2)^2},$$

where $\boldsymbol{P} = (\hbar \boldsymbol{k}, \hbar\omega/c)$ is the relativistic momentum, $\varepsilon_k$ the relativistic energy. $g_c$ is the critical coupling defined as, for $D = 3$

$$\frac{1}{g_c} = \int \frac{d^3(P/\hbar)}{(2\pi)^3} \frac{1}{(P/\hbar)^2}. \quad (4.53)$$

This result can be generalized to a $D$-space-time. In that case $g_c$ is given by:

$$\frac{1}{g_c} = \int \frac{d^D(P/\hbar)}{(2\pi)^D} \frac{1}{(P/\hbar)^2} \quad (4.54)$$

with $d^D(P/\hbar) = S_D(P/\hbar)^{D-1}d(P/\hbar)$, $D = d + 1$, $S_D = 2\pi^{D/2}/\Gamma(D/2)$ so that finally, if integrating Eq. (4.54) we retrieve the result of Chakravarty et al. [6] (expressed in $\Lambda$-unit)

$$g_c^*(d) = \frac{2(d-1)(2\pi)^d}{S_d} \Lambda^{1-d}, \quad (4.55)$$

thus finding *a posteriori* the geometrical factor that we suspected. If expressing the ratio $S_{d+1}/S_d$ i.e., $S_D/S_d = 2^{D-2}(D-1)\Gamma((D-1)/2)^2/\Gamma(D)$ we can obtain the surface of space-time $S_D$ swept by the relativistic momentum $\boldsymbol{P}$ vs $S_d$. $S_D/(2\pi)^D$ represents the integration measure $K_D$. In other words, in the $\Lambda$-scale, $g_c^*(D)$ is expressed per unit of swept surface of space-time

$$g_c^*(D) = A(D) \frac{(2\pi)^D}{S_D} \Lambda^{2-D}$$

$$A(D) = \frac{2^{D-2}(D-1)(D-2)\Gamma\left(\dfrac{D-1}{2}\right)^2}{\pi\Gamma(D)} \quad (4.56)$$

The second integral in Eq. (4.51) is badly divergent when $P$ becomes infinite (ultraviolet domain). However all divergences disappear if introducing the infinite volume contribution $\mathcal{F}_\infty(T) = \mathcal{F}(0)$ for which $m_0 = 0$ ($L_\tau \to +\infty$). As a result we have:

$$\frac{\mathcal{F}(T)-\mathcal{F}(0)}{8N^2} = \hbar cC\left[\frac{1}{L_\tau}\int \frac{d^2k}{(2\pi)^2}\ln(1-\exp(-u_k))\right.$$
$$\left. + \frac{1}{2}\int \frac{d^3(P/\hbar)}{(2\pi)^3}\ln\left(\frac{(P/\hbar)^2+(m_0c/\hbar)^2}{(P/\hbar)^2}\right) - \frac{(m_0c/\hbar)^2}{2g}\right],$$
$$\text{as } N \to +\infty. \quad (4.57)$$

This is the exact relation that Chubukov et al. have derived from a renormalization approach in the vicinity of the critical point [10]. Consequently *it brings an important verification to our result derived from $Z_N(0)$ whose expression is valid for any temperature and considered in the low-temperature limit.*

If $g < g_c$, owing to the definition of the spin stiffness $\rho_s$ given by Eq. (4.18) and after a first integration of Eq. (4.57), we get:

$$\frac{\mathcal{F}(T)-\mathcal{F}(0)}{8N^2} = \hbar cC\left[\frac{1}{L_\tau^3}\int_{X_1(x_1)}^{+\infty} \frac{\varepsilon d\varepsilon}{2\pi}\ln(1-\exp(-\varepsilon))\right.$$
$$+ \frac{1}{2}\int \frac{d^3(P/\hbar)}{(2\pi)^3}\left\{\ln\left(\frac{(P/\hbar)^2+(m_0c/\hbar)^2}{(P/\hbar)^2}\right)\right.$$
$$\left.\left. - \frac{(m_0c/\hbar)^2}{(P/\hbar)^2}\right\} - \frac{(m_0c/\hbar)^2}{2}\frac{\rho_s}{k_BT}\right],$$
$$\text{as } N \to +\infty. \quad (4.58)$$

The first integral can be expressed owing to polylogarithms (or Jonquière functions) [32]:

$$\text{Li}_s(z) = \sum_{k=1}^{+\infty} \frac{z^k}{k^s}, \quad (4.59)$$

and the Riemann zeta function $\zeta(s) = \text{Li}_s(1)$ ($s > 1$). We have:

$$\frac{\mathcal{F}(T)-\mathcal{F}(0)}{8N^2} = -\frac{\hbar cC}{2\pi}\left(\frac{k_BT}{\hbar c}\right)^3\left[X_1(x_1)\text{Li}_2\left(e^{-X_1(x_1)}\right)\right.$$
$$\left. + \text{Li}_3\left(e^{-X_1(x_1)}\right) + \frac{X_1^3(x_1)}{6} + \frac{X_1^2(x_1)}{2x_1} + ...\right],$$
(Zones 1 and 3, $T < T_c$), (4.60)

where $X_i(x_i) \approx |\zeta^*|\Lambda$ (*cf* Eq. (4.30)). If $i = 1$ $X_1(x_1)$ is thus given by Eq. (4.25) for Zone 1 and Eq. (4.27) for Zone 3.

As a result, if $x_1 \ll 1$ (Zone 1, $T < T_c$), Eq. (4.60) becomes:

$$\frac{\mathcal{F}(T)-\mathcal{F}(0)}{8N^2} = -\frac{\hbar cC}{2\pi}\left(\frac{k_BT}{\hbar c}\right)^3\left[\zeta(3) + \zeta(2)e^{-1/x_1}\right.$$
$$\left. + \frac{e^{-2/x_1}}{2x_1} + \frac{e^{-3/x_1}}{6} + ...\right],$$
$$x_1 \ll 1 \text{ (Zone 1, } T < T_c\text{). } (4.61)$$

When $x_1 \gg 1$ (Zone 3, $T < T_c$) the argument of Jonquière function can be expanded vs $1/x_1 \ll 1$ as well as the function itself. We find:

$$\frac{\mathcal{F}(T)-\mathcal{F}(0)}{8N^2} = -\frac{\hbar cC}{2\pi}\left(\frac{k_BT}{\hbar c}\right)^3\left[C_0 + \frac{C_1}{x_1} + \frac{C_2}{x_1^2} + \frac{C_3}{x_1^3} + ...\right],$$
$$x_1 \gg 1 \text{ (Zone 3, } T < T_c\text{), } (4.62)$$

with:

$$C_0 = C\text{Li}_2\left(e^{-C}\right) + \text{Li}_3\left(e^{-C}\right) + \frac{C^3}{6}, \quad C_0 = 0.961\,646,$$

$$C_1 = C\left[C'\mathrm{Li}_1\left(e^{-C}\right) + \frac{C}{2}(1-C')\right], \quad C_1 = 0.463\ 130\ ,$$

$$C_2 = \frac{C'}{2}\left[CC'\mathrm{Li}_0\left(e^{-C}\right) - C'\mathrm{Li}_1\left(e^{-C}\right) + C(C'-2)\right],$$

$$C_2 = -0.430\ 409\ , \quad (4.63)$$

$$C_3 = \frac{C'^2}{6}\left[CC'\mathrm{Li}_{-1}\left(e^{-C}\right) - 2C'\mathrm{Li}_0\left(e^{-C}\right) - C' + 3\right],$$

$$C_3 = 0.248\ 109\ ,$$

$$C' = \frac{2}{\sqrt{5}}, C' = 0.894\ 427$$

where $C = 2\ln\alpha$ is given by Eq. (4.41), $\alpha$ being the golden mean. Concerning the coefficient $C_0$ we note that the argument of the polylogarithm function is $e^{-C} = 1/\alpha^2 = 2 - \alpha$. As a result we also have $C_0 = \mathrm{Li}_3(2-\alpha) - \ln(2-\alpha)\mathrm{Li}_2(2-\alpha) - \ln^3(2-\alpha)/6$. Using relations between polylogarithms Sachdev has rigorously shown that [33]

$$\mathrm{Li}_3(2-\alpha) - \ln(2-\alpha)\mathrm{Li}_2(2-\alpha) - \frac{1}{6}\ln^3(2-\alpha) = \frac{4}{5}\zeta(3)$$

so that we derive here

$$C_0 = \frac{4}{5}\zeta(3)\ . \quad (4.64)$$

The other coefficients $C_1$, $C_2$ and $C_3$ can be also simplified through a similar work but that has no interest here.

If $g > g_c$, the following relation between the $T = 0$ gap, $\Delta$, and the coupling $g$ is useful:

$$\frac{1}{g} = \int \frac{d^3(P/\hbar)}{(2\pi)^3} \frac{1}{(P/\hbar)^2 + (\Delta/\hbar c)^2} \quad (4.65)$$

As a result and after a first integration of Eq. (4.57), we get:

$$\frac{\mathcal{F}(T) - \mathcal{F}(0)}{8N^2} = \hbar c C\left(\frac{k_B T}{\hbar c}\right)^3 \left[\int_{X_2(x_2)}^{+\infty} \frac{\varepsilon d\varepsilon}{2\pi} \ln(1-\exp(-\varepsilon))\right.$$

$$+ \frac{1}{2}\int \frac{d^3(P/\hbar)}{(2\pi)^3}\left\{\ln\left(\frac{(P/\hbar)^2 + (m_0 c/\hbar)^2}{(P/\hbar)^2}\right)\right.$$

$$\left.\left. - \frac{(m_0 c/\hbar)^2 - (\Delta/\hbar c)^2}{(P/\hbar)^2 + (\Delta/\hbar c)^2}\right\}\right]. \quad (4.66)$$

In that case too this is the exact relation that Chubukov *et al.* have derived from a renormalization approach in the vicinity of the critical point [10]. Proceeding as in the case $g < g_c$,

when expressing the first integral with polylogarithms, we get:

$$\frac{\mathcal{F}(T) - \mathcal{F}(0)}{8N^2} = -\frac{\hbar c C}{2\pi}\left(\frac{k_B T}{\hbar c}\right)^3 [X_2(x_2)\mathrm{Li}_2(e^{-X_2(x_2)})$$

$$+ \mathrm{Li}_3(e^{-X_2(x_2)}) + \frac{X_2^3(x_2)}{6} - \frac{X_2^2(x_2)}{4 x_2} + \frac{1}{12 x_2^3} + \ldots\right]$$

(Zones 2 and 4, $T > T_c$)  (4.67)

where $X_2(x_2)$ is thus given by Eq. (4.26) for Zone 2 and Eq. (4.28) for Zone 4.

When $x_2 \ll 1$ (Zone 2, $T > T_c$) Eq. (4.67) becomes:

$$\frac{\mathcal{F}(T) - \mathcal{F}(0)}{8N^2} = -\frac{\hbar c C}{2\pi}\left(\frac{k_B T}{\hbar c}\right)^3\left[\left(1 + \frac{1}{x_2}\right)e^{-1/x_2}\right.$$

$$\left. + \frac{1}{4}\left(\frac{17}{2} + \frac{5}{x_2}\right)e^{-2/x_2} + \frac{1}{9}\left(\frac{101}{6} + \frac{1}{x_2}\right)e^{-3/x_2} + \ldots\right], \quad x_2 \ll 1$$

(Zone 2, $T > T_c$).  (4.68)

When $x_2 \gg 1$ (Zone 4, $T > T_c$) the argument of Jonquière function can be expanded vs $1/x_2 \ll 1$ as well as the function itself. We find:

$$\frac{\mathcal{F}(T) - \mathcal{F}(0)}{8N^2} = -\frac{\hbar c C}{2\pi}\left(\frac{k_B T}{\hbar c}\right)^3\left[C'_0 + \frac{C'_1}{x_2} + \frac{C'_2}{x_2^2} + \frac{C'_3}{x_2^3} + \ldots\right],$$

$$x_2 \gg 1 \text{ (Zone 4, } T > T_c\text{)}, \quad (4.69)$$

with:

$$C'_0 = C_0, C'_0 = 0.961\ 646\ , C'_1 = -\frac{C_1}{2}, C'_1 = -0.231\ 565\ ,$$

$$C'_2 = \frac{C_2}{4}, C'_2 = -0.107\ 602\ , C'_3 = -\frac{C_3}{8} + \frac{1}{12}, C'_3 = 0.052\ 320$$

(4.70)

where $C_0$, $C_1$, $C_2$ and $C_3$ are given by Eq. (4.63).

If $x_1 \ll 1$ (Zone 1, $T < T_c$) the leader term of the free energy density per lattice bond is

$$\frac{\mathcal{F}(T) - \mathcal{F}_\infty(T)}{8N^2} = -\frac{\hbar c C}{2\pi}\left(\frac{k_B T}{\hbar c}\right)^3 \zeta(3) + \ldots ,$$

$$x_1 \ll 1 \text{ (Zone 1, } T < T_c\text{)}, \quad (4.71)$$

with $\mathcal{F}_\infty(T) = \mathcal{F}(0)$. As a result, coming simultaneously from Zone 3 ($T < T_c$, $x_1 \gg 1$) and Zone 4 ($T > T_c$, $x_2 \gg 1$) towards the common vertical frontier directly leading to the Néel line ($T = T_c$, case of an antiferromagnet) for which $x_1 \approx x_2 \to +\infty$, the common leader term of the free energy density per lattice bond can be written as:

$$\frac{\mathcal{F}(T)-\mathcal{F}_\infty(T)}{8N^2} = -\hbar c C \frac{\zeta(3)}{2\pi}\left(\frac{k_B T}{\hbar c}\right)^3 \widetilde{C} + \dots \qquad (4.72)$$

with owing to the result given by Eq. (4.64)

$$\widetilde{C} = \frac{C_0}{\zeta(3)} = \frac{4}{5}. \qquad (4.73)$$

In addition we retrieve the complete result of Fisher and de Gennes which states that, if hyperscaling is valid, the free energy density per lattice bond satisfies [31]

$$\frac{\mathcal{F}(T)-\mathcal{F}_\infty(T)}{8N^2} = -\hbar c C \frac{\Gamma(D/2)\zeta(D)}{\pi^{D/2}} \frac{\widetilde{C}}{L_\tau^D} + \dots . \qquad (4.74)$$

with here $D = 3$ and $L_\tau = \hbar c/k_B T$.

### C2. Specific heat density

The specific heat $C_V$ can be expressed from Eqs. (2.36) and (2.37) in the thermodynamic limit. In the vicinity of the critical point $T_c = 0$ K, *the specific heat density $\mathcal{C}_V$ can be directly obtained owing to the well-known general definition $\mathcal{C}_V = -T\partial^2 \mathcal{F}/\partial T^2$* where $\mathcal{F}$ if the free energy density. In the low-temperature range $\mathcal{F}$ has been expressed in the previous subsection owing to closed-form expressions involving parameters $x_1$ or $x_2$. As a result and due to the definition of $x_1$ and $x_2$ given by Eq. (4.21), and near $T_c$, we have the correspondence $T\partial^2/\partial T^2 = (k_B x_1/(2\pi\rho_s))\partial^2/\partial x_1^2$ for zones 1 and 3 in which $\rho_s$ is the spin stiffness and $T\partial^2/\partial T^2 = (k_B x_2/\Delta)\partial^2/\partial x_2^2$ for zones 2 and 4 in which $\Delta$ is the energy gap between the ground state and the first excited state. $\rho_s$ and $\Delta$ are defined by Eq. (4.18). Under these conditions it is possible to derive low-temperature closed-form expressions for $\mathcal{C}_V$.

For sake of simplicity we directly express the corresponding formulas *vs* parameters $x_1$ or $x_2$. We have per lattice bond

$$\frac{\mathcal{C}_V}{8N^2} = \frac{3C\zeta(3)}{\pi} k_B \left(\frac{k_B T}{\hbar c}\right)^2 \Psi_1(x_1), \text{ (Zones 1 and 3)}, \qquad (4.75)$$

with

$$\Psi_1(x_1) = 1 + \frac{\zeta(2)}{\zeta(3)}\left(1 + \frac{2}{3x_1}\right)e^{-1/x_1} + \frac{e^{-2/x_1}}{6\zeta(3)x_1} + \dots,$$
$$x_1 \ll 1 \text{ (Zone 1, } T < T_c\text{)}, \qquad (4.76)$$

$$\Psi_1(x_1) = \frac{C_0}{\zeta(3)}\left(1 + \frac{C_1}{3C_0 x_1} + \dots\right), \frac{C_0}{\zeta(3)} = \frac{4}{5},$$
$$x_1 \gg 1 \text{ (Zone 3, } T > T_c\text{)}, \qquad (4.77)$$

and

$$\frac{\mathcal{C}_V}{8N^2} = \frac{3C\zeta(3)}{\pi} k_B \left(\frac{k_B T}{\hbar c}\right)^2 \Psi_2(x_2), \text{ (Zones 2 and 4)}, \qquad (4.78)$$

with

$$\Psi_2(x_2) = \frac{1}{\zeta(3)}\left(1 + \frac{1}{x_2}\right)e^{-1/x_2} + \dots,$$
$$x_2 \ll 1 \text{ (Zone 2, } T < T_c\text{)}, \qquad (4.79)$$

and

$$\Psi_2(x_2) = \frac{C_0}{\zeta(3)}\left(1 - \frac{C_1}{6C_0 x_2} + \dots\right), \frac{C_0}{\zeta(3)} = \frac{4}{5},$$
$$x_2 \gg 1 \text{ (Zone 4, } T > T_c\text{)}. \qquad (4.80)$$

The factor common to each expression of the specific heat density per lattice bond is nothing but the specific heat of a single gapless Bose degree of freedom:

$$\frac{\mathcal{C}_{V,0}}{8N^2} = \frac{3\zeta(3)}{\pi} k_B \left(\frac{k_B T}{\hbar c}\right)^2 \qquad (4.81)$$

so that we have the general expression for $\mathcal{C}_V$

$$\frac{\mathcal{C}_V}{8N^2} = \mathcal{C}_{V,0} \Psi_i(x_i), i = 1, 2 \qquad (4.82)$$

with the following critical exponent

$$\alpha = -2. \qquad (4.83)$$

We retrieve the general scaling form given by Chubukov *et al.* [10]. $\Psi_i(x_i)$, (with $i = 1,2$) is a universal dimensionless function due to the fact that $x_i$ is dimensionless in $2 + 1$ dimensions. In the vicinity of the critical point, $\Psi_i(x_i)$ represents the *measure of the effective number of bosonic modes* in the ground state.

The examination of this number gives a first idea about the nature of magnetic arrangements. In zones 1 and 3 this number of modes is $\Psi_1(x_1)$ near the critical point. In zone 1 ($x_1 \ll 1$) $\Psi_1(x_1) \sim \Psi_1(0) = 1$ describes an ordered phase at $T = 0$ (Néel phase for an antiferromagnet). We can notably deal with Goldstone bosons when the wave vector $k$ belongs to the range $[\xi^{-1}, \xi_J^{-1}]$ where $\xi_J$ is the Josephson correlation length given by $\xi_J \sim |g_c - g|^{-\nu}$ i.e., here $\nu = 1$ and $g_c = 0$, so that $\xi_J^{-1} \approx g = k_B T/|J|$ (if $J = J_1 = J_2$): The spin dynamics is well described by rotationally averaged spin-wave fluctuations about a Néel ordered ground state (for an antiferromagnet). Thus the scale $\xi_J$ controls the crossover between a critical behavior and a Goldstone one. Chubukov *et al.* have also shown that there exists a second crossover in the Renormalized Classical Regime (RCR) when $T$ approaches $T_c = 0$ K [10]. It occurs when $k \sim \xi^{-1}$. At this scale thermal fluctuations of locally ordered regions (Kadanoff blocks of size $\xi$) destroy the long-range order. At scales larger than $\xi$ i.e., if $k < \xi^{-1}$ the ferromagnet (or the antiferromagnet) appears disordered.

In zone 3 ($x_1 \gg 1$), in the Quantum Critical Regime (QCR) with $g < g_c$, we have a single crossover between a quantum relaxational regime and a quantum critical one at $k \sim k_B T/\hbar c$ i.e., $k \sim 1/L_\tau$ [10]. However we have $\Psi_1(x_1) \sim \Psi_1(\infty) = 4/5$. That means that a residual part of the excitations has no bosonic origin. This result which is a rational number (whereas $C_0$ and $\zeta(3)$ are irrational) remains a mystery for the while and no physical understanding has been given.

In zones 2 and 4 the number of modes is given by $\Psi_2(x_2)$. In zone 2 ($x_2 \ll 1$) $\Psi_2(x_2) \sim \Psi_2(0) = 0$ describes a disordered phase (no bosonic excitations). This is due to the fact that we deal with a liquid phase composed of spin dimers or spin dimer blobs of size $\xi_\tau \approx L_\tau x_2 \ll L_\tau$ as $x_2 \ll 1$. There is a crossover between a quantum activated regime and a critical one which occurs for $k \sim \Delta/\hbar c$. In zone 4 ($x_2 \gg 1$), in the Quantum Critical Regime (QCR), $\Psi_2(x_2) \sim \Psi_2(\infty) = 4/5$ characterizes the quantum critical phase at $g > g_c$. Along the critical line, at $g = g_c$, $\Psi_2(\infty) = \Psi_1(\infty) = 4/5$. These results have been found by Chubukov et al. through a renormalization method [10].

### D. Susceptibility density

If examining the closed-form expression of the susceptibility given by Eqs. (3.23) and (3.24) the low-temperature behavior of the susceptibility is essentially ruled by that of the denominator. If dealing with a lattice characterized by two exchange energies $J_1$ (respectively, $J_2$) along each horizontal lattice line (respectively, vertical), with different Landé factors $G$ and $G'$, the denominator is given by a product of terms $1 - (P_{i,l+\varepsilon})^2$, with $i = 1,2$, where $P_{i,l+\varepsilon}$ is defined by Eq. (3.16). It is independent of the respective signs of $J_1$ and $J_2$ if $G \neq G'$ and dependent if $G = G'$.

As these quantities are close to unity the $1 - (P_{i,l+\varepsilon})^{-2}$'s diverge whereas the numerator is finite, if $G \neq G'$. When $G = G'$ the denominator exclusively diverges for ferromagnetic couplings ($J_i < 0$). Near $T_c = 0$ K, the low-temperature behavior of the total susceptibility $\chi = 3\chi$ is given the infinite $l$-limit (the $\Lambda$-limit, as $\Lambda \to +\infty$) of the corresponding series defined by Eq. (3.23).

We recall that the experimental product $\chi k_B T$ of 2$d$-ferromagnets and 2$d$-noncompensated antiferromagnets diverges near $T_c = 0$ K, in the thermodynamic limit. From a mathematical point of view, we can use the same process as the one used for evaluating the zero-field partition function $Z_N(0)$ because the corresponding polynomial structure is similar. This property has been already pointed out after Eq. (2.35).

As a result, owing to the low-temperature behaviors of the $P_i$'s previously given (cf Eq. (4.31)) and that of correlation lengths (cf Eq. (4.42)) the total susceptibility per site $\chi = 3\chi$ behaves as

$$\chi \approx \beta \xi_x \xi_y (G \pm G')^2 \text{, as } T \to 0 \quad (4.84)$$

where the sign + (respectively, −) corresponds to ferromagnetic (respectively, antiferromagnetic) couplings (if $J_1$ and $J_2$ show the same sign). As expected, near the critical point $T_c = 0$ K, for ferromagnets and noncompensated antiferromagnets, the divergence of $\chi$ is determined by that of the product of correlation lengths $\xi_x$ and $\xi_y$. They are given by Eqs. (4.38)-(4.40) for Zones 1 to 4, respectively. At the critical point we have owing to Eq. (4.42)

$$\chi \approx [C^{-1} \hbar c (G \pm G')]^2 (k_B T)^{-3} \text{, } T \to T_c = 0 \text{ K} \quad (4.85)$$

where $C$ is given by Eq. (4.41). In the $D$-space-time we derive the corresponding critical exponent

$$\gamma = 3. \quad (4.86)$$

As $\gamma = \nu(2 - \eta)$ we then obtain $\eta = -1$, thus validating the decay law 1 i.e., $|\langle S_{00} \cdot S_{0,r} \rangle| \sim \exp(-r/\xi)/r^{(D-2+\eta)}$, with $D = 3$. As a result we conclude that *critical exponents must be evaluated in the D-space-time instead of the d-space.*

More generally Eq. (4.84) can be also rewritten:

$$\chi k_B T \approx \xi_x \xi_y M(T)^2 \text{, } M(T) = (G \pm G') f(T) \text{, as } T \to 0 \quad (4.87a)$$

where $f(T)$ gives the thermal behavior of the magnetic moment per unit cell. It is notably crucial in the case of compensated antiferromagnets. As a result the elementary susceptibility per surface unit and per lattice bond is finally

$$\frac{\chi_s k_B T}{8N^2} = \frac{\chi k_B T}{\xi_x \xi_y} \approx \frac{M(T)^2}{L_\tau^2} \text{, as } T \to 0. \quad (4.87b)$$

*This law is universal i.e., it is independent of the coupling nature.* In addition, as $\chi$, $\xi_x$ and $\xi_y$ can be expanded as scaling functions near $T_c = 0$ K, it means that $M(T)$ can be also expressed as a universal (scaling) function.

We retrieve the picture of Kadanoff's blocks near the critical point. These rectangular blocks show sizes of respective lengths $\xi_x$ and $\xi_y$, the correlation lengths, and each block has a magnetic moment $M = G \pm G'$, the unit cell moment. Due to the physical meaning of the correlation length we can say that, inside each block, classical spins are strongly correlated and show the same parallel orientation ($J < 0$) or antiparallel one ($J > 0$).

When dealing with a lattice characterized by a single Landé factor $G = G'$ and even if considering different exchange energies $J_1$ and $J_2$ the denominator of susceptibility $\chi$ behaves as a product of terms $1 - P_{i,l+\varepsilon}$, with $i = 1,2$, where $P_{i,l+\varepsilon}$ is given by Eq. (3.16). Now it clearly depends on the sign of $J_1$ and $J_2$. This remains valid if $J = J_1 = J_2$. For obvious physical reasons we focus on the case of a lattice such as $G = G'$ and $J = J_1 = J_2$. The low-temperature behavior of the product $\chi k_B T$ is always given by Eq. (4.87) i.e., near $T_c = 0$ K, we deal with Kadanoff's blocks which are squares of length $\xi = \xi_x = \xi_y$ and moment $M = G$, the unit cell moment.

As a result, for a ferromagnet ($J < 0$), the reasoning used for the lattice characterized by $G \neq G'$ and $J_1 \neq J_2$ is unchanged. Thus, $\chi$ behaves as

$$\chi \approx \beta \xi^2 G^2, \; G = G', \; J < 0, \; \text{as } T \to 0. \quad (4.88)$$

The low-temperature behavior is dominated by the divergence of $\xi^2$. At the critical point $\chi$ diverges according to a $T^{-3}$ law ($\gamma = 3$).

When dealing with a compensated antiferromagnet ($J > 0$) the magnetic moment vanishes and the behavior of the product $\chi k_B T$ is given by the competition between the divergence of $\xi^2$ and the vanishing law of $M$. In other words, if the divergence of $\xi^2$ dominates, $\chi k_B T$ can diverge near $T_c = 0$ K but less strongly than in the ferromagnetic case. If the divergence of $\xi^2$ is exactly compensated by the vanishing law of $M$, $\chi k_B T$ tends towards a nonzero constant limit. Finally, if the vanishing law of $M$ is preponderant, $\chi k_B T$ vanishes near $T_c = 0$ K. This is precisely the case when considering the experimental product $\chi k_B T$ of 2d-compensated antiferromagnets.

As $\chi k_B T$ tends towards a null limit we have to consider all the contributions of the $l$-series terms as $l \to +\infty$ i.e., all the $\Lambda$-terms when $\Lambda \to +\infty$. As all the eigenvalues have a close low-temperature behavior it means that we then deal with a continuum and the $l$-series (i.e, the $\Lambda$-series) becomes an integral. Its value is essentially given in the infinite $\Lambda$-limit i.e., the integral must be evaluated by means of a steepest descent method.

In Appendix F we have expressed the field-dependent free energy density $\mathcal{F}(T, B)$ in the limit of vanishing $B$, in the case $J = J_1 = J_2$, without loss of generality. The $B$-argument can be expanded in powers of $B$. Then, if picking out the coefficient of the quartic term and considering the minimum of $\mathcal{F}(T,B)$ near the saddle point, we derive near $T_c = 0$ K the elementary susceptibility per surface unit and per lattice bond

$$\frac{\chi_s}{8N^2} = 4C\left(\frac{G}{\hbar c}\right)^2 k_B T \sum_{\omega_n = 0}^{+\infty} \int \frac{k dk}{2\pi} \frac{\varepsilon_k^2 - \omega_n^2}{(\varepsilon_k^2 + \omega_n^2)^2}, \text{as } N \to +\infty \quad (4.89)$$

where $C$ is given by Eq. (4.41) and with $kdk = \varepsilon_k d\varepsilon_k$ and $\varepsilon_k \in [|\zeta^*|\Lambda/2\pi, +\infty[$ or equivalently due to Eq. (F13) $\varepsilon_k \in [m_0 c^2/2\pi k_B T, +\infty[$. This is the exact expression differently obtained by Chubukov et al. [10]. Note the absence of direct dependence on the coupling $g$.

After summing over $\omega_n$ (see Eq. (F16)) the integration vs $\varepsilon_k$ of the first contribution appearing in Eq. (4.89) gives $[\ln(\varepsilon_k)]_{m_0c^2/k_BT}^{+\infty} \to 0$ when $T \to 0$. It means that there is *no ultraviolet divergence*. The integration of the second contribution allows one to write:

$$\frac{\chi_s}{8N^2} = C\left(\frac{G}{\hbar c}\right)^2 \frac{k_B T}{\pi}\left[\frac{m_0 c^2}{k_B T}\frac{e^{m_0 c^2/k_B T}}{e^{m_0 c^2/k_B T}-1} - \ln\left(e^{m_0 c^2/k_B T}-1\right)\right]$$
$$\text{as } N \to +\infty \quad (4.90)$$

Due to Eq. (4.30) where we have set $|\zeta^*|\Lambda \approx X_i(x_i)$, $i = 1,2$ and Eq. (4.36) for which $\xi_\tau^* \approx L_\tau(|\zeta^*|\Lambda)^{-1}$ we can write

$$\frac{m_0 c^2}{k_B T} = X_i(x_i) = \frac{L_\tau}{\xi_\tau^*}. \quad (4.91)$$

Under these conditions, if using the expression of $X_i(x_i)$ obtained by comparison of Eqs. (4.23), (4.24) and (4.30) the susceptibility per surface unit of a compensated antiferromagnet can be put under the form

$$\frac{\chi_s}{8N^2} = C\left(\frac{G}{\hbar c}\right)^2 k_B T \Omega_i(x_i), \; i = 1, 2, \; \text{as } N \to +\infty \quad (4.92)$$

where $\Omega_i(x_i)$ is a universal (scaling) function given by

$$\Omega_1(x_1) = \frac{1}{\pi}\left[\frac{1}{x_1} + \frac{\sqrt{4+e^{-2/x_1}}}{e^{-1/x_1}}\text{arcsin h}\left(\frac{e^{-1/x_1}}{2}\right)\right],$$
Zone 1 ($x_1 << 1$), Zone 3 ($x_1 >> 1$),

$$\Omega_2(x_2) = \frac{1}{\pi}\left[-\frac{1}{2x_2} + \frac{\sqrt{4+e^{1/x_2}}}{e^{1/2x_2}}\text{arcsin h}\left(\frac{e^{1/2x_2}}{2}\right)\right],$$
Zone 2 ($x_2 << 1$), Zone 4 ($x_2 >> 1$), (4.93)

Thus, if comparing Eqs. (4.88) and (4.92) we derive that the factor $k_B T \Omega_i(x_i)$ is linked to the vanishing law of the square of the magnetic moment $M$. These functions $\Omega_i(x_i)$ have also been obtained by Chubukov et al. [10]. In the Renormalized Classical Regime (RCR, $x_1 << 1$) this function has the asymptotic behavior

$$\Omega_1(x_1) = \frac{1}{\pi}\left[\frac{1}{x_1}+1\right], \text{Zone 1 } (x_1 << 1), \quad (4.94)$$

while in the Quantum-Critical Regime (QCR, $x_1 >> 1$)

$$\Omega_1(x_1) = \frac{C\sqrt{5}}{2\pi}\left[1+\frac{4}{5x_1}\right], \text{Zone 3 } (x_1 >> 1) \quad (4.95)$$

where $C$ is given by Eq. (4.41). In the Quantum-Disordered Regime (QDR, $x_2 \ll 1$) we have

$$\Omega_2(x_2) = \frac{e^{-1/x_2}}{\pi x_2}, \text{ Zone 2 } (x_2 \ll 1), \quad (4.96)$$

while in the other Quantum-Critical Regime (QCR, $x_2 \gg 1$) we find

$$\Omega_2(x_2) = \frac{C\sqrt{5}}{2\pi}\left[1 - \frac{2}{5x_2}\right], \text{ Zone 4 } (x_2 \gg 1). \quad (4.97)$$

As a result, near $T_c = 0$ K, if using the respective expressions of scaling parameters $x_1$ and $x_2$ given by Eq. (4.21), the product $\chi k_B T$ per surface unit can be put under the form for a compensated antiferromagnet

$$\frac{\chi_s k_B T}{8N^2} = 2C\rho_s\left(\frac{G}{\hbar c}\right)^2 k_B T\left[1 + \frac{k_B T}{2\pi\rho_s}\right], \text{ (RCR, } x_1 \ll 1), \quad (4.98)$$

$$\frac{\chi_s k_B T}{8N^2} = \frac{C\Delta}{\pi}\left(\frac{G}{\hbar c}\right)^2 k_B T e^{-\Delta/k_B T}, \text{ (QDR, } x_2 \ll 1), \quad (4.99)$$

$$\frac{\chi_s k_B T}{8N^2} = \frac{C^2\sqrt{5}}{2\pi}\left(\frac{G}{\hbar c}\right)^2 (k_B T)^2\left[1 + \frac{4}{5}\frac{2\pi\rho_s}{k_B T}\right], \text{ (QCR, } x_1 \gg 1), \quad (4.100)$$

$$\frac{\chi_s k_B T}{8N^2} = \frac{C^2\sqrt{5}}{2\pi}\left(\frac{G}{\hbar c}\right)^2 (k_B T)^2\left[1 - \frac{2}{5}\frac{\Delta}{k_B T}\right], \text{ (QCR, } x_2 \gg 1), \quad (4.101)$$

with

$$\Omega_i(\infty) = \frac{C\sqrt{5}}{2\pi} = 0.342\ 509, \quad \frac{C^2\sqrt{5}}{2\pi} = 0.329\ 639. \quad (4.102)$$

The result obtained for $\Omega_i(\infty)$ is exactly the same one (through the factor $C$) which has been derived by Chubukov et al. In addition these authors have found $\Omega_i(\infty) = 0.25 \pm 0.04$ owing to a Monte-Carlo method [10]. We shall comment the comparison between this result and that one given by Eq. (4.102) in Sec. V.

Thus, in addition to the experimental study of the low-temperature behavior of the correlation length, that of the product $\chi k_B T$ allows one to determine the nature of regime characterizing its vanishing behavior: If at first order in $k_B T$, the vanishing law is proportional to $k_B T$ we deal with a Renormalized Classical Regime (RCR, $x_1 \ll 1$); if it is proportional to $k_B T e^{-\Delta/k_B T}$ we deal with a Quantum-Disordered Regime (QDR, $x_2 \ll 1$); finally if the vanishing law is $(k_B T)^2$ we deal with a Quantum-Critical Regime (QCR, $x_1 \gg 1$ or $x_2 \gg 1$).

The previous results allow to determine the vanishing law of the magnetic moment $M(T)$ of a 2$d$-compensated antiferromagnet if using Eq. (4.87) and comparing with Eqs. (4.38)-(4.40) and (4.98)-(4.102). If $J = J_1 = J_2$ we find owing to Eq. (4.87b) that $M(T) \sim (\chi_s k_B T)^{1/2}/\xi$ so that it can be put under the universal form

$$M(T) = \sqrt{C} G \frac{k_B T}{\hbar c}\Omega_i(x_i)^{1/2}, J > 0, \text{ as } T \to 0. \quad (4.103)$$

If detailing for each zone of the magnetic phase diagram we have the following vanishing laws for a compensated antiferromagnet

$$M(T) = \sqrt{2C} G\left(\frac{\rho_s}{\hbar c}\right)^{1/2}\left(\frac{k_B T}{\hbar c}\right)^{1/2}, \text{ (RCR, } k_B T \ll \rho_s), \quad (4.104)$$

$$M(T) = \frac{\sqrt{C}}{\pi} G\left(\frac{\Delta}{\hbar c}\right)^{1/2}\left(\frac{k_B T}{\hbar c}\right)^{1/2} e^{-\Delta_s/2k_B T},$$
$$\text{(QDR, } k_B T \ll \Delta), \quad (4.105)$$

$$M(T) = C\left(\frac{\sqrt{5}}{2\pi}\right)^{1/2} G \frac{k_B T}{\hbar c}\left[1 + \frac{2}{5}\frac{2\pi\rho_s}{k_B T}\right],$$
$$\text{(QCR, } k_B T \gg \rho_s), \quad (4.106)$$

$$M(T) = C\left(\frac{\sqrt{5}}{2\pi}\right)^{1/2} G \frac{k_B T}{\hbar c}\left[1 - \frac{1}{5}\frac{\Delta}{k_B T}\right],$$
$$\text{(QCR, } k_B T \gg \Delta) \quad (4.107)$$

so that in the close vicinity of the critical point ($x_1 = x_2 \to +\infty$)

$$M(T) = C\left(\frac{\sqrt{5}}{2\pi}\right)^{1/2} G \frac{k_B T}{\hbar c} \to 0, \quad C\left(\frac{\sqrt{5}}{2\pi}\right)^{1/2} = 0.909\ 000,$$
$$\text{as } T \to T_c. \quad (4.108)$$

### E. Wilson ratio

We restrict the study to the case of a compensated antiferromagnet ($J = J_1 = J_2$, $J > 0$, and $G = G'$) because it is the most interesting one. As the critical temperature is $T_c = 0$ K the critical domain is characterized by temperatures $T > 0$. As a result the magnetic behaviors are essentially ruled by magnetic fluctuations described by the product $\chi k_B T$. For a compensated antiferromagnet ($J > 0$) $\chi_s k_B T/G^2 \sim \Omega_i(x_i)/L_\tau^2$ which always vanishes as $T \to 0$. The specific heat $C_v$ describes the thermal fluctuations.

Under these conditions the dimensionless Wilson ratio $R_W$ compares the magnetic and thermal fluctuations. We have

$$R_W = \frac{\pi}{3\zeta(3)}\frac{\Omega_i(x_i)}{\Psi_i(x_i)}, J > 0, i = 1,2. \quad (4.109)$$

i.e.,

$$R_W \approx \frac{1}{x_i} \to +\infty$$

($i = 1$, RCR, $k_B T \ll \rho_s$, $i = 2$, QDR, $k_B T \ll \Delta$),

$$R_W \approx \frac{C\sqrt{5}}{6C_0}, \quad \frac{C\sqrt{5}}{6C_0} = 0.372\,980,$$

$$(\text{QCR}, k_B T \gg \rho_s \text{ and } k_B T \gg \Delta) \quad (4.110)$$

where $C$ and $C_0$ are respectively given by Eqs. (4.41) and (4.63). Thus, for a compensated antiferromagnet ($J > 0$), if examining the magnetic properties, the situation is contrasted because we have a competition between the divergence of the correlation length and the evanescence of the magnetic moment $M(T)$. The magnetic fluctuations strongly dominate the thermal ones exclusively in the Renormalized Classical (RCR, $k_B T \ll \rho_s$) and in the Quantum Disordered (QDR, $k_B T \ll \Delta$) Regimes. Near criticality, for the Quantum Critical Regime ($k_B T \gg \rho_s$ and $k_B T \gg \Delta$), we have $R_W < 1$ so that the thermal fluctuations slightly dominate the magnetic ones.

## V. COMPARISON WITH EXPERIMENTAL RESULTS

In this section we compare theoretical results and available experimental data concerning susceptibilities of $2d$-compensated antiferromagnets. Of course, here we exclusively focus on magnetic compounds involving classical spins i.e., spins characterized by a spin quantum number $S \geq 5/2$. There are two classes of antiferromagnets: i) Inorganic compounds characterized by planes of magnetic ions, each one carrying a classical spin, sufficiently separated owing to nonmagnetic species (class I) and ii) mixed compounds of classical spin planes (inorganic part) which are largely separated owing to organic ligands whose length is adjustable; in addition, inside each plane, organic ligands can be also inserted between magnetic ions (class II).

In class I we have the family of $K_2MnF_4$ and $Rb_2MnF_4$ studied in the sixties by Breed [35,36]. We deal with a mechanism of *superexchange* between magnetic ions through ions $F^-$ [37]. The family of $BaMF_4$ (with M = Mn, Fe, Co) has been investigated by Eibschütz *et al*. [38]. It has notably drawn a particular attention because a $2d$-Heisenberg antiferromagnetic behavior could be observed at low temperatures. Three conditions must be respected: i) The crystallographic structure of these compounds must be a plane (for $BaMF_4$ we have puckered sheets of $MF_6$ octaedra, separated by nonmagnetic $Ba^{2+}$ ions along the *b*-axis of the crystal so that the magnetic ion lattice is characterized by a square unit cell); typical in-plane lattice constants are $a = b \sim 4$ to 5 Å and out-of-plane distance $c \sim 11$ to 13 Å; ii) within each magnetic plane, a local uniaxial anisotropy field shows a residual contribution $\alpha$ so that exchange between first-nearest neighbor can be considered as isotropic ($\alpha = 3.1 \times 10^{-4}$ for $BaMnF_4$) within a wide temperature range as it often occurs [39]; iii) if $J$ is the in-plane exchange energy between first-nearest neighbors $Mn^{2+}$ and $J'$ the out-plane exchange energy between first-nearest neighbor planes, the ratio $|J'/J|$ must be very small in order to restrict the appearance of a $3d$-magnetic ordering (characterized by a temperature $T_{3d}$) at the lowest possible temperature ($|J'/J| \sim 10^{-6}$ for $BaMnF_4$).

Unfortunately, in such kind of structures, the nature of exchange coupling can change with temperature due to anisotropy. For instance this is the case for the compound $Rb_2MnF_4$ where the coupling is isotropic from room temperature down to 46 K and becomes anisotropic (Ising type) between 46 K and $T_{3d} = 38.4$ K [40]. In addition it can be difficult to restrict dipole-dipole interactions between planes. Thus, $3d$ ordering appears when the interlayer energy between similar in-plane Kadanoff blocks of surface $\xi^2$ (where $\xi$ is the correlation length if $J_1 = J_2 = J$) belonging to first-nearest neighbor planes is such as

$$k_B T_{3d} \approx \left(\frac{\xi}{a}\right)^2 |J'| \quad (4.111)$$

where $a$ is the lattice spacing. For instance, in the case of $BaMnF_4$ Eibschütz *et al*. have found $T_{3d} = 25$ K whereas $T(\chi_{max}) = 52.6$ K [38]. As a result the temperature range $[T_{3d}, T(\chi_{max})] = 27.6$ K characterizing the $2d$-magnetic behavior is not very large.

An intermediate solution has been found in the seventies with the introduction of organic ligands between the magnetic planes. Van Amstel and de Jongh have examined the series $(C_nH_{2n+1}NH_3)_2CuX_4$ ($n = 1, 2, 3$) with X = Cl, Br in which each compound is characterized by ferromagnetic couplings [41]. The $2d$-magnetic properties find their origin in the fact that the magnetic Cu layers are separated by two sheets of alkyl ammonium groups. The crystal structure is close to that of $K_2NiF_4$ and the magnetic layers show a square unit cell. By substituting Mn ($S = 5/2$) for Cu ($S = 1/2$) the nature of coupling changes for becoming antiferromagnetic. Thus, $(CH_3NH_3)_2MnCl_4$ is characterized by the ratio $|J'/J| \sim 10^{-8}$ and a uniaxial anisotropy field contribution $\alpha = 1.1 \times 10^{-3}$ as well as $T_{3d} = 47$ K whereas $T(\chi_{max}) = 79.9$ K. In that case too, the range $[T_{3d}, T(\chi_{max})] = 32.9$ K is not large but greater than the one of $BaMnF_4$. Thus the introduction of nonmagnetic organic ligands between sheets of $Mn^{2+}$ ions has allowed to enlarge the range $[T_{3d}, T(\chi_{max})]$, as expected.

The relevant solution came at the end of the nineties with the simultaneous introduction of organic ligands between the magnetic planes but also between the in-plane magnetic ions [42-44]. We deal with class II compounds also characterized, as those of class I, by a mechanism of superexchange between in-plane magnetic ions. A complete review has been published by Escuer *et al*. [45]. In this paper the authors notably point out the central role played by the azido ligand. This ligand introduces a bridge composed of 1 to 3 nitrogen atoms between magnetic in-plane cations so that the distance between them notably increases, thus i) ensuring an isotropic character of coupling through a mechanism of superexchange and ii) drastically diminishing the dipole-dipole interactions. The shortest in-plane distance observed for the bond Mn-Mn is $a = b \sim 11,6$ Å which is the double of the similar distance observed for class I-compounds [45]. Other specific organic ligands can be similarly introduced between the sheets of magnetic ions thus allowing to control the intersheet distance along the *c*-axis and finally the value of the $3d$-ordering temperature $T_{3d}$. Typical out-of-plane distances $c$ are comparable to those ones observed for class I-compounds, with $c \sim$ 12 to 15 Å. As a result a value of $T_{3d} = 2$ K (or less) is frequently reached which provides a considerable extension of the range $[T_{3d}, T(\chi_{max})]$ in which a characteristic 2$d$-magnetic behavior appears.

For fitting the experimental points obtained for the susceptibility $\chi^z = \chi = \tilde{\chi}/3$ ($\tilde{\chi}$ being the total susceptibility whose theoretical behaviors have been examined in Subsec. IV.D) we have to determine the relevant $l$-terms that must be kept in the $l$-polynomial expansion (cf Eqs. (3.23)-(3.25)). The thermal predominance of each $l$-term is similar to that one encountered for the zero-field partition function (see Fig. 2). As a result, for fitting an experimental susceptibility, we have to take into account this study. Besides the well-known vanishing behavior of a 2$d$-compensated antiferromagnet while reaching $T_c = 0$ K the susceptibility shows a maximum at a Néel temperature $T_N$ such as $k_B T_N \sim J$ i.e., for the reduced temperature $k_B T_N / J \sim 1$. If $T > T_N$ we deal with the paramagnetic regime and the susceptibility diminishes, as expected.

We have seen in Fig. 2 that, for $k_B T/|J| \geq 0.255$ the dominant $l$-term is characterized by $l = 0$. In the thermodynamic limit, if $G = G'$ and $J_1 = J_2 = J$, the susceptibility $\tilde{\chi} = 3\chi$ where $\chi$ is given by Eq. (3.24) reduces to

$$\tilde{\chi} = 3\chi_0 = \beta G^2 \left( \frac{1 + \mathcal{L}(-\beta J)}{1 - \mathcal{L}(-\beta J)} \right)^2 , \quad \frac{k_B T}{|J|} \geq 0.255 . \quad (4.112)$$

$\mathcal{L}(-\beta J)$ is the Langevin function and $\chi_0$ the $l$-polynomial expansion of $\chi$ restricted to $l = 0$. Differentiating the susceptibility with respect to temperature for obtaining its maximum leads to the following equation

$$(4x - 1)(1 - \coth^2 x) + \frac{2}{x}(2 - \coth x) + \frac{1}{x^2} = 0 ,$$

$$x = \frac{JS(S+1)}{k_B T} \quad (4.113)$$

where we have renormalized $J$ by the factor $S(S + 1)$. This equation can only be solved numerically with a great accuracy [18]. The unique solution is given by the following relation

$$\frac{k_B T(\chi_{max})}{JS(S+1)} = 1.26254250 \pm 0.00000001 . \quad (4.114)$$

As $\chi$ and $\tilde{\chi} = 3\chi$ show the same maximum we have replaced $\tilde{\chi}$ by $\chi$ for practical reasons. This study has already been achieved by Lines for various $S$-values and notably in the classical limit ($S \to +\infty$), from Rushbrooke high-temperature (HT) expansions [46]. He found $k_B T(\chi_{max})/JS(S+1) = 1.12 \pm 0.10$. Finally, if calculating the theoretical value of the susceptibility $\chi_{max}$ as well as the product $\chi_{max} k_B T(\chi_{max})$ we find the exact results

$$\frac{\chi_{max} JS(S+1)}{N_A G^2} = 0.0936 , \quad \frac{\chi_{max} k_B T(\chi_{max})}{C} = 0.3545 \quad (4.115)$$

where $N_A$ is the Avogadro number and $C$ is the Curie constant i.e., $C = G^2/3k_B$ (we have restricted the precision of the result given by Eq. (4.114) for practical reasons). de Jongh has calculated these quantities from refined Rushbrooke HT expansions (stopped at the ninth rank). He respectively found closed values i.e., $\chi_{max} JS(S+1)/N_A G^2 = 0.1122 \pm 0.0002$ and $\chi_{max} k_B T(\chi_{max})/C = 0.3380 \pm 0.0002$ [47,48].

Thus, from the knowledge of the experimental value of $T(\chi_{max})$, $\chi_{max}$ or $\chi_{max} k_B T(\chi_{max})$ the value of $J$ can be directly determined. In addition, once $J$ known, if $T_{3d}$ is the 3$d$-ordering temperature, the ratios $k_B T_{3d}/JS(S+1)$ and $k_B T(\chi_{max})/JS(S+1)$ as well as the intermediate ratios $k_B T/JS(S+1)$ can allow to determine the $l$-values which must be used in the $l$-polynomial expansion of the susceptibility, in the temperature range $[T_{3d}, +\infty[$. This work can be easily extended to the general case $J_1 \neq J_2$. We can derive equations similar to Eq. (4.113)-(4.115) but now they are functions of $J_1$ and $J_2$. It is then possible to determine $J_1$ and $J_2$ from the experimental value of $T(\chi_{max})$ and $\chi_{max}$ or $T(\chi_{max})$ and $\chi_{max} k_B T(\chi_{max})$.

We have previously fitted the experimental susceptibility of BaMnF$_4$ (class I-compound) [18] measured by Holmes et al. (see Fig. 13a) [39]. The 3$d$-ordering temperature is $T_{3d} = 25.0$ K and $T(\chi_{max}) = (52.6 \pm 2.5)$ K because $\chi$ shows a broad maximum. We have a 2$d$-magnetic behavior within the range $[T_{3d}, T(\chi_{max})]$ of width $T(\chi_{max}) - T_{3d} = 27.6$ K. If $T > T(\chi_{max})$ we have a paramagnetic behavior. From Eq. (4.114) we derive $J/k_B = (4.76 \pm 0.24)$ K. The Landé factor is $G = (1.97 \pm 0.10)$ $\mu_B/\hbar$ which is very close to the theoretical value $G = 2$ $\mu_B/\hbar$. The HT expansion derived from Eq. (4.112) also gives the same value of $J/k_B$. de Jongh et al. have fitted the HT susceptibility [47]. They found $J/k_B = (5.50 \pm 0.17)$ K. However we have shown that these HT expansions [47,48] do not cover enough the region of maximum where $k_B T(\chi_{max})/JS(S+1) \sim 1$ whereas ours do it [18]. Due to Eq. (4.114) we know that $k_B T(\chi_{max})/JS(S+1) = 1.2625$ and from the experimental value of $T_{3d}$ we have $k_B T_{3d}/JS(S+1) = 0.60$. As it appears on Figs. 2a and 2b we must have $k_B T/JS(S+1) \geq 0.255$ for justifying the use of Eq. (4.112) for which the $l$-polynomial of the susceptibility is restricted to the term $l = 0$ in the range $[T_{3d}, +\infty[$. We verify that this restriction is duly valid.

We have also done a similar work for the experimental susceptibility of (CH$_3$NH$_3$)$_2$MnCl$_4$ (an intermediate compound between classes I and II) [18] studied by Van Amstel et al. [41] (see Fig. 13b). The 3$d$-ordering temperature is $T_{3d} = 47.0$ K and $T(\chi_{max}) = (79.9 \pm 3.3)$ K because $\chi$ shows a broad maximum. We have a 2$d$-magnetic behavior within the range $[T_{3d}, T(\chi_{max})]$ of width $T(\chi_{max}) - T_{3d} = 32.9$ K. From Eq. (4.114) we derive $J/k_B = (7.23 \pm 0.30)$ K. The HT expansion derived from Eq. (4.112) also gives the same value of $J/k_B$ (see Fig. 13b). The Landé factor is $G = (1.89 \pm 0.10)$ $\mu_B/\hbar$ which is very close to the theoretical value $G = 2$ $\mu_B/\hbar$.

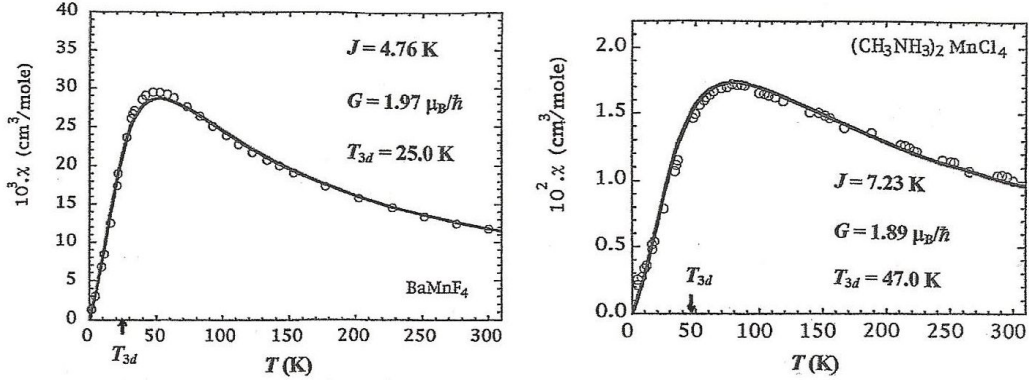

FIG. 13. (a) Fits of the experimental susceptibility of the class I-compounds BaMnF$_4$ [39] and (b) (CH$_3$NH$_3$)$_2$MnCl$_4$ [41] with the theoretical expression of the static susceptibility given by Eq. (4.112) [18]. The fits have been achieved for temperatures $T > T_{3d}$. The continuation of the corresponding curves below $T_{3d}$ and the coincidence with experimental points has no physical meaning because they have been excluded from the fits.

Due to Eq. (4.114) we know that $k_B T(\chi_{max})/JS(S + 1) =$ 1.2625 and from the experimental value of $T_{3d}$ we have $k_B T_{3d}/JS(S + 1) = 0.743$ which justifies the use of Eq. (4.112) for which the $l$-polynomial of the susceptibility is restricted to the term $l = 0$. Indeed we always have $k_B T/JS(S + 1) \geq 0.255$ for any temperature $T$ belonging to the range $[T_{3d}, +\infty[$.

Finally we have examined the experimental susceptibility of a series of class II-compounds [42,45]. We just report here the case of [Mn(DENA)$_2$(N$_3$)$_2$]$_n$. DENA is the organic ligand diethylnicotinamide [44] which allows an important separation between sheets of Mn$^{2+}$ ions ($c \sim 13$ Å). Within each sheet Mn$^{2+}$ ions are linked through 4 azido bridges, each one involving 3 nitrogen atoms so that the path length of superexchange between magnetic ions varies ($a \sim 14$ Å, $b \sim 6.7$ Å). This permits to ensure a good isotropic character of couplings between Mn$^{2+}$ ions. Due to this crystallographic construction the 3$d$-ordering temperature is such as $T_{3d} < 2.0$ K. Under these conditions we have fitted the experimental susceptibility (see Fig. 14a) as well as the product $\chi k_B T$ (see Fig. 14b) between 2 K and the room temperature. $\chi$ shows a relatively sharp maximum at 44 K. We have a 2$d$-magnetic behavior within the range $[T_{3d}, T(\chi_{max})]$ of width $T(\chi_{max}) - T_{3d} = 42$ K. Due to Eq. (4.114) we have $k_B T(\chi_{max})/JS(S + 1) = 1.2625$ and from the experimental value of $T_{3d}$ $k_B T_{3d}/JS(S + 1) < 0.057$. It means that, that times, we can use the $l$=0-term from room temperature down to $k_B T/JS(S + 1) = 0.255$ and the $l$=1-term in the range $0.043 < k_B T/JS(S + 1) < 0.255$. From Eqs. (3.23) and (3.24) giving the $l$-polynomial expansion of $\chi$ restricted to $l = 0$ and $l = 1$ we derive $J/k_B = (4.00 \pm 0.02)$ K and a Landé factor $G = (1.98 \pm 0.02)$ µ$_B$/ℏ which is very close to the theoretical value. From a practical point the $l$=0-term has been used down to 8.925 K and the $l$=1-term between 8.925 K and 2 K (this expansion is valid down to 1.995 K but, in this range, we deal with a 3$d$-magnetic behavior). We observe a remarkable agreement on Fig. 14 between the experimental points and the theoretical curves for the value of $J/k_B$ measured for the same kind of bridging ligand involved in other class II-compounds [44,45].

For all the experimental susceptibilities of class I-compounds that we have fitted the $l$=0-term of the $l$-polynomial expansion of the susceptibility has been retained on the range $[T_{3d}, T(\chi_{max})]$, thus validating the use of Eq. (4.112). It can be expanded in the low $T$-limit if $0.255 < k_B T_{3d} /JS(S + 1) < 1$. Under these conditions we have for the corresponding surface density of susceptibility per bond expressed near $T_{3d}$ in the low-temperature range

$$\frac{\chi_s k_B T}{8N^2} \approx \frac{1}{4}\left(\frac{G}{\hbar c}\right)^2 (k_B T)^2, J > k_B T > k_B T_{3d}, (l = 0)$$

as $T \rightarrow T_{3d} \sim 0$ K. (4.116)

If comparing with the low-temperature behaviors of the product $\chi_s k_B T$ (cf Eqs. (4.98)-(4.101)) the unique case for which the $T$-cancellation law has the form $(k_B T)^2$ concerns the Quantum Critical Regime (QCR). Thus, if comparing Eqs. (4.116), (4.100) and (4.101), we identify the factor to $\Omega(\infty) = 0.25$. It means that this factor is associated with the term $l = 0$ in the $l$-polynomial expansion of the susceptibility. Due to the fact that all the $l$-eigenvalues has the same low-temperature behavior while approaching $T_c = 0$ K (including the one corresponding to $l = 0$) just differing by a factor exclusively depending on $l$, it now clearly appears that the value $\Omega(\infty) = 0.25$ obtained by Chubukov et al through a Monte Carlo method [10] is a lower bound of the exact value of $\Omega(\infty)$ given by Eq. (4.102) and also calculated by these authors through a renormalization process.

Finally, the remarkable good agreement between the $J$-values of the exchange energy derived from the fits and the corresponding ones previously measured as well as a value of the Landé factor close to the theoretical one has allowed to characterize the magnetic properties of a new class of compounds obeying a quantum critical behavior in the low-temperature domain.

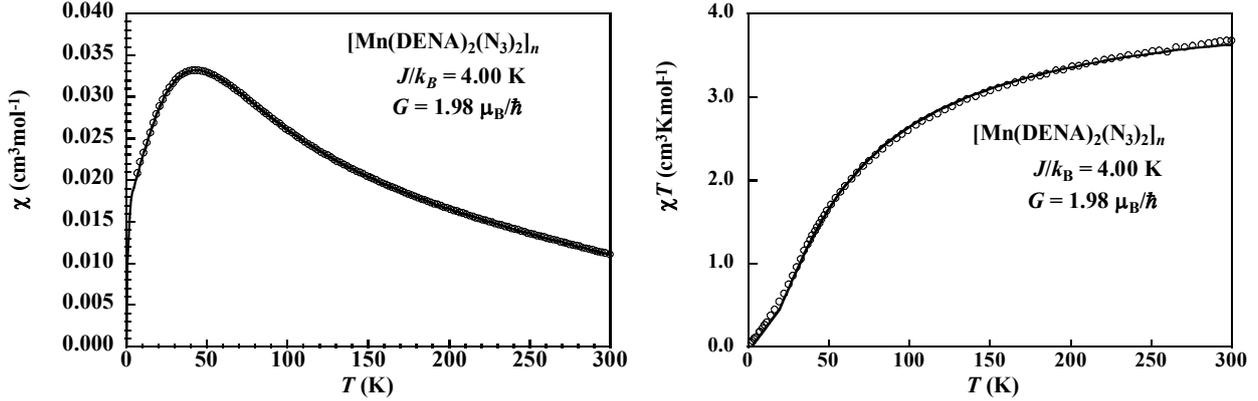

FIG. 14. (a) Fits of the experimental susceptibility $\chi$ of the class II-compounds $[Mn(DENA)_2(N_3)_2]_n$ (DENA = diethylnicotinamide ligand) with the theoretical expression of the static susceptibility given by Eq. (4.112). The fits have been achieved for temperatures $T > 2$ K. Magnetization measurements have allowed to prove that $T_{3d} < 2$ K [44]. The continuation of the theoretical curves below $T_{3d}$ has no physical meaning because experimental points have been excluded from the fits.

## VI. CONCLUSION

In this paper we have presented the exact general theory of the two-dimensional Heisenberg square lattice composed of classical spins. In the thermodynamic limit a numerical study has allowed to select the higher-degree term of the characteristic $l$-polynomial associated with the zero-field partition function $Z_N(0)$. It is characterized by $m_{i,j} = 0$ and $l_{i,j} = l \geq 0$ for the whole lattice so that a very simple exact closed-form expression has been derived, thus permitting to express the free energy $F$ and the specific heat $C_V$, *for any temperature*. We have reported a thermal study of the basic term appearing in the higher-degree one of $Z_N(0)$. We have shown that it appears crossovers between two consecutive terms. Thus, if coming from high temperatures where the term characterized by $l = 0$ is dominant, near absolute zero, eigenvalues showing increasing $l$-values are more and more selected when the temperature is cooling down. When $T$ reaches zero, all the successive dominant eigenvalues become equivalent so that the critical temperature is $T_c = 0$ K. Owing to a similar method applied in the thermodynamic limit we have derived an exact expression *valid for any temperature* for the spin-spin correlations, the correlation length $\xi$ and the susceptibility $\chi$ which show the same thermal crossovers as $Z_N(0)$.

Near $T_c = 0$ K we have obtained a diagram of magnetic phases which is similar to the one derived through a renormalization approach. The low-temperature magnetic properties are described in terms of universal parameters $k_B T/\rho_s$ and $k_B T/\Delta$ where $\rho_s$ and $\Delta$ are the spin stiffness and the $T=0$-energy gap between the ground state and the first exited one, respectively. We have given the low-temperature expressions of the correlation length $\xi$ and the surface density of $F$, $C_V$ and $\chi$: They are exactly similar to the corresponding ones derived through a renormalization process, for each zone of the magnetic phase diagram. We have retrieved the correct value of the correlative critical exponents. In addition, we have shown that, near $T_c = 0$ K, the lattice is composed of quasi rigid quasi independent Kadanoff blocks of length $\xi$ and magnetic moment $M(T)$, the unit cell moment, so that $\chi k_B T \sim \xi^2 M(T)^2$. For a compensated antiferromagnet the expression of the susceptibility has necessitated a special treatment due to the fact that all the eigenvalues gives a small thermal contribution. In that case too we have retrieved the result of Chubukov et al. [10] obtained through a renormalization process. This work has allowed to identify the discrepancy between this result and that one obtained by these authors owing to a Monte Carlo method: We have shown that the corresponding result $\Omega = 0.25$ must be attributed to the $l$=0-term of the $l$-polynomial expansion of the susceptibility.

Finally we have compared experimental susceptibilities to the theoretical expression of $\chi$ derived in this article, for two types of $2d$-compounds characterized by the absence of organic ligands (class I) between $Mn^{2+}$ ions assembled in sheets and/or between these magnetic planes but also in presence of these ligands (class II). In the whole temperature range i.e., between room temperature down to $T_{3d}$, the $3d$-ordering temperature, we have obtained a remarkable good agreement between the $J$-values of the exchange energy derived from the fits and the corresponding ones previously measured as well as a value of the Landé factor close to the theoretical one. These fits have allowed to show that all these compounds are characterized by a quantum critical regime.

## APPENDIX A: EXPRESSION OF THE ZERO-FIELD PARTITION FUNCTION OF A SQUARE LATTICE IN THE THERMODYNAMIC LIMIT

For $T > 0$ K, between two consecutive crossover temperatures $T_{i,<}$ and $T_{i,>}$, we have shown in the main text that, in the thermodynamic limit, $Z_N(0)$ can be written as

$$Z_N(0) \approx (4\pi)^{8N^2}[u_{\max}(T)]^{4N^2}, \; T \in [T_{l_i,<}, T_{l_i,>}],$$
$$u_{\max}(T) = F_{l_i,l_i}\lambda_{l_i}(-\beta J)^2, \text{ as } N \to +\infty, \quad (A1)$$

if $J = J_1 = J_2$; $\lambda_{l_i}(-\beta J)$ is the dominant eigenvalue characterizing the range $[T_{l_i,<}, T_{l_i,>}]$, with $l_{\max} = l_i$.

More generally the zero-field partition function of an infinite square lattice can be rewritten as:

$$Z_N(0) = (4\pi)^{8N^2} \prod_{i=-(N-1)}^{N} \prod_{j=-(N-1)}^{N} \sum_{l_i=0}^{+\infty} \sum_{l_j=0}^{+\infty} u_{l_i,l_j}(T) \quad (A2)$$

with:

$$u_{l_i,l_j}(T) = F_{l_i,l_j}\lambda_{l_i}(-\beta J)\lambda_{l_j}(-\beta J),$$
$$l_i = l_j \text{ or } l_i \neq l_j, T \in [T_{l_i,<}, T_{l_i,>}]. \quad (A3)$$

Inside each range $[T_{l_i,<}, T_{l_i,>}]$ a numerical study achieved in the main text (*cf* Fig. 2) has led to a classification of terms $u_{l_i,l_j}(T)$, with $l_i = l_j$ or $l_i \neq l_j$. If factorizing the dominant term $u_{\max}(T)$ of the polynomial expression giving $Z_N(0)$ Eq. (A2) can be written under the following form when $N \to +\infty$:

$$Z_N(0) = (4\pi)^{8N^2}[u_{\max}(T)]^{4N^2}\{1 + S(N,T)\},$$
$$S(N,T) = S_1(N,T) + S_2(N,T), T \in [T_{l_i,<}, T_{l_i,>}] \quad (A4)$$

where $S_1(N,T)$ and $S_2(N,T)$ are defined by Eq. (2.34).

For $S_1(N,T)$ the current term is $[u_{l,l}(T)/u_{\max}(T)]^{4N^2} < 1$. wheras for $S_2(N,T)$ it is composed of a product of terms $[u_{l_i,l_j}(T)/u_{\max}(T)]^{n_{i,j}} < 1$, with $n_{i,j} < 4N^2$. As a result $S_1(N,T)$ and $S_2(N,T)$ are *absolutely convergent series* characterized by positive current terms lower than unity.

Due to the numerical property of $u_{l_i,l_i}(T)$ and $u_{l_i,l_j}(T)$, for a given $T \in [T_{l_i,<}, T_{l_i,>}]$, a classification of the various terms of $S_1(N,T)$ and $S_2(N,T)$ can be achieved. As a result the series $S(N,T) = S_1(N,T) + S_2(N,T)$ is also an absolutely convergent series characterized by an infinite sum of positive vanishing current terms symbolically labeled $X_k(N,T)$. These terms can be written in the decreasing modulus order so that

$$S(N,T) = \sum_{k=0}^{+\infty} X_k(N,T), \; 0 < X_k(N,T) < 1, \quad (A5)$$

with:

$$X_1(N,T) > X_2(N,T) > \ldots > X_\infty(N,T), \; T \in [T_{l_i,<}, T_{l_i,>}]. \; (A6)$$

Now we artificially share the infinite series $S(N,T)$ into two parts:

$$S(N,T) = S_{k_i}^{\text{B}}(N,T) + S_{k_i}^{\text{E}}(N,T), T \in [T_{l_i,<}, T_{l_i,>}], \; (A7)$$

with:

$$S_{k_i}^{\text{B}}(N,T) = \sum_{k=0}^{k_i} X_k(N,T), \; S_{k_i}^{\text{E}}(N,T) = \sum_{k=k_i}^{+\infty} X_k(N,T) \; (A8)$$

where $S_{k_i}^{\text{B}}(N,T)$ and $S_{k_i}^{\text{E}}(N,T)$ are the beginning and the end of $S(N,T)$, respectively.

We have the natural inequalities $S_{k_i}^{\text{B}}(N,T) < S(N,T)$ and $S_{k_i}^{\text{E}}(N,T) < S(N,T)$. As we deal with an infinite (absolutely convergent) series made of positive vanishing current terms $X_k(N,T) < 1$ it is always possible to find a particular value $k_i = k_1$ of the general index $k$ such as:

$$S_{k_1}^{\text{B}}(N,T) = S_{k_1}^{\text{E}}(N,T) = \frac{\varepsilon}{2}, \; S(N,T) = \varepsilon, 0 < \varepsilon < 1,$$
$$T \in [T_{l_i,<}, T_{l_i,>}]. \quad (A9)$$

If increasing $N \gg 1$ of $n > 0$ we automatically have

$$S_{k_1}^{\text{K}}(N+n,T) < S_{k_1}^{\text{K}}(N,T) = \frac{\varepsilon}{2}, \text{K} = \text{B, E}, T \in [T_{l_i,<}, T_{l_i,>}]$$
$$(A10)$$

because the inequality $0 < X_k(N,T) < 1$ imposes $0 < X_k(N+n,T) < X_k(N,T) < 1$. Finally, if calling $S(N+n,T)$ the sum $S_{k_1}^{\text{B}}(N+n,T) + S_{k_1}^{\text{E}}(N+n,T)$ we have

$$S(N+n,T) < S(N,T) = \varepsilon, T \in [T_{l_i,<}, T_{l_i,>}]. \quad (A11)$$

As a result we derive

$$S(N,T) = S_1(N,T) + S_2(N,T) \to 0, \text{ as } N \to +\infty,$$
$$\text{for } T \in [T_{l_i,<}, T_{l_i,>}], \quad (A12)$$

and

$$Z_N(0) \approx (4\pi)^{8N^2}[u_{\max}(T)]^{4N^2}, \text{ as } N \to +\infty,$$
$$\text{for } T \in [T_{l_i,<}, T_{l_i,>}]. \quad (A13)$$

This reasoning can be repeated for each new range of temperature $[T_{l_j,<}, T_{l_j,>}]$, with $j \neq i$. Consequently, if summing Eq. (A13) over all the ranges $[T_{l_i,<}, T_{l_i,>}]$ so that $T = \sum_{i=0}^{i_{\max}}(T_{l_i,>} - T_{l_i,<}), T_{l_0,<} = 0, \; T_{l_{i-1},>} = T_{l_i,<}$ ($i \neq 0$) and $T_{l_{i_{\max}},>} = T$ we always have

$$\sum_{l=0}^{+\infty} u_{l,l}(T)^{4N^2} \gg \prod_{i=-(N-1)}^{N} \prod_{j=-(N-1)}^{N} \sum_{l_i=0}^{+\infty} \sum_{l_j=0, l_j \neq l_i}^{+\infty} u_{l_i,l_j}(T),$$

$$\text{as } N \to +\infty \quad (A14)$$

so that finally

$$Z_N(0) \approx (4\pi)^{8N^2} \sum_{l=0}^{+\infty} {}^t \left[ F_{l,l} \lambda_l(-\beta J_1) \lambda_l(-\beta J_2) \right]^{4N^2},$$

$$\text{as } N \to +\infty. \quad (A15)$$

In other words this expression is valid for any temperature $T = \sum_{i=0}^{i_{max}} (T_{l_i,>} - T_{l_i,<})$. However the special notation $\sum_{l=0}^{+\infty} {}^t$ recalls that the summation is truncated: In each temperature range $[T_{l_i,<}, T_{l_i,>}]$ the eigenvalue $\lambda_{l_i}(-\beta J_k)$ with $k = 1, 2$ is dominant. At the frontier $T = T_{l_i,<}$ $\lambda_{l_i}(-\beta J_k)$ and $\lambda_{l_i-1}(-\beta J_k)$ must be taken into account whereas if $T = T_{l_i,>}$ we have to consider $\lambda_{l_i}(-\beta J_k)$ and $\lambda_{l_i+1}(-\beta J_k)$.

## APPENDIX B: EXPRESSION OF $|\zeta|\Lambda$ NEAR THE CRITICAL POINT

The expression of the thermodynamic functions derived from the spin-spin correlations involves ratios of Bessel functions $I_l(lz)$. Near the critical point $T_c = 0$ K, these functions have to be evaluated in the double limit $\beta|J| = l|z| \to +\infty$ and $l \to +\infty$. In this limit Olver has shown [24] that the argument $\beta|J| = l|z|$ must be replaced by $l|\zeta|$ where

$$\zeta = -\frac{J}{|J|} \left[ \sqrt{1+z^2} + \ln\left(\frac{|z|}{1+\sqrt{1+z^2}}\right) \right], \quad z = \frac{\beta J}{l}. \quad (B1)$$

In the main text, we have seen that, near $T_c = 0$ K, $l$ is replaced by $\Lambda = 2l$. $l|\zeta|$ then becomes $|\zeta^*|\Lambda$. At the fixed point $z_c^* = 1/4\pi$ we exactly have $|\zeta^*| = 0$. Near this critical point (see Fig. 8), $\sqrt{1+z^{*2}} \approx \left|\ln\left(z^* / \left[1+\sqrt{1+z^{*2}}\right]\right)\right|$ for any Zone 1 to 4. As a result Eq. (B1) reduces to

$$|\zeta^*| \approx 2\left|\ln\left(\frac{1}{|z^*|} + \sqrt{1+\frac{1}{z^{*2}}}\right)\right|, \quad z^* = \frac{\beta J}{\Lambda}. \quad (B2)$$

or equivalently

$$|\zeta^*| \approx 2\left|\text{argsh}\left(\frac{1}{|z^*|}\right)\right|, \text{ as } |z^*| \to z_c^*. \quad (B3)$$

When $|z^*| < 1$ or $|z^*| > 1$, near $z_c^* = 1/4\pi$, Eq. (B2) finally reduces to $2|\ln(|z^*|/2)|$ which depends on $\Lambda$. As the ratio $|z^*|/z_c^* = T_c/T$ is independent of $\Lambda \gg 1$ we can write

$$|\zeta^*|\Lambda \approx 2\ln\left(\left|\frac{1}{2}\frac{z^*}{z_c^*}\right|^\Lambda\right), \text{ as } |z^*| \to z_c^*. \quad (B4)$$

If $|\zeta^*|\Lambda$ is a scaling parameter we must show that $(|z^*|/z_c^*)^\Lambda$ is independent of $\Lambda$. The comparison with Eqs. (B3) and (B4) allows one to write

$$|\zeta^*|\Lambda \approx 2\,\text{argsh}\left(\left(\frac{1}{2}\frac{|z^*|}{z_c^*}\right)^\Lambda\right), \text{ as } |z^*| \to z_c^*. \quad (B5)$$

In Zones 1 ($x_1\Lambda \ll 1$, $x_1 \ll 1$) and 3 ($x_1\Lambda \gg 1$, $x_1 \gg 1$), $|z^*| > z_c^*$ so that, from the definition of $\rho_s$ (cf Eq. (4.19)), we have

$$\Lambda(|z^*| - z_c^*) = \frac{\rho_s}{k_B T} \quad (B6)$$

or equivalently by introducing $x_1$ given by Eq. (4.21):

$$\Lambda(|z^*| - z_c^*) = \frac{2z_c^*}{x_1}, \quad z_c^* = \frac{1}{4\pi}. \quad (B7)$$

Using the well-known relation $(1 \pm u/\Lambda)^\Lambda = \exp(\pm u)$, as $\Lambda \to +\infty$, we derive that, near $z_c^*$,

$$\frac{|z^*|}{z_c^*} = \exp(2/x_1\Lambda), \left(\frac{|z^*|}{z_c^*}\right)^\Lambda = \exp(2/x_1). \quad (B8)$$

In Zones 2 ($x_2\Lambda \sim 1$, $x_2 \ll 1$) and 4 ($x_2\Lambda \gg 1$, $x_2 \gg 1$) $|z^*| < z_c^*$. We similarly have from the definition of $\Delta$ (cf Eq. (4.19))

$$\Lambda(z_c^* - |z^*|) = \frac{\Delta}{4\pi k_B T} \quad (B9)$$

or equivalently by introducing $x_2$ given by Eq. (4.21):

$$\Lambda(z_c^* - |z^*|) = \frac{z_c^*}{x_2} \quad (B10)$$

so that, near $z_c^*$,

$$\frac{|z^*|}{z_c^*} = \exp(-1/x_2\Lambda), \left(\frac{|z^*|}{z_c^*}\right)^\Lambda = \exp(-1/x_2). \quad (B11)$$

As $x_1$ and $x_2$ are scaling parameters the ratios $\left(|z^*|/z_c^*\right)^\Lambda$ and $\left(z_c^*/|z^*|\right)^\Lambda$ appear as scaling parameters. Thus $|\zeta^*|\Lambda$ given by Eq. (B4) or (B5) is a scaling parameter.

Now, for expressing $|\zeta^*|\Lambda$, we have to choose between the ratios $|z^*|/z_c^*$ and $z_c^*/|z^*|$, respectively given by Eqs. (B8) and (B11). The behavior of $|\zeta^*|\Lambda$ as depicted by Fig. 9 imposes the following properties: i) If $|z^*| > z_c^*$ ($T < T_c$) $|\zeta^*|\Lambda$ increases with $|z^*|$ ($T$ diminishes) and must vary as a function of argument $-1/x_1$, thus imposing to consider the ratio $z_c^*/|z^*|$; ii) if $|z^*| < z_c^*$ ($T > T_c$) $|\zeta^*|\Lambda$ increases when $|z^*|$ diminishes ($T$ increases) and must vary as a function of argument $1/x_2$, thus imposing to consider the ratio $z_c^*/|z^*|$.

As a result, if taking into account Eqs. (B5), (B8) and (B11), we can write

$$|\zeta^*|\Lambda \approx 2\,\mathrm{argsh}\left(\frac{\exp(-1/x_1)}{2}\right) \text{ (Zones 1 and 3)}, \quad \text{(B12)}$$

$$|\zeta^*|\Lambda \approx 2\,\mathrm{argsh}\left(\frac{\exp(1/2x_2)}{2}\right) \text{ (Zones 2 and 4)}. \quad \text{(B13)}$$

As a result, in Zone 1 ($x_1\Lambda \ll 1$, $x_1 \ll 1$) we have

$$|\zeta^*|\Lambda \approx \exp(-1/x_1),\; x_1 \ll 1 \text{ (Zone 1)}, \quad \text{(B14)}$$

so that $|\zeta^*|\Lambda \ll 1$ as $x_1 \ll 1$. In Zone 2 ($x_2\Lambda \sim 1$, $x_2 \ll 1$)

$$|\zeta^*|\Lambda \approx \frac{1}{x_2} + 2\exp(-1/x_2),\; x_2 \ll 1 \text{ (Zone 2)} \quad \text{(B15)}$$

so that $|\zeta^*|\Lambda \gg 1$ as $x_2 \ll 1$. In Zones 3 ($x_1\Lambda \gg 1$, $x_1 \gg 1$) and 4 ($x_2\Lambda \gg 1$, $x_2 \gg 1$) where $1/x_1$ and $1/x_2$ vanish, a Taylor expansion of Eqs. (B12) and (B13) near $z_c^*$ gives:

$$|\zeta^*|\Lambda \approx C - \frac{2}{\sqrt{5}x_1},\; x_1 \gg 1 \text{ (Zone 3)},$$

$$|\zeta^*|\Lambda \approx C + \frac{1}{\sqrt{5}x_2},\; x_2 \gg 1 \text{ (Zone 4)}, \quad \text{(B16)}$$

where $C = 2\ln\left((1+\sqrt{5})/2\right) = 0.962\,424$ (cf Eq. (4.41)).

Thus, in Zones 3 and 4, we have $|\zeta^*|\Lambda \approx C \approx 1$ as $1/x_1 \ll 1$ and $1/x_2 \ll 1$. At the frontier between Zones 3 and 4 i.e., along the vertical line reaching the Néel line at $T_c$, $x_1$ and $x_2$ become infinite so that:

$$|\zeta_c^*|\Lambda \approx C,\; T = T_c. \quad \text{(B17)}$$

# APPENDIX C: ASYMPTOTIC EXPANSIONS OF LARGE ORDER MODIFIED BESSEL FUNCTIONS OF THE FIRST KIND. APPLICATION TO THE LOW-TEMPERATURE SPIN-SPIN CORRELATIONS

The expression of spin-spin correlations involves ratios such as $\lambda_{l\pm1}(lz)/\lambda_l(lz)$ if considering $J = J_1 = J_2$ for sake of simplicity, with $z = -\beta J/l$. Near the critical point $T_c = 0$ K, these ratios have to be evaluated in the double limit $\beta|J| = l|z| \to +\infty$ and $l \to +\infty$. Olver has shown [24] that the Bessel function $I_l(lz)$ can be expanded in the infinite $l$- and $l|z|$-limits as the following series

$$I_l(l|z|) \approx \frac{(1+z^2)^{-1/4}}{\sqrt{2\pi l}}\left\{\exp(l|\zeta|)\sum_{s=0}^{+\infty}\frac{U_s(u)}{l^s} + \exp(-l|\zeta|)\sum_{s=0}^{+\infty}\frac{U_s(u)}{(-l)^s}\right\} \quad \text{(C1)}$$

where the coefficients $U_s(u)$ are detailed in [24,25]. $\zeta$ is given by Eq. (4.4) and $u = 1/\sqrt{1+z^2}$ with $|z| = \beta|J|/l$. Olver has also shown that the derivative of $I_l(lz)$ with respect to $lz$ i.e., $I'_l(l|z|) = dI_l(l|z|)/d(l|z|)$, with $l \gg 1$, can be similarly expressed as

$$I'_l(l|z|) \approx \frac{(1+z^2)^{1/4}}{\sqrt{2\pi l}|z|}\left\{\exp(l|\zeta|)\sum_{s=0}^{+\infty}\frac{V_s(u)}{l^s}\right.$$
$$\left. - \exp(-l|\zeta|)\sum_{k=0}^{+\infty}\frac{V_s(u)}{(-l)^s}\right\} \quad \text{(C2)}$$

where the coefficients $V_s(u)$ are polynomials in $u$ given by $V_s(u) = U_s(u) - u(1-u^2)\left(U_{s-1}(u)/2 + udU_{s-1}(u)/du\right)$, $s \geq 1$.

In a previous paper [19] we have extended Olver's work to large $l$-values (i.e., not necessarily infinite) and for any real argument $z$ varying from a finite value to infinity. If setting:

$$X_\pm = 1 + \mathcal{X}_\pm(u),\; \mathcal{X}_\pm(u) = \sum_{s=1}^{+\infty}\frac{X_s(u)}{(\pm l)^s},$$

$$X_\pm(u) = U_\pm(u) \text{ or } V_\pm(u),\; X_s(u) = U_s(u) \text{ or } V_s(u), \quad \text{(C3)}$$

and using the recurrence relations of modified Bessel functions i.e., in our case relations between functions $\lambda_l(lz)$, $\lambda_{l+1}(lz)$ and $\lambda_{l-1}(lz)$ (cf Eq. (2.15) in which $\lambda_l(lz) \sim (\pi/2lz)^{1/2}I_l(lz)$ as $l \to +\infty$) we can define the following ratio

$$\frac{\lambda_{l\pm1}(lz)}{\lambda_l(lz)} \approx -\frac{J}{|J|}\left\{\mp\left(\frac{1}{|z|} + \frac{1}{2l|z|}\right) + \frac{I'_l(l|z|)}{I_l(l|z|)}\right\},\; \text{as } T \to 0. \quad \text{(C4)}$$

Due to the fact that $|z|^{-1} \gg (2l|z|)^{-1}$ as $l \to +\infty$, Eq. (C4) can be expanded by means of Eqs. (C1)-(C3)

$$\frac{\lambda_{l\pm1}(lz)}{\lambda_l(lz)} \approx -\frac{J}{|J|}\left\{\mp\frac{1}{|z|}+\frac{1}{u|z|}\frac{1+V_+(u)}{1+U_+(u)}\times\right.$$
$$\left.\times\left[1-\exp(-2l|\zeta|)\left(\frac{1+V_-(u)}{1+V_+(u)}+\frac{1+U_-(u)}{1+U_+(u)}\right)\right]\right\}$$
$$+ O(\exp(-4l|\zeta|)), \text{ as } T \to 0. \quad (C5)$$

Using the expansion of series $U_\pm(u)$ and $V_\pm(u)$ we have:

$$\frac{I'_l(l|z|)}{I_l(l|z|)} \approx \frac{1}{u|z|}\left[1-\frac{u}{2l}-\frac{u^2}{8l^2}-2\exp(-2l|\zeta|)\left(1-\frac{u}{4l}-\frac{3u^2}{32l^2}\right)\right]$$
$$+ O(\exp(-4l|\zeta|)), \text{ as } T \to 0. \quad (C6)$$

In the main text, we have seen that, near $T_c = 0$ K, $l$ must be replaced by $\Lambda = 2l$ and more generally by any new scale $\Lambda'$. Under these conditions we must establish a relation between the ratios $\lambda_{l\pm1}(l|z|)/\lambda_l(l|z|)$, $\lambda_{\Lambda\pm1}(|z^*|\Lambda)/\lambda_\Lambda(|z^*|\Lambda)$, with $|z^*| = \beta|J|/\Lambda$ and $\lambda_{\Lambda'\pm1}(|z'|\Lambda')/\lambda_{\Lambda'}(|z'|\Lambda')$, with $|z'| = \beta|J|/\Lambda'$ and the imposed condition

$$l|z| = |z^*|\Lambda = |z'|\Lambda' = \beta|J|, \Lambda = 2l, \Lambda' = \alpha l, \text{ as } l \to +\infty. \quad (C7)$$

Owing to the multiplication theorem of the modified Bessel functions of the first kind $I_l(\alpha x)$ for finite $x > 0$, $\alpha > 0$ we have $I_l(\alpha x) = \alpha^l \sum_{k=0}^{+\infty}((\alpha^2-1)x/2)^k/k!)I_{l+k}(x)$ [25]. This series is absolutely convergent. If considering the ratio $I_{l+1}(\alpha x)/I_l(\alpha x)$ i.e., $\lambda_{l+1}(\alpha x)/\lambda_l(\alpha x)$, we have in the infinite $k$-limit

$$\frac{\lambda_{\Lambda/\alpha+1}(\Lambda|z|/\alpha)}{\lambda_{\Lambda/\alpha}(\Lambda|z|/\alpha)} \approx \frac{\lambda_{\Lambda+1}(\Lambda|z|)}{\lambda_\Lambda(\Lambda|z|)}, \text{ as } l = \Lambda/\alpha \to +\infty,$$
$$|z^*|\Lambda \to +\infty \quad (C8)$$

and a similar relation if $l+1$ is replaced by $l-1$ on condition that $\lambda_{l-1}(\alpha x)/\lambda_l(\alpha x) < 1$ (see Fig. 4).

As a result we can write

$$\frac{\lambda_{\Lambda\pm1}(z^*\Lambda)}{\lambda_\Lambda(z^*\Lambda)} \approx \frac{\lambda_{l\pm1}(lz)}{\lambda_l(lz)} \approx -\frac{J}{|J|}\left\{\mp\frac{1}{|z^*|}+\frac{I'_\Lambda(|z^*|\Lambda)}{I_\Lambda(|z^*|\Lambda)}\right\},$$
$$\frac{I'_\Lambda(|z^*|\Lambda)}{I_\Lambda(|z^*|\Lambda)} \approx \frac{1}{u^*|z^*|}-\frac{1}{|z^*|\Lambda}-2\exp(-|\zeta^*|\Lambda)\left(\frac{1}{u^*|z^*|}-\frac{1}{2|z^*|\Lambda}+..\right),$$
$$\text{as } T \to 0. \quad (C9)$$

$|\zeta^*|\Lambda$ has been calculated near $z_c^* = 1/4\pi$ in Appendix B: $|\zeta^*|\Lambda \sim 1$ (Zones 3 and 4, cf Eqs. (B16) and (B17)) but $|\zeta^*|\Lambda \ll 1$ (Zone 1, cf Eq. (B14)) and $|\zeta^*|\Lambda \gg 1$ in Zone 2 (cf Eq. (B15)). A Taylor expansion gives at first order

$$\exp(-|\zeta^*|\Lambda) \approx 1-|\zeta^*|\Lambda+..., \text{ Zones 1, 3 and 4,}$$
$$\exp(-|\zeta^*|\Lambda) \approx 0, \text{ Zone 2, as } T \to 0. \quad (C10)$$

If considering the general $\Lambda$-expression of the spin-spin correlation i.e., that given by Eq. (3.19) where, near $T_c = 0$ K, $l$ is replaced by $\Lambda \to +\infty$, we have $C_{l\pm1} \sim C_{\Lambda\pm1} \to 1/2$ and $K_{l\pm1} \sim K_{\Lambda\pm1} \to 1/2$ due to Eqs. (3.4) and (3.16). As a result we can write the spin-spin correlation as

$$<\mathbf{S}_{00}.\mathbf{S}_{k,k'}> \approx \frac{1}{2}\left[\left(\frac{\lambda_{\Lambda+1}(z^*\Lambda)}{\lambda_\Lambda(z^*\Lambda)}\right)^{k+k'} + \left(\frac{\lambda_{\Lambda-1}(z^*\Lambda)}{\lambda_\Lambda(z^*\Lambda)}\right)^{k+k'}\right],$$
$$\text{as } T \to 0. \quad (C11)$$

i.e., if setting

$$P_{\Lambda\pm1} \approx \frac{\lambda_{\Lambda\pm1}(|z^*|\Lambda)}{\lambda_\Lambda(|z^*|\Lambda)}, \quad (C12)$$

as $F_{l\pm1,l\pm1}/F_{l,l} \sim F_{\Lambda\pm1,\Lambda\pm1}/F_{\Lambda,\Lambda} \sim 1$, we have

$$<\mathbf{S}_{00}.\mathbf{S}_{k,k'}> \approx \frac{1}{2}\left(-\frac{J}{|J|}\right)^{k+k'}\left[P_{\Lambda+1}^{k+k'} + P_{\Lambda-1}^{k+k'}\right], \text{ as } T \to 0. \quad (C13)$$

Thus the spin-spin correlation can be expressed owing to Eq. (C9). In the first of this equation, and in Zone 1 exclusively, if using Eq. (B8), we have $|z^*| \approx z_c^* \exp(2/x_1\Lambda)$ so that $|z^*|^{-1} \approx (z_c^*)^{-1}\exp(-2/x_1\Lambda) \ll (z_c^*)^{-1}$. As a result, near $T_c = 0$ K, the factor $|z^*|^{-1}$ can be neglected with respect to the leader term $(u|z^*|)^{-1} \approx (z_c^*)^{-1}$ of $|I'_\Lambda(|z^*|\Lambda)/I_\Lambda(|z^*|\Lambda)|$. As a result $P_{\Lambda+1} \approx P_{\Lambda-1}$ so that

$$<\mathbf{S}_{00}.\mathbf{S}_{k,k'}> \approx \left(-\frac{J}{|J|}\right)^{k+k'}\left|\frac{I'_\Lambda(|z^*|\Lambda)}{I_\Lambda(|z^*|\Lambda)}\right|^{k+k'},$$
$$\text{as } T \to 0, \text{ Zone 1 } (x_1 \ll 1). \quad (C14)$$

In Zones 2 and 4 (cf Eq. (B10)) and in Zone 3 (cf Eq. (B8)), we have $|z^*|^{-1} \approx (z_c^*)^{-1}$ so that $|z^*|^{-1}$ cannot be neglected with respect to the leader term of $|I'_\Lambda(|z^*|\Lambda)/I_\Lambda(|z^*|\Lambda)|$. As a result, if expanding $P_{\Lambda\pm1}^{k+k'}$ in Eq. (C13), the factor $|z^*|^{-a} \sim |z_c^*|^{-a}$ ($a$ varying between 1 and $k+k'$) only disappears when $a$ is odd:

$$<\mathbf{S}_{00}.\mathbf{S}_{k,k'}> = 2\left(-\frac{J}{|J|}\right)^K \sum_{v=0}^{\lfloor K/2 \rfloor}\binom{K}{2v}\frac{1}{|z^*|^{K-2v}}\left|\frac{I'_\Lambda(|z^*|\Lambda)}{I_\Lambda(|z^*|\Lambda)}\right|^{2v}$$
$$K = k+k', \text{ as } T \to 0,$$
$$\text{Zone 2 } (x_2 \ll 1),$$
$$\text{Zone 3 } (x_1 \gg 1), \text{ Zone 4 } (x_2 \gg 1), \quad (C15)$$

where $\lfloor K/2 \rfloor$ is the floor function which gives the integer part of $K/2$.

In Zones 1, 3 and 4, if using the relation $\exp(-|\zeta^*|\Lambda) \approx 1 - |\zeta^*|\Lambda + \ldots$ (*cf* Eq. (C10)) we can write

$$\frac{I'_\Lambda(|z^*|\Lambda)}{I_\Lambda(|z^*|\Lambda)} = S_1 + S_2,$$

$$S_1 = \frac{1}{u^*|z^*|} \frac{1+V_+(u^*)}{1+U_+(u^*)} \left(1 - \frac{1+V_-(u^*)}{1+V_+(u^*)} - \frac{1+U_-(u^*)}{1+U_+(u^*)}\right),$$

$$S_2 = \frac{1}{u^*|z^*|} \frac{1+V_+(u^*)}{1+U_+(u^*)} \left(\frac{1+V_-(u^*)}{1+V_+(u^*)} + \frac{1+U_-(u^*)}{1+U_+(u^*)}\right) |\zeta^*|\Lambda,$$

$$\text{as } T \to 0, \quad (C16)$$

where $|\zeta^*|\Lambda$ has been expressed near $T_c = 0$ K in Appendix B. $U_\pm(u^*)$ and $V_\pm(u^*)$ have been calculated in a previous paper [19]. We found

$$\begin{pmatrix} U_+(z^*_c) \\ V_+(z^*_c) \end{pmatrix} \approx 1 \pm \left(\frac{z^*_c}{2}\right)^2, \quad \begin{pmatrix} U_-(z^*_c) \\ V_-(z^*_c) \end{pmatrix} \approx \frac{1}{e} \begin{pmatrix} U_+(z^*_c) \\ V_+(z^*_c) \end{pmatrix}. \quad (C17)$$

In the left formula the sign + of the right-hand side (respectively, the sign –) refers to $U_+$ (respectively, $V_+$).

Finally, if comparing Eqs. (C6) and (C16) we have to express the current term $u^*/\Lambda$ (with $\Lambda = 2l$) appearing in the series $U_\pm(u^*)$ and $V_\pm(u^*)$ and their combinations. For Zone 1 ($x_1\Lambda \ll 1$, $x_1 \ll 1$), $|z^*| \gg z^*_c = 1/4\pi$; we derive from Eq. (B7) that $u^*/\Lambda \sim 1/|z^*|\Lambda \sim x_1/2 z^*_c$. For Zone 2 ($x_2\Lambda \sim 1$, $x_2 \ll 1$) $|z^*| \ll z^*_c = 1/4\pi$ and $u^*/\Lambda \sim x_2$ due to Eq. (B10). In Zones 3 ($x_1\Lambda \gg 1$, $x_1 \gg 1$) and 4 ($x_2\Lambda \gg 1$, $x_2 \gg 1$), the current term $u^*/\Lambda \sim 1/\Lambda$ vanishes as $\Lambda \to +\infty$. As a result, after taking into account the previous remarks we have

$$S_1 = -1 + O((x_1/2z^*_c)^2), \text{ as } T \to 0, \text{ Zone 1 } (x_1 \ll 1)$$

$$S_1 = \frac{1}{z^*_c}[1 - x_2] + O(x_2^2), \text{ as } T \to 0, \text{ Zone 2 } (x_2 \ll 1)$$

$$S_1 = -\frac{1}{z^*_c}, \text{ Zone 3 } (x_1 \gg 1), \text{ Zone 4 } (x_2 \gg 1). \quad (C18)$$

For $S_2$ the situation is more complicated in Zones 1 and 2 because we have already expressed $|\zeta^*|\Lambda$ vs $x_1$ (respectively, $x_2$) in Appendix B. We set

$$S_2 = 2f_1 f_2, \quad f_1 = \frac{|\zeta^*|\Lambda}{u^*|z^*|}, \text{ as } T \to 0, \text{ Zone 1 } (x_1 \ll 1),$$

$$f_2 = \frac{1+V_+}{1+U_+}\left(\frac{1+V_-}{1+V_+} + \frac{1+U_-}{1+U_+}\right). \quad (C19)$$

If considering $\ln(f_1 f_2)$ we have $\ln f_1 = \ln(|\zeta^*|\Lambda) - \ln(u^*|z^*|)$ where $\ln(u^*|z^*|) \approx \ln(z^*_c)$ so that $\ln f_1 \approx \ln(|\zeta^*|\Lambda/z^*_c)$ near $z^*_c$. It remains to evaluate $\ln f_2$. Calculations are tedious but not difficult. We skip the intermediate steps and give the final result for Zone 1:

$$S_2 = \frac{2|\zeta^*|\Lambda}{e}\left(\frac{1}{z^*_c} - \frac{x_1}{2z^*_c}\right) + O((x_1/2z^*_c)^2), \text{ as } T \to 0,$$

$$\text{Zone 1 } (x_1 \ll 1),$$

$$S_2 = 2\exp(-|\zeta^*|\Lambda)\left(\frac{1}{z^*_c} - \frac{x_2}{2z^*_c}\right) + O((x_2/z^*_c)^2), \text{ as } T \to 0,$$

$$\text{Zone 2 } (x_2 \ll 1) \quad (C20)$$

where $|\zeta^*|\Lambda$ is given by Eqs. (B14) (Zone 1) and (B15) (Zone 2). Indeed, for Zone 2, as $\exp(-|\zeta^*|\Lambda) \approx 0$, that has not been necessary to consider the full series $U_\pm$ and $V_\pm$ near $z^*_c$ so that only their first terms are sufficient. Thus, for Zones 1 and 2, we can write:

$$\left|\frac{I'_\Lambda(|z^*|\Lambda)}{I_\Lambda(|z^*|\Lambda)}\right| = 1 - \frac{2}{ez^*_c}\exp(-1/x_1)\left(1 - \frac{x_1}{2}\right) + \ldots, \quad (C21a)$$

$$\left|\frac{I'_\Lambda(|z^*|\Lambda)}{I_\Lambda(|z^*|\Lambda)}\right| = \frac{1}{z^*_c}\left[1 - x_2 - 2\exp(-1/x_2)\left(1 - \frac{x_2}{2}\right)\right] + \ldots \quad (C21b)$$

with $x_2 \gg 2\exp(-1/x_2)(1 - x_2/2)$. If factorizing $x_2$ and recalling that $|\zeta^*|\Lambda \sim 1/x_2$ we finally have in Zone 2

$$\left|\frac{I'_\Lambda(|z^*|\Lambda)}{I_\Lambda(|z^*|\Lambda)}\right| = \frac{x_2}{z^*_c}[|\zeta^*|\Lambda - 1] + \ldots \text{ Zone 2 } (x_2 \ll 1). \quad (C22)$$

For Zones 3 and 4, as $u^*/\Lambda \sim 0$, we simply have

$$\left|\frac{I'_\Lambda(|z^*|\Lambda)}{I_\Lambda(|z^*|\Lambda)}\right| = \frac{1}{z^*_c}[2|\zeta^*|\Lambda - 1] + \ldots, \text{ as } T \to 0,$$

$$\text{Zone 3 } (x_1\Lambda \gg 1, x_1 \gg 1),$$
$$\text{Zone 4 } (x_2\Lambda \gg 1, x_2 \gg 1), \text{ as } T \to 0, \quad (C23)$$

Under these conditions, we can write

$$P_{\Lambda\pm1} \approx \frac{\lambda_{\Lambda\pm1}(z^*\Lambda)}{\lambda_\Lambda(z^*\Lambda)} \approx -\frac{J}{|J|}\left[1 - \frac{8\pi}{e}|\zeta^*|\Lambda\left(1 - \frac{x_1}{2}\right) + \ldots\right],$$

$$\text{as } T \to 0, \text{ Zone 1 } (x_1\Lambda \ll 1, x_1 \ll 1),$$

$$P_{\Lambda\pm1} \approx -\frac{J}{|J|}\frac{1}{z^*_c}\left[\mp 1 + x_2(|\zeta^*|\Lambda - 1) + \ldots\right], \text{ as } T \to 0,$$

$$\text{Zone 2 } (x_2\Lambda \sim 1, x_2 \ll 1),$$

$$P_{\Lambda\pm1} \approx -\frac{J}{|J|}\frac{1}{z^*_c}\left[\mp 1 + 2|\zeta^*|\Lambda - 1 + \ldots\right], \text{ as } T \to 0,$$

$$\text{Zone 3 } (x_1\Lambda \gg 1, x_1 \gg 1),$$
$$\text{Zone 4 } (x_2\Lambda \gg 1, x_2 \gg 1), \text{ as } T \to 0, \quad (C24)$$

where $|\zeta^*|\Lambda$ is respectively given by Eqs. (B14)-(B16). In Zones 1 and 3 we always have $|\zeta^*|\Lambda < 1$ near $T_c = 0$ K. In Zone 4 we must have $1/x_2 < \sqrt{5}(1-C) = 0.084$.

## APPENDIX D: RENORMALIZED SPIN-SPIN CORRELATIONS. ORNSTEIN-ZERNIKE LAW AND CORRELATION LENGTH NEAR $T_c = 0$ K

The expression of spin-spin correlations just given in Appendix C does not tend towards unity in Zones 2 to 4 notably because the factor $1/u^*|z^*|$ appearing in the low-temperature expansion of $I_{\Lambda\pm1}(|z^*|\Lambda)/I_\Lambda(|z^*|\Lambda)$ does not tend to unity near $z_c^* = 1/4\pi$. As a result, we have to make a renormalization of the temperature scale. We define a new coefficient $\widetilde{\Lambda}$ such as $\beta|J| = |z^*|\Lambda = |\widetilde{z}|\widetilde{\Lambda}$ with $\widetilde{\Lambda} = \alpha\Lambda$. Owing to the multiplication theorem cited before Eq. (C8) the dilation factor $\alpha$ is such as

$$\frac{\alpha}{u^*|z^*|} = 1, \quad \alpha = \frac{z_c^*}{\sqrt{1+z_c^{*2}}} \approx z_c^*, \quad \widetilde{\Lambda} = \alpha\Lambda. \quad (D1)$$

Under these conditions we can define the renormalized spin-spin correlation near $T_c = 0$ K:

$$<\widetilde{S_{00}\cdot S_{k,k'}}> \approx \frac{1}{2}\left[\widetilde{P}_{\Lambda+1}^{k+k'} + \widetilde{P}_{\Lambda-1}^{k+k'}\right], \quad \widetilde{P}_{\Lambda\pm1} \approx \frac{\lambda_{\Lambda\pm1}(\widetilde{z}\widetilde{\Lambda})}{\lambda_\Lambda(\widetilde{z}\widetilde{\Lambda})},$$
$$\text{as } T \to 0. \quad (D2)$$

Near $z_c^*$ we finally have

$$\widetilde{P}_{\Lambda+1} \approx \widetilde{P}_{\Lambda+1} \approx -\frac{J}{|J|}\left[1 - \frac{8\pi}{e}|\zeta^*|\Lambda\left(1 - \frac{x_1}{2}\right) + ...\right], \quad \alpha = 1,$$
$$\text{as } T \to 0, \text{ Zone 1 } (x_1\Lambda \ll 1, x_1 \ll 1),$$

$$\widetilde{P}_{\Lambda\pm1} \approx -\frac{J}{|J|}\left|\mp|\zeta^*|\Lambda + (|\zeta^*|\Lambda - 1) + ...\right|, \quad \alpha = \frac{z_c^*}{x_2},$$
$$\text{as } T \to 0, \text{ Zone 2 } (x_2\Lambda \sim 1, x_2 \ll 1).$$

In Zones 3 ($x_1\Lambda \gg 1$, $x_1 \gg 1$) and 4 ($x_2\Lambda \gg 1$, $x_2 \gg 1$)

$$\widetilde{P}_{\Lambda\pm1} \approx -\frac{J}{|J|}\left|\mp 1 + (2|\zeta^*|\Lambda - 1) + ...\right|, \quad \alpha = z_c^* = \frac{1}{4\pi},$$
$$\text{as } T \to 0, \text{ for Zone 3 } (x_1\Lambda \gg 1, x_1 \gg 1),$$
$$\text{Zone 4 } (x_2\Lambda \gg 1, x_2 \gg 1), \quad (D3)$$

where $|\zeta|\Lambda$ is respectively given by Eqs. (B14) for Zone 1, (B15) for Zone 2 and (B16)-(B17) for Zones 3 and 4.

As seen in Appendix C the spin-spin correlation between first-nearest neighbors $<S_{00}\cdot S_{0,1}>$ or $<S_{00}\cdot S_{1,0}>$ reduces to $(P_{\Lambda+1} + P_{\Lambda-1})/2$. Thus, its renormalized expression is

$$<\widetilde{S_{00}\cdot S_{01}}> \approx -\frac{J}{|J|}\left[1 - \frac{8\pi}{e}|\zeta^*|\Lambda\left(1-\frac{x_1}{2}\right) + ...\right], \text{ as } T \to 0,$$
$$\text{Zone 1 } (x_1\Lambda \ll 1, x_1 \ll 1),$$

$$<\widetilde{S_{00}\cdot S_{01}}> \approx -\frac{J}{|J|}\left[1 - |\zeta^*|\Lambda + ...\right], \text{ as } T \to 0,$$
$$\text{Zone 2 } (x_2\Lambda \sim 1, x_2 \ll 1),$$

$$<\widetilde{S_{00}\cdot S_{01}}> \approx -\frac{J}{|J|}\left[1 - 2|\zeta^*|\Lambda + ...\right], \text{ as } T \to 0,$$
$$\text{for Zone 3 } (x_1\Lambda \gg 1, x_1 \gg 1),$$
$$\text{Zone 4 } (x_2\Lambda \gg 1, x_2 \gg 1). \quad (D4)$$

In Zone 1 and 3 we always have $|\zeta^*|\Lambda < 1$ near $T_c = 0$ K. In Zone 4 we recall that we must have $1/x_2 < \sqrt{5}(1-C) = 0.084$.

As a result, near $T_c = 0$ K, the renormalized spin-spin correlation can be expressed under the general form:

$$<\widetilde{S_{00}\cdot S_{01}}> \approx -\frac{J}{|J|}\left[1 - f(x_i) + ...\right], \text{ as } T \to 0, \quad (D5)$$

with

$$f(x_1) = \frac{8\pi}{e}|\zeta^*|\Lambda\left(1 - \frac{x_1}{2}\right), \text{ Zone 1 } (x_1\Lambda \ll 1, x_1 \ll 1),$$
$$f(x_2) = |\zeta^*|\Lambda \text{ Zone 2 } (x_2\Lambda \sim 1, x_2 \ll 1),$$
$$f(x_i) = 2|\zeta^*|\Lambda, i = 1, \text{ Zone 3 } (x_1\Lambda \gg 1, x_1 \gg 1),$$
$$i = 2, \text{ Zone 4 } (x_2\Lambda \gg 1, x_2 \gg 1). \quad (D6)$$

For $k$ and $k'$ finite but not too large i.e., if $(k + k')f(x_i) \ll 1$ the spin-spin correlation given by Eqs. (D2), (D3) and (D5) can be reduced to the first order so that the renormalized spin-spin correlation can be written under the generic form

$$\left|<\widetilde{S_{00}\cdot S_{k,k'}}>\right| \approx 1 - (k + k')f(x_i) \text{ } (i = 1 \text{ or } 2),$$
$$\text{as } T \to 0, \quad (D7)$$

except for Zone 2 for which $|\zeta^*|\Lambda \gg 1$ (no long-range order).

When $k$ and $k'$ become infinite, the spin-spin correlation still contains the ratio $I'_\Lambda(|z|\Lambda)/I_\Lambda(|z|\Lambda) \to [1 - f(x_i)]/z_c^*$, with $i = 1$ or 2. According to Eqs. (C15) and (C24) we can artificially write

$$\left|<\widetilde{S_{00}\cdot S_{k,k'}}>\right| \approx 2\sum_{v=0}^{\lfloor K/2 \rfloor}\binom{K}{2v}\frac{1}{z_c^{*K-2v}}\left[\frac{1}{z_c^*}\left(1 - \frac{f(x_i)}{2}\right)\right]^{2v},$$
$$i = 1, 2, K = k + k' \to +\infty, \text{ as } T \to 0,$$
$$\text{Zone 3 } (x_1\Lambda \gg 1, x_1 \gg 1),$$
$$\text{Zone 4 } (x_2\Lambda \gg 1, x_2 \gg 1), \quad (D8)$$

with $f(x_i) < 1$. When $K \to +\infty$ $\lfloor K/2 \rfloor \approx K/2$. Due to the fact that $z_c^* \gg 1$ the higher degree term $V_\nu = z_c^{*-2\nu}(1 - f(x_i)/2)^{2\nu}$ is obtained when $\nu = K/2$. The $\nu$-summation which becomes

$$\binom{K}{0}V_{-K/2} + \binom{K}{2}V_{-(K-2)/2} + \binom{K}{4}V_{-(K-4)/2} + \ldots + \binom{K}{K-2}V_{-1}$$

$$+ \binom{K}{K}$$

reduces to unity because $V_{-\nu} \ll 1$ and the first term $V_{K/2} = (1 - f(x_i)/2)^K$ is selected. As a result, if introducing the dilation factor $z_c^{*K}/2$ in Eq. (D8) the renormalized spin-spin correlation can be written as

$$\left| < \widetilde{S_{00}.S_{k,k'}} > \right| \approx \left(1 - \frac{f(x_i)}{2}\right)^{K/2}, K = k + k' \to +\infty,$$

Zones 1, 3 and 4, as $T \to 0$. (D9)

For Zone 1 this result can be directly obtained without renormalization because we have seen in Appendix C that the term $|z|^{-1}$ has been neglected.

Now let us consider the special case of Zone 2 ($x_2 \Lambda \sim 1$, $x_2 \ll 1$) for which $f(x_2) = |\zeta^*| \Lambda \gg 1$. We use the trick $|\zeta^*|\Lambda - 1 \approx |\zeta^*|\Lambda - 2 \approx 2(|\zeta^*|\Lambda/2 - 1)$. Then the spin-spin correlation can be also given by slightly modifying Eq. (D8)

$$\left| < \widetilde{S_{00}.S_{k,k'}} > \right| \approx \frac{2}{z_c^{*K}} \sum_{\nu=0}^{K/2} 2^{2\nu} \binom{K}{2\nu} \left(1 - \frac{f(x_2)}{2}\right)^{2\nu},$$

$K = k + k' \to +\infty$, as $T \to 0$,
Zone 2 ($x_2 \Lambda \sim 1$, $x_2 \ll 1$). (D10)

In that case it is easier to show that the previous sum reduces to its highest-degree term and, after a new convenient renormalization, the spin-spin correlation is again given by Eq. (D9).

Consequently, if expanding Eq. (D9), we have the current $\nu$-term $\binom{K/2}{\nu}(-f(x_i)/2)^\nu$. In the infinite-$K$ limit $\binom{K/2}{\nu}$ reduces to $(K/2)^\nu/\nu!$ so that

$$\left| < \widetilde{S_{00}.S_{k,k'}} > \right| \approx \sum_{\nu=0}^{K/2} \frac{(K/2)^\nu}{\nu!} \left(-\frac{f(x_i)}{2}\right)^\nu \quad (D10)$$

i.e.,

$$\left| < \widetilde{S_{00}.S_{k,k'}} > \right| \approx \exp\left(-\frac{k+k'}{2}\frac{f(x_i)}{2}\right), i = 1, 2,$$

$K = k + k' \to +\infty$, as $T \to 0$. (D11)

Thus we retrieve the Ornstein-Zernike law

$$\left| < \widetilde{S_{00}.S_{k,k'}} > \right| \approx \exp\left(-\frac{k+k'}{\sqrt{2}}\frac{1}{\xi\sqrt{2}}\right), K = k + k' \to +\infty,$$

as $T \to 0$. (D12)

where $\xi$ is the scaling correlation length expressed near $T_c = 0$ K along one of the three axes of the $D$-space time with

$$\xi = \frac{\xi^*}{2a\sqrt{2}} = \frac{\xi_\tau^*}{L_\tau} = \frac{2}{f(x_i)} = (|\zeta^*|\Lambda)^{-1}, \text{ as } T \to 0 \quad (D13)$$

where $L_\tau = \hbar c/k_B T$.

## APPENDIX E: EXPRESSION OF THE FREE ENERGY DENSITY NEAR $T_c = 0$ K

Rosenstein *et al.* have shown that, in the vanishing-field limit, the partition function $Z_N(0)$ can be written [34]

$$Z_N(0) = \int D\boldsymbol{S}\, \delta(\boldsymbol{S}^2 - 1) \exp\left(-\frac{1}{2\hbar g} \int d^3x (\partial_\mu \boldsymbol{S})^2\right) \quad (E1)$$

where the exchange Hamiltonian given by Eq. (2.2) has been expressed in the continuum limit, for a lattice characterized by $J = J_1 = J_2$. $g$ is the coupling constant given by Eq. (4.6). The constraint $\boldsymbol{S}^2 = 1$ is conveniently implemented by introducing a Lagrange multiplier field $\gamma$. We have

$$Z_N(0) = \int D\boldsymbol{S} D\gamma \exp\left(-\frac{1}{2\hbar} \int d^3x \left[(\partial_\mu \boldsymbol{S})^2 + \gamma^2\left(\boldsymbol{S}^2 - \frac{1}{g}\right)\right]\right)$$

(E2)

where $\boldsymbol{S}^2$ has been rescaled so that the kinetic energy term is conventionally renormalized. The integrand becomes gaussian in $\boldsymbol{S}$ and can formally be integrated. This provides an effective action for the $\gamma$-field such as:

$$Z_N(0) = \int D\gamma \exp\left(-\frac{1}{\hbar} S_{\text{eff}}(\gamma)\right) \quad (E3)$$

where

$$S_{\text{eff}}(\gamma) = -\frac{\gamma(m)^2}{2g} + \frac{\hbar}{2} \text{tr} \ln(\hat{p}^2 + \gamma(m)^2 \mathbb{1}). \quad (E4)$$

$\hat{p}$ is the momentum operator, tr the trace operator and $\mathbb{1}$ the identity matrix. $\gamma(m)$ and $m$ will be defined later.

In the main text $Z_N(0)$ is expressed with a $l$-summation over eigenvalues $\lambda_l(-\beta J)$ given by Eq. (2.37). For a given temperature $T$, each eigenvalue $\lambda_{l_i}(-\beta J)$ is dominant within the range $\delta T_{l_i} = T_{l_i,>} - T_{l_i,<}$ with $T = \sum_{i=0}^{i_{max}} (T_{l_i,>} - T_{l_i,<})$, $T_{l_0,<} = 0$, $T_{l_{i-1},>} = T_{l_i,<}$ ($i \neq 0$) and $T_{l_{i_{max}},>} = T$ or equivalently $\sum_{i=0}^{+\infty} \delta T_{l_i}/T = 1$. Thus $\delta T_{l_i}/T$ appears as the corresponding weight of the eigenvalue $\lambda_{l_i}(-\beta J)$ at temperature $T$.

In this range the $l$-summation giving $Z_N(0)$ reduces to a single term $F(l,l,0)\lambda_l(-\beta J)^2$ so that we have $\ln Z_N(0) \approx \ln(F(l,l,0)\lambda_l(-\beta J)^2)$. Near the critical point $T_c = 0$ K all the eigenvalues have a close thermal behavior and the dominant ones are characterized by an index $l \gg 1$. In addition the argument $\beta|J|$ is then replaced by $|z^*|\Lambda$ (cf Eq. (4.12)) with $\Lambda = 2l \gg 1$. Thus, in the low-temperature limit, we always have $F(l_i,l_i,0) \ll |\lambda_{l_i}(-\beta J)|$: Its contribution to $\mathcal{F}(T)$ can be neglected in the thermodynamic limit ($N \to +\infty$).

As a result the elementary contribution per bond to $Z_N(0)$ is $\delta Z_N(0)_{l_i} = \lambda_{l_i}(\beta|J|)\delta T_{l_i}/T$, as $Z_N(0)$ is an even function of $\beta J$. Similarly we can define the elementary contribution to $\ln Z_N(0)$ by $\delta \ln Z_N(0)_{l_i} = \ln(\lambda_{l_i}(\beta|J|))\delta T_{l_i}/T$.

While approaching $T_c = 0$ K the ranges $T_{l_i,>} - T_{l_i,<}$ are tighter and tighter and the corresponding weight is very small. We tend towards a continuum: There is an increasing number of eigenvalues characterized by a higher and higher index $l$ (replaced by $\Lambda$ near $T_c = 0$ K). In other words we deal with an infinity of eigenvalues showing an infinite argument so that their total contribution dominates the ones characterized by a lower value of $l$ in spite of the fact that they have a larger weight (see Fig. 2). As the dominant eigenvalue changes between two consecutive ranges $\delta T_\Lambda$ and $\delta T_{\Lambda+1}$, with $\delta T_\Lambda \approx \delta T_{\Lambda+1}$, if $\Lambda \to +\infty$, we can formally write for a given temperature $T$ close to $T_c = 0$ K

$$\delta \ln(Z_N(0)) \approx \lim_{M \to +\infty} \sum_{\Lambda=0}^{M} \ln(\lambda_\Lambda(|z^*|\Lambda))\frac{\delta T_\Lambda}{T}, \quad \lim_{M \to +\infty} \sum_{\Lambda=0}^{M} \frac{\delta T_\Lambda}{T} = 1 \tag{E5}$$

where from Eq. (C1)

$$\lambda_\Lambda(|z^*|\Lambda) \approx \frac{1}{|z^*|\Lambda}\left(\frac{|z^*|}{\sqrt{1+z^{*2}}}\right)^{1/2}\left(\exp(|\zeta^*|\Lambda)\sum_{s=0}^{+\infty}\frac{U_s}{\Lambda^s}\right.$$
$$\left. + \exp(-|\zeta^*|\Lambda)\sum_{s=0}^{+\infty}\frac{U_s}{(-\Lambda)^s}\right), \text{ as } \Lambda \to +\infty. \tag{E6}$$

The coefficients $U_s$ have been calculated by Olver [24] and $|\zeta^*|\Lambda$ is given by Eqs. (4.23)-(4.29). As $\Lambda \to +\infty$ the previous series can be restricted to the first term $U_0 = 1$. Consequently, near $T_c$, we have $\lambda_\Lambda(|z^*|\Lambda) \approx 2K_1 \cosh(|\zeta^*|\Lambda)$ where $K_1 \approx 1/z_c^{*1/2}\Lambda$.

From Eqs. (E3)-(E4) involving the effective action $S_{\text{eff}}$ the effective $\Lambda$-contribution to $Z_N(0)_{\text{eff},\Lambda}$ is $\exp(-S_{\text{eff},\Lambda}(\gamma)/\hbar)$ and that one contributing to $\ln(Z_N(0)_{\text{eff},\Lambda})$ can be written $\ln(Z_N(0)_{\text{eff},\Lambda}) = -S_{\text{eff},\Lambda}(\gamma)/\hbar$ inside the range $\delta T_\Lambda$. As a result we have:

$$-\frac{1}{\hbar}\delta S_{\text{eff}}(\gamma) = -\frac{1}{\hbar}\lim_{M \to +\infty}\sum_{\Lambda=0}^{M} S_{\text{eff},\Lambda}(\gamma)\frac{\delta T_\Lambda}{T}. \tag{E7}$$

Combining Eqs. (E4)-(E6) we obtain the total effective contribution

$$\lim_{M \to +\infty}\sum_{\Lambda=0}^{M}\ln(\lambda_\Lambda(|\widetilde{z}|\Lambda))\frac{\delta T_\Lambda}{T}\bigg|_{\text{eff}} = \frac{\gamma(m)^2}{2\hbar g}$$
$$-\frac{1}{2}\lim_{M \to +\infty}\sum_{\Lambda=-M}^{M}\ln(\hat{p}^2 + \gamma(m)^2\mathbb{1})_\Lambda \frac{\delta T_\Lambda}{T} \tag{E8}$$

if using the second part of Eq. (E5). $\mathbb{1}$ is the identity matrix (with $\dim \mathbb{1} = 2M + 1$, $M \to +\infty$). In Eq. (E8) we have taken into account the fact that $\lambda_\Lambda(|\widetilde{z}|\Lambda)$ is the dominant eigenvalue over the range $\delta T_\Lambda$: It means that the trace of the eigenmatrix associated with the operator $\ln(\hat{p}^2 + \gamma(m)^2\mathbb{1})$ reduces to the single dominant $\Lambda$-element $\ln(\hat{p}^2 + \gamma(m)^2\mathbb{1})_\Lambda$ i.e., $\ln(p_\Lambda^{*2} + \gamma(m)^2)$, as $\Lambda \gg 1$.

In the thermodynamic limit, the elementary free energy density per lattice bond can be expressed as

$$\frac{\delta\mathcal{F}(T)}{8N^2} = -\hbar c C^2\left(\frac{k_B T}{\hbar c}\right)^3 \lim_{M \to +\infty}\sum_{\Lambda=0}^{M}\ln(\lambda_\Lambda(|z^*|\Lambda))\frac{\delta T_\Lambda}{T},$$
$$\text{as } N \to +\infty \tag{E9}$$

where $C$ is given by Eq. (4.41). As we deal with a surface free energy density $\delta\mathcal{F}$, the dimensionless weight $\delta T_\Lambda/T$ must be linked to a surface element through a factor that we shall identify below. Owing to Eqs. (E7) and (E8) $\delta\mathcal{F}(T)$ must appear as

$$\frac{\delta\mathcal{F}(T)}{8N^2} = -\hbar c C^2\left(\frac{k_B T}{\hbar c}\right)^3\left[\frac{\gamma(m)^2}{2\hbar g}\right.$$
$$\left. -\frac{1}{2}\lim_{M \to +\infty}\sum_{\Lambda=-M}^{M}\ln(p_\Lambda^2 + \gamma(m)^2)\frac{\delta T_\Lambda}{T}\right], \text{ as } N \to +\infty. \tag{E10}$$

As a result, in a first step, we have to show that the $\Lambda$-sum over $\ln(\lambda_\Lambda(|z^*|\Lambda))$ in Eq. (E9) can be transformed into a $\Lambda$-sum over $\ln(p_\Lambda^{*2} + \gamma(m)^2)$ in Eq. (E10) when $\delta\mathcal{F}(T)$ is expressed near its extremum: This is the *effective contribution*. While approaching $T_c = 0$ K, as $\Lambda \to +\infty$, all the temperature ranges $\delta T_\Lambda$ tend towards the same common limit $\delta T$ so that the weight $\delta T_\Lambda/T$ can be written as $\delta T/T$.

Now we wish to show that the $\Lambda$-sum in Eqs. (E9) and (E10) can be written as an identity recalled by Sachdev [33] (i.e., a Poisson summation):

$$\lim_{M\to+\infty}\left[\frac{1}{L}\sum_{n=1}^{M}\ln\left(\frac{4\pi^2 n^2}{L^2}+a^2\right)\right]=\frac{1}{2}\lim_{M\to+\infty}\int_{-2\pi M/L}^{+2\pi M/L}\frac{d\omega}{2\pi}\ln(\omega^2+a^2)$$
$$+\frac{1}{L}\ln(1-\exp(-L|a|))-\frac{\ln a^2}{2L},\quad \omega=\frac{2\pi}{L}.\quad (E11)$$

$a$ will be specified below.

The bound of the integral appearing in Eq. (E11) i.e., $2\pi M/L$, represents the $M$-th current value of the momentum $p/\hbar$ (in fact $pc/\hbar$) along the slab width. Along the $i\tau$-direction we have a periodic structure due to the propagation of spin waves. $pc/\hbar$ is quantized in integer multiples of $1/L$, $M$ being the corresponding quantum number. For each mode $M$, $p/M = h/\lambda_{DB}$ where $\lambda_{DB} = 2\pi L_\tau$ is the thermal de Broglie wavelength characterizing the spin waves and $L_\tau = \hbar c/k_B T$. As a result Eq. (E11) appears as a summation over Matsubara frequencies $(p/\hbar)_M = 2\pi\omega_M/c$. As $M$ is integer, we then deal with *bosonic excitations*. Finally we have $L = L_\tau$ where $L_\tau$ is the slab thickness.

If defining the renormalized quantities $X^*$ in the $\Lambda$-scale
$$\lambda_{DB}^* = 2\pi L_\tau^*,\ L_\tau^*\Lambda = L_\tau,\ p_\Lambda^*\Lambda = \frac{\hbar}{L_\tau^*\Lambda},\ \frac{p_\Lambda^*}{\hbar} = 2\pi\frac{\omega_\Lambda^*}{c},$$
$$p_\Lambda^*\Lambda = p,\ \omega_\Lambda^*\Lambda = \omega,\ a_\Lambda^*\Lambda = a \quad (E12)$$

and if expanding $\lambda_\Lambda(|z^*|\Lambda)/K_1 L_\tau^{*2} \approx 2\cosh(|\zeta^*|\Lambda)/L_\tau^{*2}$ we can write for the mode $M = \pm\Lambda$

$$\ln(\lambda_{\pm\Lambda}(|z^*|\Lambda)) = \ln\left(\left(\frac{p_{\pm\Lambda}^*}{\hbar}\right)^2 + a_{\pm\Lambda}^{*2}\right) + \ln(K_1 L_\tau^{*2}) \quad (E13)$$

where
$$\frac{1}{L_\tau^{*2}} = \left(\frac{p_{\pm\Lambda}^*}{\hbar}\right)^2,\ \frac{p_{\pm\Lambda}^*}{\hbar} = \pm\frac{\omega}{c}\Lambda,\ a_{\pm\Lambda}^{*2} = \left(\frac{p_{\pm\Lambda}^*}{\hbar}\right)^2 + \left(\frac{|\zeta^*|\Lambda}{L_\tau^*}\right)^2.$$
$$(E14)$$

In Eq. (E11), the upper bound of the integral is $2\pi M/L$. If using the bound $M/L$ $p_{\pm\Lambda}^*$ and $a_{\pm\Lambda}^*$ must be divided by the factor $2\pi$ and $L_\tau$ must be multiplied by $2\pi$. This will be restored in Appendix F. If $E = ((pc)^2 + (m_0 c^2)^2)^{1/2}$ is the relativistic energy and $m_0 c^2$ the rest energy we can identify $a_{\pm\Lambda}$ owing to Eqs (E12) and (E14)

$$|a_{\pm\Lambda}^*| = \frac{E_{\pm\Lambda}}{\hbar c},\ \frac{m_0 c^2}{\hbar c} = \frac{|\zeta^*|\Lambda}{L_\tau^*}. \quad (E15)$$

As a result, in Eq. (E10), the $\Lambda$-summation between $\Lambda \to -\infty$ and $\Lambda \to +\infty$ can be written as $\delta T_\Lambda/T \sim \delta T/T$

$$\lim_{M\to+\infty}\frac{1}{L_\tau}\left[\sum_{\Lambda=0}^{M}\ln(\lambda_\Lambda(|z^*|\Lambda))\right]\frac{\delta T}{T}\approx$$

$$\lim_{M\to+\infty}\frac{1}{L_\tau}\left[\sum_{\Lambda=1}^{M}\ln\left(\frac{(p_\Lambda/\hbar)^2 + a_\Lambda^2}{\Lambda^2}\right)\right]\frac{\delta T}{T} \quad (E16)$$

The term $\ln(K_1 L_\tau^{*2})$ can be omitted when calculating $\mathcal{F}(T) - \mathcal{F}(0)$. As $\lambda_0(|z^*|\Lambda) = \sinh(|z^*|\Lambda)/|z^*|\Lambda$, with $|z^*|\Lambda = \beta|J|$, we have $\ln(\lambda_0(|z^*|\Lambda)) \approx \beta|J| - \ln(2\beta|J|)$ i.e., $1/g - \ln(2/g)$. Thus, near $T_c = 0$ K, $\ln(\lambda_0(|z^*|\Lambda))/L_\tau \approx |J|/\hbar c - \ln(2/g)/L_\tau$. The first term $|J|/\hbar c$ is constant and jointly appears in $\mathcal{F}(T)$ and $\mathcal{F}(0)$; it can be omitted when calculating $\mathcal{F}(T) - \mathcal{F}(0)$. In addition, near $T_c = 0$ K, we have $\ln(2/g)/L_\tau = \ln(1/g)/L_\tau + \ln 2.k_B T/\hbar c \approx \ln(1/g)/L_\tau$ because $\ln(2)/L_\tau$ vanishes. As a result $\ln(\lambda_0(|z^*|\Lambda))/L_\tau \approx -\ln(1/g)/L_\tau$.

Finally it becomes possible to write the right-hand side of Eq. (E16) by means of a Poisson summation so that the effective contribution to $\mathcal{F}(T)$ is

$$\lim_{M\to+\infty}\left[\frac{1}{L_\tau}\sum_{\Lambda=0}^{M}\ln(\lambda_\Lambda(|z^*|\Lambda))\frac{\delta T}{T}\right]=$$
$$\left[\frac{1}{2}\lim_{M\to+\infty}\int_{-M/L_\tau}^{+M/L_\tau}\frac{d(p/\hbar)}{2\pi}\ln\left(\frac{(p/\hbar)^2 + a^2}{\Lambda^{-2}}\right)\right.$$
$$\left.+\frac{1}{L_\tau}\left\{\ln(1-\exp(-L_\tau|a|))-\frac{\ln a^2}{2}-\frac{\ln(1/g)}{2}\right\}\right]\frac{\delta T}{T}. \quad (E17)$$

The value of the $\Lambda$-summation essentially comes from the larger and larger values of $\Lambda$. In other words it means that the integral must be calculated owing to a steepest descent method. This imposes the search of the extremum of the integral argument $(p/\hbar)^2 + a^2$ i.e., $(p/\hbar)^2 + (E/\hbar c)^2$ (due to Eq. (E15)) with respect to $p/\hbar$, where $E$ is the relativistic energy. We must have $\partial E/\partial(p/\hbar) = 0$ so that the argument is minimized when $p/\hbar = 0$ i.e., $L_\tau \to +\infty$ (or $T \to 0$ K) which is a reasonable result. We then derive that $E_{\min}/\hbar c = |a|_{\min}$ with, by comparison between Eqs. (E10) and (E17), $|a|_{\min} = \|\gamma\|$ where $\|\gamma\|$ is the saddle-point value of the auxiliary $\gamma$-field. As a result, due to Eqs. (E9)-(E10) and (E14), we then derive that $\delta\mathcal{F}(T)$ is evaluated near its extremum (i.e., at the saddle point) with

$$\|\gamma\| = \min(|a|) = \frac{|\zeta^*|\Lambda}{L_\tau} = \frac{m_0 c^2}{\hbar c},\ \text{as } T\to 0. \quad (E18)$$

Due to Eq. (4.30) where we have set that $|\zeta^*|\Lambda \approx X_i(x_i)$, $i = 1, 2$, we can write

$$\frac{|\zeta^*|\Lambda}{L_\tau} = \frac{X_i(x_i)}{L_\tau} = \frac{m_0 c^2}{\hbar c} = \frac{1}{\xi_\tau^*}. \quad (E19)$$

We directly retrieve the relation of Eq. (4.37) as well as the result of Chubukov *et al.* [10] obtained by the technique of a Lagrange multiplier. In the infinite volume system ($L_\tau \to +\infty$ i.e., at $T = 0$ K rigorously) $\|\gamma\|_\infty = 0$ i.e., $m_0 = 0$.

In a last step we have to express the infinitesimal element $\delta T/T$ in Eq. (E17). From the conservation of the modulus of the relativistic momentum $\boldsymbol{P} = (\hbar \boldsymbol{k}, \hbar\omega/c)$, for a fixed $k$-value at the equilibrium, we derive that $\hbar^2 k dk = -\delta|\zeta^*|\Lambda.|\zeta^*|\Lambda/L_\tau^2$. From Eq. (4.4) we have $\delta|\zeta^*| = \sqrt{1+z^{*2}}\delta|z^*|/|z^*|$ with, near the critical point, $\sqrt{1+z_c^{*2}} \approx 1$ as $z_c^* \ll 1$ so that $\delta|\zeta^*| \approx \delta T/T$ and $|\zeta_c^*|\Lambda \approx C$ due to Eq. (B17). As a result, we obtain the following relation

$$L_\tau^2 C^{-1} \hbar k d(\hbar k) = L_\tau^2 C^{-1} \varepsilon_k d\varepsilon_k \approx -\frac{dT}{T}, \text{ as } T \to 0 \quad (E20)$$

where $\varepsilon_k = \sqrt{(\hbar k)^2 + (m_0 c^2)^2}$ is the relativistic energy.

If introducing the critical coupling $g_c$ given by Eq. (4.53), in the thermodynamic limit, we find near $T_c = 0$ K

$$\frac{\mathcal{F}(T)}{8N^2} = \hbar c C \left[ \frac{1}{L_\tau} \int \frac{d^2 k}{(2\pi)^2} \ln(1 - \exp(-u_k)) \right.$$
$$\left. + \frac{1}{2} \int \frac{d^3(P/\hbar)}{(2\pi)^3} \ln\left(\frac{(P/\hbar)^2 + (m_0 c/\hbar)^2}{\Lambda^{-2}}\right) - \frac{(m_0 c/\hbar)^2 - 1}{2g} \right]$$
$$\text{as } N \to +\infty. \quad (E21)$$

with

$$\frac{d^3(P/\hbar)}{(2\pi)^3} = \frac{d^2 k}{(2\pi)^2} \frac{d(\omega/c)}{2\pi}, \quad u_k = \frac{L_\tau \varepsilon_k}{\hbar c} = \frac{\varepsilon_k}{k_B T}. \quad (E22)$$

### APPENDIX F: EXPRESSION OF THE SUSCEPTIBILITY DENSITY NEAR $T_c = 0$ K, FOR A COMPENSATED ANTIFERROMAGNET

Now we have to take into account the field contribution. The following treatment is exclusively valid for a compensated antiferromagnet. Indeed, as $\chi k_B T \sim \xi^2 M(T)^2$, near $T_c = 0$ K, $\chi k_B T$ diverges for a ferromagnet but vanishes for a compensated antiferromagnet. As a result, in this latter case, the elementary field-dependent surface free energy density $\delta \mathcal{F}(B,T)$ shows a minimum and the corresponding integral can be evaluated through a steepest descent method around a saddle point. $\chi$ can then be derived from $\mathcal{F}(B,T)$ through the adequate $B$-derivation.

As for $\exp(-\beta H^{ex})$ we must know express $\exp(-\beta H^{mag})$. We have

$$\exp(\beta GB \cos\theta) = \sqrt{4\pi} \sum_{l_B=0}^{+\infty} \sqrt{2l_B + 1} \lambda_{l_B}(\beta GB) Y_{l_B,0}^*(\boldsymbol{S}), \quad (F1)$$

with

$$\lambda_{l_B}(\beta GB) = \left(\frac{\pi}{2\beta GB}\right)^{1/2} I_{l_B+1/2}(\beta GB). \quad (F2)$$

and the corresponding eigenvalue of $\exp(-\beta H_{i,j})$, with $H_{i,j} = H_{i,j}^{ex} + H_{i,j}^{mag}$, becomes if $J = J_1 = J_2$

$$\lambda_l(-\beta J, B) = \lambda_l(-\beta J)\lambda_{l_B}(\beta GB). \quad (F3)$$

Near $T_c = 0$ K $l$ is replaced by $\Lambda = 2l \gg 1$. In the thermodynamic limit $l \sim l_B \to +\infty$ ($\Lambda \sim \Lambda_B \to +\infty$) so that we have to consider

$$\lambda_\Lambda(|z^*|\Lambda, B) = \left(\frac{\pi}{2|z^*|\Lambda}\right)^{1/2} I_\Lambda(|z^*|\Lambda) \left(\frac{\pi}{2z_B^*\Lambda}\right)^{1/2} I_\Lambda(z_B^*\Lambda),$$

$$z_B^* = \frac{\beta GB}{\Lambda} \quad (F4)$$

with $\lambda_\Lambda(|z^*|\Lambda, 0) \approx 2K_1 \cosh(|\zeta^*|\Lambda)$ where $K_1 \approx 1/z_c^{*1/2}\Lambda$. But, as we consider the vanishing-field limit for expressing the susceptibility, the argument $\beta GB$ must be considered as small with respect to $\beta|J|$. It also means that, in this limit exclusively, $\lambda_\Lambda(|z^*|\Lambda, B) = \lambda_\Lambda(\beta|J|, \beta GB)$ is the dominant eigenvalue in the same temperature range $\delta T_\Lambda$ where $\lambda_\Lambda(|z^*|\Lambda)$ is dominant so that the corresponding weight is $\delta T_\Lambda/T$.

Eq. (E6) has allowed to derive $\lambda_\Lambda(|z^*|\Lambda)$ in the double limit $|z^*|\Lambda = \beta|J| \to +\infty$, $\Lambda \to +\infty$. Thus we can write $\lambda_{l_B}(\beta GB) \approx \lambda_\Lambda(\beta GB) = \lambda_\Lambda(z_B^*\Lambda) \approx 2K_2 \cosh(z_B^*\Lambda)$ with $K_2 = 1/z_B^{*1/2}\Lambda$. As for $\lambda_\Lambda(|z^*|\Lambda)$ we similarly introduce the dimensionless argument for $\lambda_\Lambda(z_B^*\Lambda)$

$$\zeta_B = \sqrt{1 + z_B^{*2}} + \ln\left(\frac{|z_B^*|}{1 + \sqrt{1 + z_B^{*2}}}\right)$$

whose scaling form reduces to

$$|\zeta_B|\Lambda \approx z_B^*\Lambda = \beta GB \text{ as } \beta GB \ll 1.$$

As a result we can finally write

$$\lambda_\Lambda(|z^*|\Lambda, B) = 4K_1 K_2 \cosh(|\zeta^*|\Lambda) \cosh(\beta GB), \text{ as } T \to 0. \quad (F5)$$

If considering $\ln(\lambda_\Lambda(|z^*|\Lambda, B))$ the factor $\ln(4K_1 K_2)$ will disappear when deriving $\delta \mathcal{F}(B,T)$ with respect to $B$. We can then define an effective eigenvalue $\lambda_\Lambda(|z^*|\Lambda, B)_{eff} = \lambda_\Lambda(|z^*|\Lambda, B)/K_1 K_2$ with

$$\lambda_\Lambda(|z^*|\Lambda, B)_{\text{eff}} = 2[\cosh(|\zeta^*|\Lambda + \beta GB) + \cosh(|\zeta^*|\Lambda - \beta GB)] \quad (F6)$$

with $\beta GB \ll |\zeta^*|\Lambda$. Under these conditions $\ln(\lambda_\Lambda(|z^*|\Lambda, B))$ can be expanded owing to a work similar to that one achieved for $\lambda_\Lambda(|z^*|\Lambda)$ when $B = 0$ (cf Eqs. (E13)-(E15)). As a result we can similarly write for the mode $M = \pm\Lambda$

$$\ln(\lambda_{\pm\Lambda}(|\tilde{z}|\Lambda, B)) = \ln\left(\left(\frac{p^*_{\pm\Lambda}}{2\pi\hbar}\right)^2 + \left(\frac{a^*_{\pm\Lambda}(B)}{2\pi}\right)^2\right) \quad (F7)$$

with $p^*_{\varepsilon\Lambda}\Lambda = p_{\varepsilon\Lambda}$, $a^*_{\varepsilon\Lambda}(B)\Lambda = a_{\varepsilon\Lambda}(B)$ and

$$\left(\frac{a_{\varepsilon\Lambda}(B)}{2\pi}\right)^2 = \left(\frac{p_{\varepsilon\Lambda}}{2\pi\hbar}\right)^2 + \left(\frac{|\zeta^*|\Lambda + \varepsilon\beta GB}{2\pi L_\tau}\right)^2, \frac{\beta G}{L_\tau} = \frac{G}{\hbar c},$$
$$\varepsilon = \pm 1. \quad (F8)$$

In the thermodynamic limit, near $T_c = 0$ K, the elementary free energy density per lattice bond can be expressed as

$$\frac{\delta\mathcal{F}(T,B)}{8N^2} = -\hbar c C^2 \left(\frac{k_B T}{\hbar c}\right)^3 \lim_{M\to+\infty} \sum_{\Lambda=0}^{M} \ln(\lambda_\Lambda(|z^*|\Lambda), B) \frac{\delta T_\Lambda}{T},$$
$$\text{as } N \to +\infty \quad (F9)$$

i.e., owing to Eq. (F8)

$$\frac{\delta\mathcal{F}(T,B)}{8N^2} = -\hbar c C^2 \left(\frac{k_B T}{\hbar c}\right)^3 \sum_{\varepsilon=\pm 1}$$
$$\times \sum_{\omega_{\varepsilon\Lambda}=0}^{+\infty} \ln\left(\left(\frac{\omega_{\varepsilon\Lambda}}{2\pi}\right)^2 + \left(\frac{a^*_{\varepsilon\Lambda}(B)}{2\pi}\right)^2\right) \frac{\delta T_\Lambda}{T}, \text{ as } N \to +\infty \quad (F10)$$

where $C$ is given by Eq. (4.41) and

$$\frac{\omega_{\varepsilon\Lambda}}{2\pi} = \frac{p_{\varepsilon\Lambda}}{2\pi\hbar}, a_{\varepsilon\Lambda}(B) = \frac{\varepsilon_\Lambda(B)}{\hbar c},$$
$$a_\Lambda(B) = \frac{1}{2\pi}\left[k_\Lambda^2 + \left(\frac{|\zeta^*|\Lambda}{L_\tau} + \varepsilon\frac{G}{\hbar c}B\right)^2\right]^{1/2}. \quad (F11)$$

Near $T_c = 0$ K we deal with a continuum of eigenvalues of quasi equal weight $\delta T_\Lambda/T$ as previously explained. We then have the common limit $\delta T_\Lambda/T = \delta T/T$ as $\Lambda \to +\infty$.

If wishing to express the susceptibility, we have to derive $\mathcal{F}(T,B)$ in the vanishing $B$-limit. Then, if picking out the coefficient of the quartic term of the $B$-expansion and considering the minimum of $\mathcal{F}(T,B)$ near the saddle point, we derive the elementary susceptibility per surface unit near $T_c = 0$ K

$$\frac{\delta\chi_s}{8N^2} = -4\hbar c C^2 \left(\frac{G}{\hbar c}\right)^2 \left(\frac{k_B T}{\hbar c}\right)^3 \sum_{\omega_n=0}^{+\infty} \frac{\omega_n^2 - \varepsilon_k^2}{(\omega_n^2 + \varepsilon_k^2)^2} \frac{\delta T}{T},$$
$$\text{as } N \to +\infty \quad (F12)$$

after summing over $\varepsilon$. $\varepsilon_k$ is the zero-field relativistic energy such as

$$\varepsilon_k = \frac{1}{2\pi}\left[(\hbar k)^2 + (m_0 c^2)^2\right]^{1/2}, \frac{|\zeta^*|\Lambda}{L_\tau} = \frac{m_0 c^2}{\hbar c} \quad (F13)$$

with $|\zeta^*|\Lambda = X_i(x_i) = m_0 c^2/k_B T$ due to Eqs. (4.30) and (E20) as well as (cf Eq. (E20))

$$L_\tau^2 C^{-1} \hbar k d(\hbar k) = L_\tau^2 C^{-1} \varepsilon_k d\varepsilon_k \approx -\frac{dT}{T}, \text{ as } T \to 0. \quad (F14)$$

Finally, in the thermodynamic limit, we can write the susceptibility per surface unit (i.e., per surface of Kadanoff block) and per lattice bond

$$\frac{\chi_s}{8N^2} = 4C\left(\frac{G}{\hbar c}\right)^2 k_B T \sum_{\omega_n=0}^{+\infty} \int \frac{kdk}{2\pi} \frac{\varepsilon_k^2 - \omega_n^2}{(\varepsilon_k^2 + \omega_n^2)^2}, \text{as } N \to +\infty \quad (F15)$$

with $kdk = \varepsilon_k d\varepsilon_k$ and $\varepsilon_k \in [|\zeta^*|\Lambda/2\pi, +\infty[$ or equivalently due to Eq. (F13) $\varepsilon_k \in [m_0 c^2/2\pi k_B T, +\infty[$. This is the exact expression differently obtained by Chubukov et al. [10].

Mapple allows one to calculate the $\omega_n$-summation:

$$\sum_{\omega_n=0}^{+\infty} \frac{\varepsilon_k^2 - \omega_n^2}{(\omega_n^2 + \varepsilon_k^2)^2} = \frac{1}{2}\left[\frac{1}{\varepsilon_k^2} + \frac{\pi^2}{\sinh^2(\pi\varepsilon_k)}\right]. \quad (F16)$$

The integration vs $\varepsilon_k$ of the first part appearing in Eq. (F16) gives $[\ln(\varepsilon_k)]_{m_0 c^2/k_B T}^{+\infty} \to 0$ when $T \to 0$. It means that there is *no ultraviolet divergence*. The integration of the second part allows one to write:

$$\frac{\chi_s}{8N^2} = C\left(\frac{G}{\hbar c}\right)^2 \frac{k_B T}{\pi}\left[\frac{m_0 c^2}{k_B T} \frac{e^{m_0 c^2/k_B T}}{e^{m_0 c^2/k_B T} - 1} - \ln(e^{m_0 c^2/k_B T} - 1)\right]$$
$$\text{as } N \to +\infty. \quad (F17)$$